\documentclass[aps,prb,reprint,superscriptaddress,10pt,longbibliography]{revtex4-2}

\usepackage[main=english,dutch]{babel}

\makeatletter
\@ifundefined{l@en}{\let\l@en\l@english}{}
\@ifundefined{l@da}{\let\l@da\l@danish}{}
\@ifundefined{l@nl}{\let\l@nl\l@dutch}{}
\makeatother

\usepackage{amsmath,amssymb,mathtools}
\usepackage{bm}

\usepackage{graphicx}
\usepackage{xcolor}
\usepackage[caption=false]{subfig}

\usepackage{dcolumn}
\usepackage{booktabs}
\usepackage{tabularx}
\usepackage{makecell}
\usepackage{siunitx}
\newcolumntype{Y}{>{\centering\arraybackslash}X}

\usepackage{microtype}
\usepackage[section]{placeins}

\begin{document}
\title{Anatomy of Spin--Orbit Torques in Monolayer Fe$_3$GeTe$_2$ and Fe$_3$GaTe$_2$: Insights from atomistic and momentum-space decompositions}

\author{Gusthavo M. S. Brizolla}
\email{Gusthavo.Brizolla@physik.uni-regensburg.de}
\affiliation{Institute of Theoretical Physics,
             University of Regensburg, 93053 Regensburg, Germany}

\author{Stepan S. Tsirkin}
\affiliation{Institute of Physics,
             \'Ecole Polytechnique F\'ed\'erale de Lausanne (EPFL),
             CH-1015 Lausanne, Switzerland}

\author{Yaroslav Zhumagulov}
\affiliation{Institute of Physics,
             \'Ecole Polytechnique F\'ed\'erale de Lausanne (EPFL),
             CH-1015 Lausanne, Switzerland}

\author{Jaroslav Fabian}
\affiliation{Institute of Theoretical Physics,
             University of Regensburg, 93053 Regensburg, Germany}
\affiliation{\text{Halle-Berlin-Regensburg Cluster of Excellence CCE, University of Regensburg, 93040 Regensburg, Germany}}

\date{\today}

\begin{abstract}
We present a systematic first-principles study of the spin-orbit
torques in the ferromagnetic monolayers
Fe$_3$GeTe$_2$ (FGT) and Fe$_3$GaTe$_2$ (FGaT).
Despite sharing the same crystal structure (point group $D_{3h}$)
and predominantly Fe~$3d$ spin-polarized bands, the two materials exhibit markedly different current-induced torques. We reveal these differences by computing the full angular dependence of the torkance---the torque per unit applied electric field---using linear-response theory with symmetry-adapted spin--orbit-coupled Wannier functions. FGaT may be viewed as a hole-doped analogue of FGT, since Ga contributes one valence electron fewer than Ge. Although the work functions differ by only about $28$~meV, the band filling near $K$ and $K'$ changes substantially: the density of states at $\varepsilon_F$ is reduced by a factor of three and its spin polarization reverses from minority in FGT to majority in FGaT. These electronic changes are reflected in the torques resolved by time-reversal parity, sublattice, and momentum. In particular, we identify pronounced hidden torques in FGaT and relate the suppression of its fourth-harmonic Fermi-sea component to the evolution of momentum-space pockets. Finally, we discuss the emergence of such self-torques, which are not captured by the conventional picture of current-induced spin accumulation, within a symmetry-based phenomenological framework. Our results provide microscopic insight into current-induced torques in two-dimensional ferromagnets and offer guidance for defect and van der Waals engineering of layered magnetic materials.

\end{abstract}
 
\maketitle

\section{Introduction}
\label{sec:intro}

Electrical control of magnetization is a central goal of
spintronics~\cite{RevModPhys.76.323}. Particularly important is
current-induced magnetization switching, achieved first through the spin-transfer torque predicted by Slonczewski and
Berger~\cite{SlONCZEWSKI1996L1,PhysRevB.54.9353,PhysRevLett.84.3149}. An appealing alternative is the spin--orbit torque (SOT), which relies on spin--orbit coupling to induce magnetization dynamics: an electric current generates a spin accumulation~\cite{EDELSTEIN1990233}, which exerts a torque on the magnetization~\cite{RevModPhys.87.1213} and, above a critical current, switches it~\cite{mihai2010a,miron2011a,RevModPhys.91.035004}.

Van der Waals (vdW) magnets are natural hosts for SOT physics: they remain magnetically ordered down to the monolayer~\cite{huang2017a,gong2017a,Fei2018}, their carrier density is gate-tunable~\cite{deng2018a}, and their low crystal symmetry admits torque components forbidden in high-symmetry
metals~\cite{macneill2017a,kao2022a}. Among layered magnetic metals, Fe$_3$GeTe$_2$ (FGT) and Fe$_3$GaTe$_2$ (FGaT) are prominent, combining itinerant ferromagnetism with strong perpendicular anisotropy. The Curie temperature of FGT, $T_C\approx200$--$220$~K in the bulk, drops to ${\sim}70$--$130$~K in the monolayer~\cite{Fei2018,deng2018a,roemer2020a} but can be restored to room temperature by ionic gating in few-layer samples~\cite{deng2018a}, while FGaT retains room-temperature ferromagnetism down to a few layers~\cite{zhang2022a,doi:10.1021/acs.nanolett.4c01019}. SOT
switching has been demonstrated for both in heterostructures with heavy metals, topological insulators, and low-symmetry
semimetals~\cite{doi:10.1021/acs.nanolett.9b01043,doi:10.1126/sciadv.aaw8904,wang2023a,Kajale2024,SHARMA2025100847,zhang2025a,pandey2025a}.

Lacking inversion symmetry in monolayer form, both compounds support a \emph{self-torque}: an SOT generated by the spin--orbit fields of the material itself, with no interface required. However, the crystal symmetry makes their self-torque unconventional. Indeed, the horizontal mirror forbids the magnetization-independent Rashba-Edelstein field induced by electric current, so the leading
spin polarization must be linear in the in-plane magnetization (which breaks the horizontal mirror symmetry) and acts as a current-induced in-plane anisotropy~\cite{PhysRevLett.122.217203}. 
In centrosymmetric bulk FGT and FGaT, the corresponding fields remain ``hidden'', alternating from
layer to layer and canceling in the net response while producing layer-resolved torques~\cite{Martin02012023,PhysRevResearch.4.L042022} that trace back to the local symmetry of each magnetic site~\cite{PhysRevLett.113.157201,PhysRevB.105.064412,PhysRevB.104.224414}.

First-principles calculations for monolayer FGT have mapped the
angular dependence of the torque for both time-reversal-even and time-reversal-odd channels, 
revealing harmonics well beyond the leading fieldlike and dampinglike forms---with direct consequences for field-free switching---and coefficients varying strongly with the chemical potential~\cite{PhysRevB.108.144422}. However, the microscopic origin of this filling dependence remains
unresolved. Which electronic states near the Fermi level $\varepsilon_F$
carry the dominant torque contributions? Why are the harmonic components
so sensitive to band filling? Can such changes be realized intrinsically
through stoichiometric modification rather than electrostatic gating?
More broadly, do inversion-asymmetric monolayers host hidden torques that
are not layer-resolved, as in centrosymmetric systems, but instead emerge
from the site or sublattice degrees of freedom within a single layer?

In this work, we address these questions through a comprehensive first-principles
and symmetry analysis of the SOTs in monolayer FGT and FGaT. These two
ferromagnets form a controlled comparison: they share the same crystal
structure with point group $D_{3h}$, and therefore the same symmetry-allowed
torque forms, as well as the same dominant Fe~$3d$ character at
$\varepsilon_F$. However, they differ by one valence electron, with FGaT
effectively representing a hole-doped analogue of FGT. Any qualitative
difference in their torques can therefore be traced primarily to the change
in band filling.

Specifically, we compute the angular dependence of the intrinsic SOTs of both
monolayers using Kubo--Bastin linear response~\cite{PhysRevB.90.174423}
with symmetry-adapted Wannier functions constructed from fully relativistic
density-functional theory. Thus, spin--orbit coupling is incorporated from
the outset, rather than introduced as an additional term during
Wannierization. We employ a rigid rotation of the exchange field, benchmarked
against self-consistent noncollinear calculations, to enable dense angular
sampling of the torque landscape.

The electronic structures of FGT and FGaT are qualitatively similar, but the
hole doping in FGaT leads to reduced band filling, most prominently around
the $K$ and $K'$ points. Individual band features are shifted by up to
$\sim100$~meV between the two compounds, exceeding the difference in work
functions of only $\sim30$~meV. These energy scales are accessible by
electrostatic or ionic gating~\cite{SHARMA2025100847,zhang2025a}. The two
materials also exhibit comparable valley-Zeeman splittings of about
$40$~meV.

Despite these electronic differences, the overall torque magnitudes remain
similar. However, the pronounced fourth-harmonic contribution to the
time-reversal-even torque in FGT is strongly reduced in FGaT, which we trace
to the suppression of the $K$-point Fermi pockets. In contrast, the
time-reversal-odd torques are largely transferable between the two
compounds. A particularly promising route for torque engineering emerges
from the hidden Rashba--Edelstein-like torques associated with the two Fe
sublattices related by the horizontal mirror symmetry. In FGaT, these
site-resolved torques can exceed the symmetry-allowed components by nearly
a factor of 30. Although they cancel exactly in the ideal monolayer, breaking
the sublattice symmetry provides a route to activate these large hidden
contributions.

We organize the paper as follows. In Sec.~\ref{sec:methods} we present 
the computational framework, including
the first-principles setup, symmetry-adapted Wannier construction,
Kubo--Bastin torkance formalism, and rigid-rotation approach.
Section~\ref{sec:results} compares the electronic structures of 
the two monolayers (Sec.~\ref{sec:symmetry_bands})
and then analyzes their angular-dependent torkances
(Sec.~\ref{sec:angular_dependence}), including 
atom- and momentum-resolved contributions. 
Section~\ref{sec:discussion} discusses the symmetry-based phenomenology
of the torques, followed by the conclusions in Sec.~\ref{sec:conclusion}. 
We leave technical details, including benchmarking, to the Appendixes.

\section{Computational methods}
\label{sec:methods}
\subsection{Density-functional theory}
\label{sec:dft}
\begin{figure}[tbp]
  \centering
  \includegraphics[width=\columnwidth]{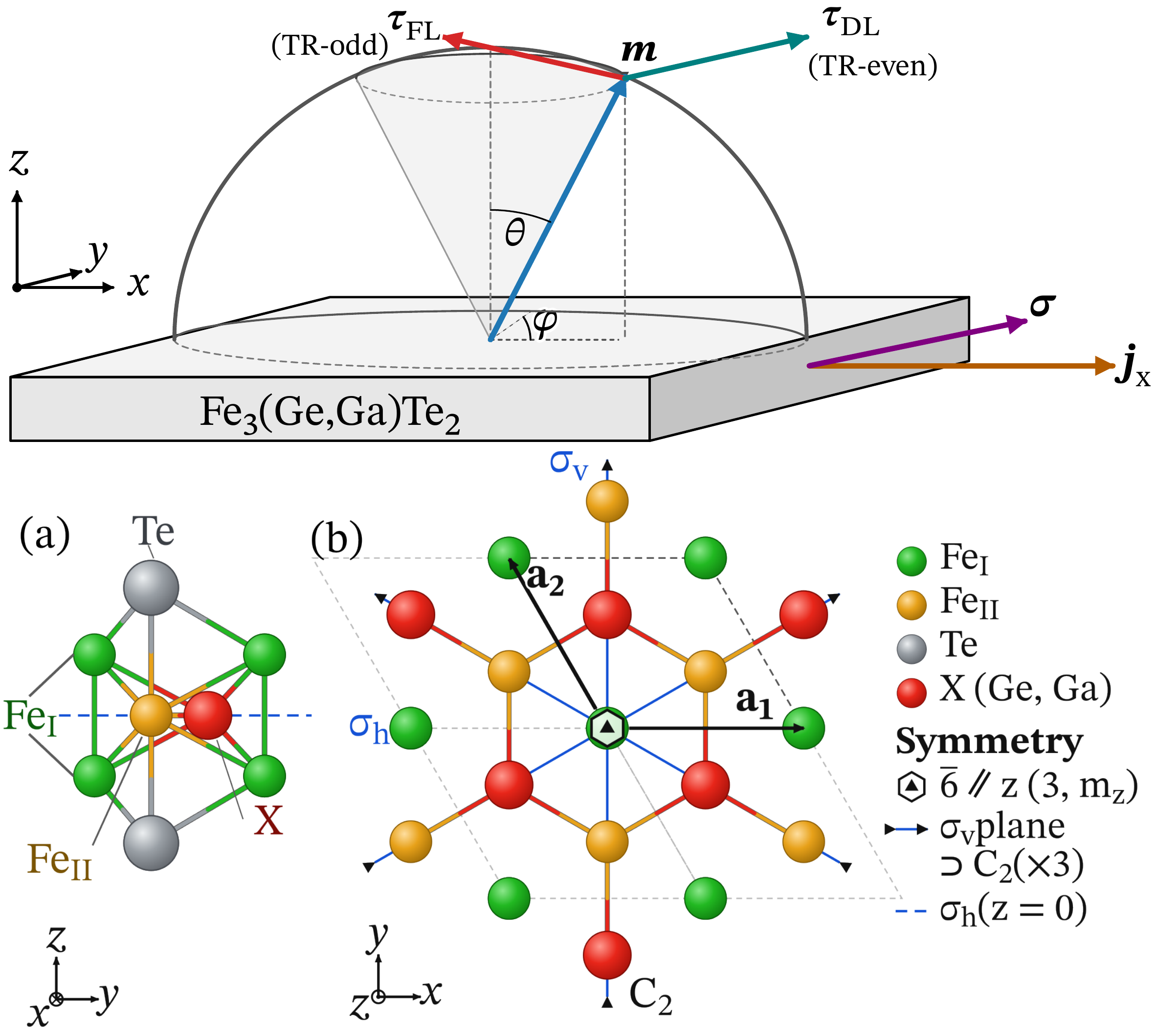}
\caption{\label{fig:structure} Geometry and crystal structure.
\textit{Top:} response geometry. An in-plane field
$\mathbf{E}\parallel\hat{\mathbf{x}}$ drives a current $j_x$; the
magnetization direction $\hat{\mathbf{m}}$ is parametrized by
spherical angles $(\theta,\phi)$. The axis
$\hat{\boldsymbol{\sigma}}\equiv\hat{\mathbf{z}}\times\hat{\mathbf{E}}$
and the conventional bilayer torques---fieldlike
$\boldsymbol{\tau}_{\mathrm{FL}}\parallel\hat{\mathbf{m}}\times\hat{\boldsymbol{\sigma}}$
(time-reversal odd) and dampinglike
$\boldsymbol{\tau}_{\mathrm{DL}}\parallel\hat{\mathbf{m}}\times(\hat{\boldsymbol{\sigma}}\times\hat{\mathbf{m}})$
(time-reversal even)---are drawn only to fix axes and nomenclature: in
the $D_{3h}$ monolayer both rigid forms are symmetry-forbidden, and the
leading self-torque is magnetization-dependent
(Sec.~\ref{sec:symmetry_bands}).
\textit{Bottom:} (a) side and (b) top views of Fe$_3X$Te$_2$, with the
inequivalent sites Fe$_{\mathrm{I}}$ (green, outer planes) and
Fe$_{\mathrm{II}}$ (amber, central plane). The nonmagnetic space group
is $P\bar{6}m2$ (No.~187, point group $D_{3h}$); panel (b) marks the
$\bar{6}$ axis and the three $\sigma_v$ mirrors, whose traces contain
the in-plane $C_2$ axes, and panel (a) the horizontal mirror $\sigma_h$
(dashed).}
\end{figure}

First-principles calculations were performed with \textsc{Quantum
ESPRESSO}~\cite{Giannozzi_2009,Giannozzi_2017,10.1063/5.0005082} using
fully relativistic optimized norm-conserving Vanderbilt (ONCV)
pseudopotentials~\cite{PhysRevB.88.085117} from the \textsc{PseudoDojo}
library~\cite{VANSETTEN201839}. Spin--orbit coupling (SOC) was treated
self-consistently in the noncollinear framework, with the magnetization
initialized along $\hat{\mathbf{z}}$. The response geometry, crystal
structure, unit cell, and point symmetries ($D_{3h}$) of monolayer
Fe$_3X$Te$_2$ ($X=$~Ge, Ga) are sketched in
Fig.~\ref{fig:structure}. The monolayer is a five-plane stack
Te--Fe$_{\mathrm{I}}$--(Fe$_{\mathrm{II}}$,\,$X$)--Fe$_{\mathrm{I}}$--Te
with two inequivalent iron sites: the $\sigma_h$-related pair
Fe$_{\mathrm{I}}$ in the outer planes, and the single Fe$_{\mathrm{II}}$
sharing the central plane with $X$. This
Fe$_{\mathrm{I}}$/Fe$_{\mathrm{II}}$ partition underlies the
atom-resolved analysis of the SOT (Sec.~\ref{sec:layer3}). Periodic
images were separated by 30~\AA\ of vacuum, and spurious inter-image
Coulomb interactions were removed by a two-dimensional
cutoff~\cite{PhysRevB.73.233103,PhysRevB.96.075448}.

The in-plane lattice constant and internal coordinates were relaxed
within the PBE-GGA~\cite{PhysRevLett.77.3865} (residual forces
$<10^{-5}$~Ry/Bohr, total-energy changes $<10^{-5}$~Ry), giving
$a=3.95$~\AA\ (FGT) and $a=3.88$~\AA\ (FGaT), $1.0\%$ and $2.7\%$ below
the bulk experimental values of $3.991$ and $3.986$~\AA,
respectively~\cite{https://doi.org/10.1002/ejic.200501020,PhysRevB.93.014411,wu2024a}. All subsequent electronic-structure and response calculations used the Perdew--Wang LDA~\cite{PhysRevB.45.13244} at this geometry: for bulk FGT, LDA yields Fe moments within the experimental range, whereas PBE
overestimates them by ${\sim}30$--$40\%$~\cite{PhysRevB.93.134407,PhysRevB.93.144404}, and LDA was likewise the choice of the previous first-principles SOT study of monolayer FGT~\cite{PhysRevB.108.144422}.

Plane waves were expanded to a kinetic-energy cutoff of 120~Ry. The
self-consistent cycle used a $24\times24\times1$ $\mathbf{k}$ mesh,
first-order Methfessel--Paxton smearing~\cite{PhysRevB.40.3616} of
width $0.02$~Ry, and a self-consistency threshold of $10^{-8}$~Ry; the
same mesh provides the coarse grid for the Wannier construction of
Sec.~\ref{sec:wannier}.

\subsection{Wannier interpolation}
\label{sec:wannier}
Brillouin-zone integrals on the dense $\mathbf{k}$ meshes required for
linear response were evaluated by Wannier interpolation, using
symmetry-adapted Wannier functions (SAWFs)~\cite{PhysRevB.87.235109}
built in \textsc{WannierBerri}~\cite{tsirkin2021a}, which combines
Marzari--Vanderbilt localization~\cite{PhysRevB.56.12847} and
Souza--Marzari--Vanderbilt disentanglement~\cite{PhysRevB.65.035109}
with symmetry constraints on the projectors. The SAWFs are constructed
directly from the spinor Bloch states of Sec.~\ref{sec:dft}, in which
SOC is included self-consistently from the outset; the interpolated
model therefore inherits the \emph{ab initio} spin--orbital
entanglement rather than acquiring SOC as an added term. The
construction enforces the full magnetic space group of the
$\hat{\mathbf{m}}\parallel\hat{\mathbf{z}}$ ground state,
$P\bar{6}m'2'$: the symmetry elements of Fig.~\ref{fig:structure}(b),
with the $\sigma_v$ mirrors and the in-plane $C_2$ axes entering only
in combination with time reversal. Irreducible representations and
multiplicities were verified with \textsc{IrRep}~\cite{IRAOLA2022108226}.

Each monolayer is described by $N_w=48$ spinor functions ($24$ spatial
orbitals $\times\,2$ spin), projected onto Fe~$3d$, $X$~$p$, and
Te~$p$ orbitals respecting the site symmetries of the corresponding
Wyckoff positions. This set spans the states within several eV of the
Fermi level $\varepsilon_F$: the exchange splitting resides in the
Fe~$3d$ shell, the strongest atomic SOC on the heavy Te atoms. The
$s$-derived bands (Te~$5s$, $X$~$s$) lie well below this $p$--$d$
manifold, separated by a gap, and contribute negligibly to the
torkance. The Wannier subspace is extracted by disentanglement within
an outer window of $[-8.0,\,+5.45]$~eV about $\varepsilon_F$ for FGT
and $[-6.5,\,+6.7]$~eV for FGaT, with bands frozen up to $+1.5$ and
$+1.0$~eV, respectively: the interpolated bands coincide with the
\emph{ab initio} ones throughout the occupied $p$--$d$ manifold and to
at least $1$~eV above $\varepsilon_F$.
Appendix~\ref{app:hodd_soc_contamination} quantifies both the
sensitivity to the Wannierization mesh and the residual symmetry
breaking introduced by the rigid rotation of the exchange field used
to sample magnetization directions away from the ground state; the
interpolated operators recover the exact symmetry-protected zeros of
the torkance at $\hat{\mathbf{m}}\parallel\pm\hat{\mathbf{z}}$
(Appendix~\ref{app:rigid_validation}).

The converged Wannier charge centers follow the pattern dictated by
site symmetry: for each symmetry-related set of functions the centroid
obeys the site group exactly, pinned to the nucleus on the $\bar{6}m2$
sites (Fe$_{\mathrm{II}}$, $X$) and free to shift only along
$\hat{\mathbf{z}}$ on the trigonal-axis sites (Fe$_{\mathrm{I}}$, Te),
where it moves by at most $41$~m\AA\ (${\approx}1\%$ of the in-plane
lattice constant $a$). Individual functions within a set may displace
further while preserving the set centroid: the $X$~$p_z$-derived pair
splits symmetrically by $\pm73$~m\AA\ about the nucleus, and the
largest single-function deviation is an outward shift of the
Te~$p_z$-like functions into the vacuum, ${\approx}0.2$~\AA\
(${\approx}5\%$ of $a$), identical in the two compounds. Every Wannier
function thus remains essentially attached to its parent atom, as
required by the sublattice projectors of
Appendix~\ref{app:operators_resolved}. The same basis carries all
operators entering the linear response --- the Hamiltonian, the
velocity including its Berry-connection contribution, and the spin
operator from which the exchange torque is built --- as detailed in
Appendixes~\ref{app:wannier} and~\ref{app:operators}.

\subsection{Torkance: Kubo formulas}
\label{sec:LR}
A dc electric field $\mathbf{E}$ applied in the plane of the monolayer
drives the electronic system into a current-carrying state that exerts
a net torque $\boldsymbol{\tau}=(\tau_x,\tau_y,\tau_z)$ on the
magnetization---the nonequilibrium expectation,
$\boldsymbol{\tau}=\langle\hat{\boldsymbol{\tau}}\rangle$, expectation value of the
exchange-torque operator $\hat{\boldsymbol{\tau}}$ defined below. In what follows, the unhatted $\boldsymbol{\tau}$ and its components
$\tau_a$ denote this macroscopic torque, whereas the microscopic
operator, $\hat{\tau}_a$, and its band-basis matrix elements,
$\hat{\tau}^{a}_{nm}$, carry a hat. To linear order in the field,
\begin{equation}
  \tau_a(\hat{\mathbf{m}})
  = \sum_b t_{ab}(\hat{\mathbf{m}})\,E_b,
  \qquad a \in \{x, y, z\},
  \quad b \in \{x, y\},
\label{eq:torkance_def}
\end{equation}
which defines the torkance tensor $t_{ab}$, the torque per unit
applied electric field.
We split $t_{ab}$ into parts even and odd under magnetization reversal
$\hat{\mathbf{m}}\to-\hat{\mathbf{m}}$, following the Kubo--Bastin
formulation of Freimuth~\emph{et~al.}~\cite{PhysRevB.90.174423},
evaluated with a constant band broadening $\Gamma$ in the Green's
functions. With the shorthands
$x_n=\varepsilon_{n\mathbf{k}}-\varepsilon_F$ and
$g_n=x_n^2+\Gamma^2$, where $\varepsilon_{n\mathbf{k}}$ are the band
energies, the $T=0$ even (Fermi-sea) torkance is
\begin{align}
  t^{\mathrm{even}}_{ab}
  &= \frac{e\hbar}{2\pi N_k}\sum_{\mathbf{k}}\sum_{n \neq m}
    \mathrm{Im}\!\bigl[\hat{\tau}^{a}_{nm}\,v^{b}_{mn}\bigr]
    \nonumber\\
  &\quad\times
    \biggl[
      \frac{\Gamma\,(x_m - x_n)}{g_n\,g_m}
    + \frac{2\Gamma}{(x_n - x_m)\,g_m}
    \nonumber\\
  &\quad\phantom{\times\biggl[}
    + \frac{2}{(x_n - x_m)^2}\,
      \mathrm{Im}\ln\!\frac{x_m - i\Gamma}{x_n - i\Gamma}
    \biggr],
\label{eq:tork_even_main}
\end{align}
and the odd (Fermi-surface) torkance is
\begin{equation}
  t^{\mathrm{odd}}_{ab}
  = \frac{e\hbar}{\pi N_k}\sum_{\mathbf{k}}\sum_{n,m}
    \frac{\Gamma^2\,
      \mathrm{Re}\!\bigl[\hat{\tau}^{a}_{nm}\,v^{b}_{mn}\bigr]}
         {g_n\,g_m}.
\label{eq:tork_odd_main}
\end{equation}
Here $N_k$ is the number of $\mathbf{k}$ points, and
$v^{b}_{nm}=\langle n\mathbf{k}|\hat{v}_b|m\mathbf{k}\rangle$ and
$\hat{\tau}^{a}_{nm}=\langle n\mathbf{k}|\hat{\tau}_a|m\mathbf{k}\rangle$
are the matrix elements of the velocity and torque operators in the
band basis $|m\mathbf{k}\rangle$.
The odd sum runs over all ordered band pairs, including $n=m$; the
ordered $n\neq m$ sum in Eq.~\eqref{eq:tork_even_main} retains only
the part of the bracket antisymmetric in $n\leftrightarrow m$,
matching the antisymmetry of
$\mathrm{Im}[\hat{\tau}^{a}_{nm}v^{b}_{mn}]$, so the second term is
unambiguous despite its asymmetric form.

The torque operator $\hat{\tau}_a$ is the exchange torque: the
commutator of the exchange field---the time-reversal-odd part of the Hamiltonian, which carries the dependence on the magnetization direction---with the spin operator, both in Wannier-interpolated form (Appendix~\ref{app:operators_torque}). Decomposing the interpolated Hamiltonian into its two time-reversal components makes the separation explicit [Fig.~\ref{fig:bands_decomposition_FGT_FGaT}]: the odd part has a spectrum centered on zero and confined to the exchange scale, narrow compared with the full $p$--$d$ dispersion, and carries no filling of its own.
Spin--orbit coupling adds to this exchange field small spin-mixing terms that do not transform as a rigid rotation of the magnetization; they are removed from the torque operator by a spin projection (Appendix~\ref{app:operators_ssproj}). The projection enters the torque operator only: band energies, states,
and velocities are those of the full, unprojected Hamiltonian at every magnetization direction (Sec.~\ref{sec:rotation}). The velocity operator contains, besides the band-dispersion (Peierls)
term, the Berry-connection contribution required for gauge covariance of the response (Appendix~\ref{app:velocity}). The torkance is expressed in units of $e\,a_0$.

In a generic magnet the leading harmonics of the torkance---those
generated by a rigid, magnetization-independent effective field---are
the conventional dampinglike ($\boldsymbol{\tau}_{\mathrm{DL}}$) and
fieldlike ($\boldsymbol{\tau}_{\mathrm{FL}}$) forms of
Fig.~\ref{fig:structure}. In the $D_{3h}$ monolayer this shell is symmetry-empty---no such rigid field exists---so every allowed harmonic is intrinsically $\hat{\mathbf{m}}$-dependent and the angular dependence carries higher harmonics (Sec.~\ref{sec:symmetry_bands})~\cite{PhysRevLett.122.217203,PhysRevB.108.144422}. We therefore use even and odd---the time-reversal parity---as the
primary channel labels, reserving fieldlike and dampinglike for the
dynamical action.

As the broadening $\Gamma\to 0$, $t^{\mathrm{even}}$ approaches its
intrinsic interband (Berry-curvature--type) value while
$t^{\mathrm{odd}}$ scales as $1/\Gamma$, the hallmark of a
scattering-limited Fermi-surface response (Appendix~\ref{app:kubo}).
The results reported below use $\Gamma=10$~meV, $T=0$~K, and an
$800\times800$ Monkhorst--Pack $\mathbf{k}$ mesh, whose density is
dictated by the Fermi-surface Lorentzians of width $\Gamma$ that
dominate $t^{\mathrm{odd}}$; at this broadening, all torkances
reported below are converged with respect to the mesh.

\subsection{Magnetization rotation and angular dependence}
\label{sec:rotation}
The magnetization orientation is parametrized by spherical angles as
$\hat{\mathbf{m}}(\theta,\phi)=(\sin\theta\cos\phi,\sin\theta\sin\phi,\cos\theta)$,
with the polar angle $\theta$ measured from $\hat{\mathbf{z}}$ and the
azimuthal angle $\phi$ from $\hat{\mathbf{x}}$; the field is fixed
along $\mathbf{E}\parallel\hat{\mathbf{x}}$, as indicated in
Fig.~\ref{fig:structure}. Because SOC ties the electronic structure to
the magnetization direction, the angular dependence of
Eq.~\eqref{eq:torkance_def} would in principle require a separate
self-consistent calculation and Wannierization for every orientation,
which is impractical for the dense angular grids used below. One route
computes the electronic structure without SOC and adds it a posteriori
as an atomic term in the Wannier basis \cite{PhysRevB.108.144422}. We, instead, retain the self-consistently coupled SOC of Sec.~\ref{sec:dft} and rotate only the exchange part of the Hamiltonian.

In this rigid-exchange-rotation scheme the \emph{ab initio}
Hamiltonian is computed once, for
$\hat{\mathbf{m}}\parallel\hat{\mathbf{z}}$, and split into its
time-reversal-even (TR-even) and time-reversal-odd (TR-odd) parts: the
former collects the orbital and spin--orbit terms, the latter is the
exchange field of Sec.~\ref{sec:LR}. A rotated magnetization
$\hat{\mathbf{m}}$ is described by the spinor transformation
\begin{equation}
  H_{\mathrm{odd}}(\mathbf{k};\hat{\mathbf{m}})
  = U_R\,H_{\mathrm{odd}}(\mathbf{k};\hat{\mathbf{z}})\,U_R^{\dagger},
\label{eq:rotation_main}
\end{equation}
where $U_R$ rotates the spinor index by $\theta$ about
$\hat{\mathbf{n}}=(\hat{\mathbf{z}}\times\hat{\mathbf{m}})/|\hat{\mathbf{z}}\times\hat{\mathbf{m}}|$---the
SU(2) rotation that maps $\hat{\mathbf{z}}\to\hat{\mathbf{m}}$---and
acts as the identity on the orbital index; the TR-even Hamiltonian is
left unchanged. The TR-odd contributions to the velocity---Peierls and
Berry-connection---are rotated by the same $U_R$, while the TR-even
velocity is not (Appendixes~\ref{app:operators_ssproj},
\ref{app:operators_velocity_rotation}). A single Wannierization thus
serves all orientations, which avoids gauge mismatches between
independent constructions and yields $t_{ab}(\theta,\phi)$ efficiently
on the dense $\mathbf{k}$ meshes of Sec.~\ref{sec:LR}.

Exact at the reference orientation by construction, the scheme is
benchmarked away from it, point by point along a $yz$-plane sweep of
the magnetization, against fully self-consistent noncollinear DFT: the
two agree to within ${\sim}5\%$ over most of the sweep, with the
largest deviation, ${\sim}17\%$, confined to $t^{\mathrm{even}}_{zx}$
at $\hat{\mathbf{m}}\parallel\pm\hat{\mathbf{y}}$. The deviation
reflects the residual SOC-induced dependence of the Hamiltonian on
$\hat{\mathbf{m}}$ that a rigid rotation of its exchange part cannot
capture---an error that grows with the SOC strength, so the scheme
should be applied with care to more strongly spin--orbit-coupled
magnets (Appendix~\ref{app:rigid_validation}).

\section{Results}\label{sec:results}
\subsection{Electronic structure of monolayer Fe$_3$GeTe$_2$ and
Fe$_3$GaTe$_2$}
\label{sec:symmetry_bands}
Monolayers Fe$_3X$Te$_2$ are noncentrosymmetric hexagonal layers
cleaved from the centrosymmetric $P6_3/mmc$
bulk~\cite{deng2018a,zhang2022a}. The point group is $D_{3h}$
($\bar{6}m2$), generated by the roto-reflection axis $\bar{6}$ along $\hat{\mathbf{z}}$, which comprises the threefold rotation $C_3$ and the horizontal mirror $\sigma_h$ (the layer plane), together with three vertical mirrors $\sigma_v$, each containing an in-plane twofold axis $C_2$ along $\langle 210\rangle$, i.e., through Fe$_{\mathrm{II}}$ and the in-plane projections of the Fe$_{\mathrm{I}}$ and $X$ sites [Fig.~\ref{fig:structure}(b)]. The Fe$_\mathrm{I}$--Fe$_\mathrm{I}$ dumbbells stand perpendicular to the
central Fe$_\mathrm{II}$--$X$ plane [Fig.~\ref{fig:structure}(a)], and the broken inversion makes a self-torque symmetry-allowed under an in-plane electric field.

\begin{figure*}[tbp]
  \centering
  \includegraphics[width=\textwidth]{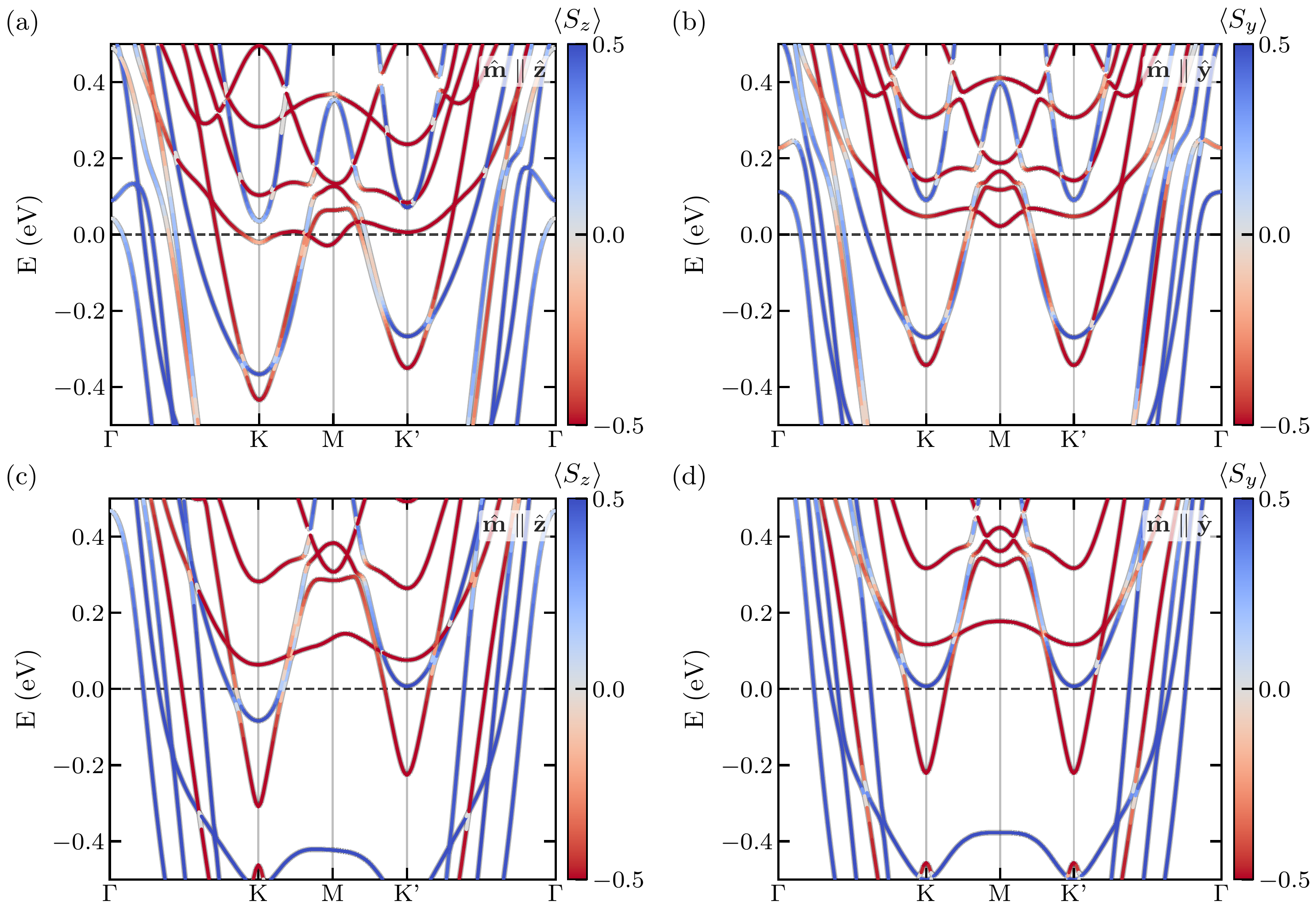}
  \caption{Spin-projected band structures along
$\Gamma$--K--M--K$'$--$\Gamma$ for Fe$_3$GeTe$_2$ with
$\hat{\mathbf{m}}\parallel\hat{\mathbf{z}}$~(a) and
$\hat{\mathbf{m}}\parallel\hat{\mathbf{y}}$~(b), and for Fe$_3$GaTe$_2$
with $\hat{\mathbf{m}}\parallel\hat{\mathbf{z}}$~(c) and
$\hat{\mathbf{m}}\parallel\hat{\mathbf{y}}$~(d), obtained
self-consistently for each magnetization orientation with the fully
relativistic LDA+SOC setup of Sec.~\ref{sec:dft}.
Color encodes the spin expectation value projected onto
$\hat{\mathbf{m}}$: $\langle S_z\rangle$ for
$\hat{\mathbf{m}}\parallel\hat{\mathbf{z}}$, $\langle S_y\rangle$ for
$\hat{\mathbf{m}}\parallel\hat{\mathbf{y}}$ (in units of $\hbar$).
Energies in each panel are referenced to the Fermi level of the
respective compound (horizontal dashed line, $\varepsilon_F=0$.)}
  \label{fig:bands}
\end{figure*}

Spin--orbit torque relies on the interplay of magnetic exchange and SOC and we now establish how the two monolayers differ in this respect. They share the point group $D_{3h}$, so the symmetry-allowed structure of their electronic and magnetic responses is identical, and the difference is band filling: replacing Ge with Ga removes one valence electron, hole-doping the monolayer. Referenced to the vacuum level, the calculated Fermi level of FGaT lies ${\sim}28$~meV below that of FGT for $\hat{\mathbf{m}}\parallel\hat{\mathbf{z}}$, so its work function is $28$~meV larger. More consequential than this global offset are the accompanying shifts of individual band
features relative to $\varepsilon_F$, typically $30$--$130$~meV.
Figure~\ref{fig:bands} shows the fully relativistic band structures for out-of-plane and in-plane magnetization, colored by the spin projection onto $\hat{\mathbf{m}}$; comparing panels (a) and (c), the minority-spin pocket at K rises from $-0.44$~eV in FGT to $-0.31$~eV in FGaT. The spin polarization of the density of states at $\varepsilon_F$ is likewise opposite, minority in FGT and majority in FGaT(Appendix~\ref{app:dos_valleys}).

The dependence of the bands on the magnetization \emph{direction} is entirely due to SOC. The absence of space inversion together with $\sigma_h$ induces out-of-plane valley-Zeeman (Ising) spin--orbit fields of opposite sign at K and K$'$. For $\hat{\mathbf{m}}\parallel\hat{\mathbf{z}}$ these add to the exchange with opposite signs in the two valleys, making K and K$'$ inequivalent [panels (a),(c)]. Rotating $\hat{\mathbf{m}}$ into the plane restores a twofold rotation that exchanges the valleys, and the corner spectra equalize [panels (b),(d)]. Half the K--K$'$ splitting of the deep minority pocket gives a valley-Zeeman coupling of $42$~meV in both compounds, with neighboring deep corner bands yielding $40$--$50$~meV. Their near identity across the substitution
shows the Ising fields to be set by the Te environment, which the Ge${\to}$Ga replacement leaves intact, in contrast to the filling shifts of the band features themselves. The in-plane orientation also breaks $\sigma_h$ and lifts its parity protection, so band crossings present at $\hat{\mathbf{m}}\parallel\hat{\mathbf{z}}$ open into avoided crossings.

The calculated LDA ground state reproduces the experimental
site-moment hierarchy and the out-of-plane easy axis of mono- and few-layer FGT~\cite{PhysRevB.93.014411,ghosh2023a}. At
$\hat{\mathbf{m}}\parallel\hat{\mathbf{z}}$ the FGT moments are
$2.155\,\mu_B$ on each Fe$_\mathrm{I}$ and $1.253\,\mu_B$ on
Fe$_\mathrm{II}$, an average of $1.854\,\mu_B$/Fe against the
experimental $1.62$--$1.70\,\mu_B$/Fe~\cite{doi:10.7566/JPSJ.82.124711,PhysRevB.93.144404}, a smaller overestimate than PBE. In FGaT the Fe$_\mathrm{I}$ moment decreases to $1.996\,\mu_B$ and Fe$_\mathrm{II}$ increases to $1.312\,\mu_B$, close to the monolayer LDA values of Ruiz \emph{et~al.}~\cite{doi:10.1021/acs.nanolett.4c01019}; $X$ and Te carry small antiparallel moments, below $0.05\,\mu_B$ throughout. Rotating $\hat{\mathbf{m}}$ into the plane changes the moments by ${\sim}0.1\%$ on average and the chemical potential by $<0.1$~meV in FGT but ${\approx}6$~meV in FGaT, reflecting the stronger sensitivity of the FGaT corner pockets to the magnetization direction.

\subsection{Angular dependence of the torkances in Fe$_3$GeTe$_2$ and
Fe$_3$GaTe$_2$}
\label{sec:angular_dependence}
We now present the main results of this work: the full angular
dependence of the torkances, sampled over the magnetization sphere with the rigid-exchange-rotation scheme of Sec.~\ref{sec:rotation} applied to the Wannier models of Sec.~\ref{sec:wannier} and the Kubo formulas of Sec.~\ref{sec:LR}. We first fix the notation used to extract the angular components. With the field direction fixed, $\mathbf{E}=E_x\hat{\mathbf{x}}$ (Fig.~\ref{fig:structure}), the torkance is a single vector field on the magnetization sphere,
\begin{equation}
\boldsymbol{\tau}(\hat{\mathbf{m}})=E_x\,\mathbf{t}_x(\hat{\mathbf{m}}),
\qquad
\mathbf{t}_x\equiv\bigl(t_{xx},\,t_{yx},\,t_{zx}\bigr),
\label{eq:tx_definition}
\end{equation}
everywhere tangent to it, since $\boldsymbol{\tau}\perp\hat{\mathbf{m}}$. Each element decomposes as $t_{ab}=t_{ab}^{\mathrm{even}}+t_{ab}^{\mathrm{odd}}$ into the Fermi-sea (TR-even) and Fermi-surface (TR-odd) channels of
Eqs.~\eqref{eq:tork_even_main} and~\eqref{eq:tork_odd_main}. We
further project $\mathbf{t}_x$ onto the tangent frame of the sphere, spanned by the polar and azimuthal unit vectors
$\hat{\mathbf{e}}_\theta=\partial_\theta\hat{\mathbf{m}}$ and
$\hat{\mathbf{e}}_\phi = (\sin\theta)^{-1}\partial_\phi\hat{\mathbf{m}}$ of the parametrization $\hat{\mathbf{m}}(\theta,\phi)$ of Sec.~\ref{sec:rotation}:
\begin{equation}
\label{eq:proj_general}
t_\theta=\mathbf{t}_x\cdot\hat{\mathbf{e}}_\theta,
\qquad
t_\phi=\mathbf{t}_x\cdot\hat{\mathbf{e}}_\phi.
\end{equation}
The frame is undefined at the poles
$\hat{\mathbf{m}}=\pm\hat{\mathbf{z}}$, but the torkance vanishes there exactly. Finally, because $\hat{\mathbf{e}}_\theta\to+\hat{\mathbf{e}}_\theta$ and
$\hat{\mathbf{e}}_\phi\to-\hat{\mathbf{e}}_\phi$ under reversal
$(\theta,\phi)\to(\pi-\theta,\phi+\pi)$, the superscripts in
$t_{\theta,\phi}^{\mathrm{even/odd}}
=\mathbf{t}_x^{\mathrm{even/odd}}\cdot\hat{\mathbf{e}}_{\theta,\phi}$
refer to the Cartesian channels, not to the parity of the scalar
components: $t_\theta^{\mathrm{even}}$ and $t_\phi^{\mathrm{odd}}$
are even scalar fields on the sphere, while
$t_\phi^{\mathrm{even}}$ and $t_\theta^{\mathrm{odd}}$ are odd and these last two are precisely the combinations that vanish
identically on the equator.

\subsubsection{Full-sphere overview}
\label{sec:layer1_sphere}
\begin{figure*}[tbp]
  \centering
  \includegraphics[width=\textwidth]{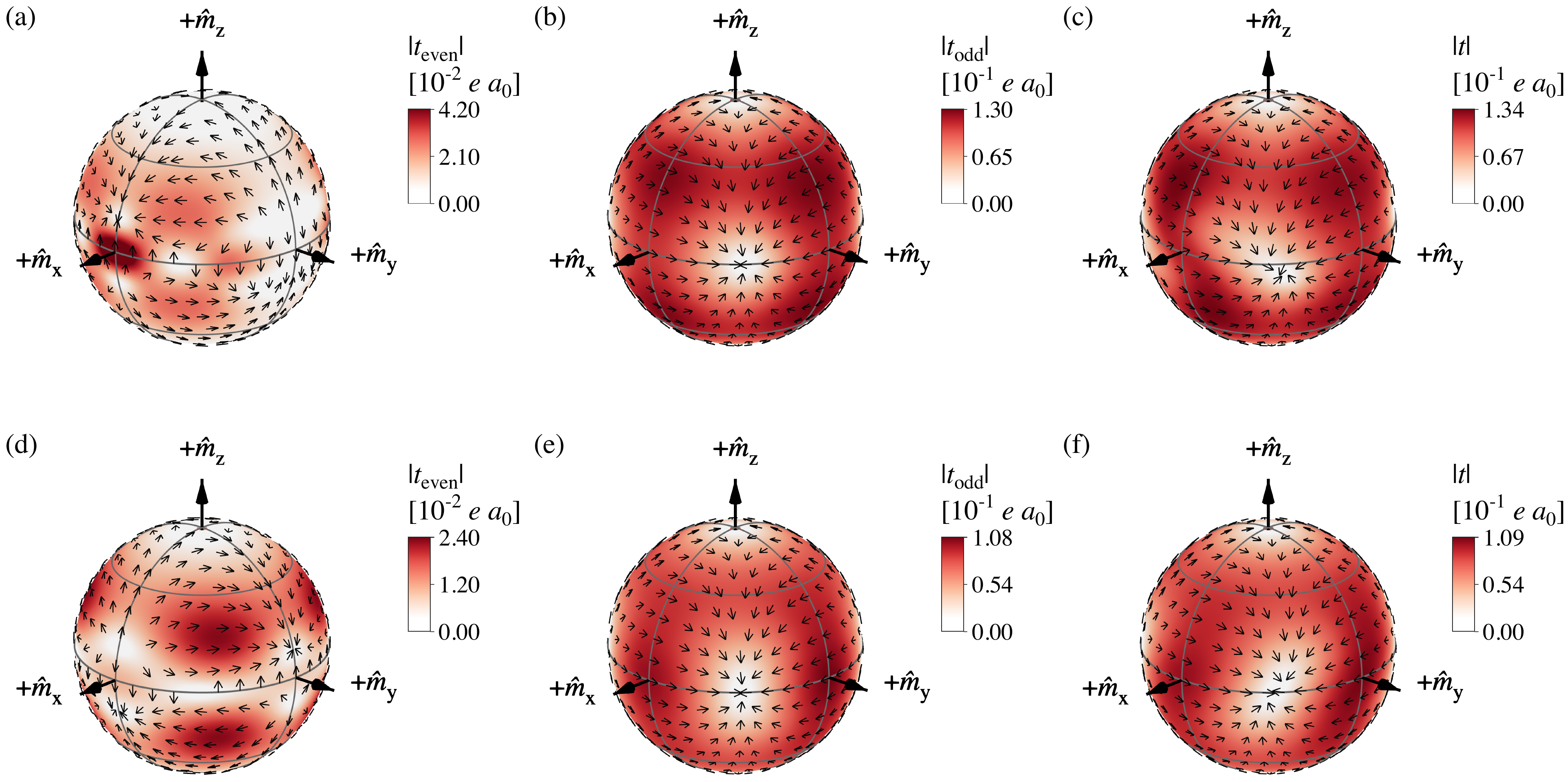}
\caption{Angular dependence of the spin-orbit torkance on the unit sphere of magnetization directions for monolayer Fe$_3$GeTe$_2$ [(a)--(c)] and Fe$_3$GaTe$_2$ [(d)--(f)], for
$\mathbf{E}\parallel\hat{\mathbf{x}}$, $\Gamma=10$~meV, on an
$800\times800$ Monkhorst--Pack mesh, at the Fermi level of each
compound. Panels (a),(d) show the magnitude $|\mathbf{t}_x^{\mathrm{even}}|
=[(t_\theta^{\mathrm{even}})^2+(t_\phi^{\mathrm{even}})^2]^{1/2}$ of the TR-even (Fermi-sea) channel; (b),(e) the TR-odd (Fermi-surface) $|\mathbf{t}_x^{\mathrm{odd}}|$; (c),(f) the total
$|\mathbf{t}_x|$ [Eqs.~\eqref{eq:tx_definition} and~\eqref{eq:proj_general}]. Color encodes the magnitude; arrows give the tangent-plane direction of the displayed torkance. }
  \label{fig:angular_maps_FGT_FGaT}
\end{figure*}

We plot the calculated torkances for FGT and FGaT in Fig.~\ref{fig:angular_maps_FGT_FGaT}.
For magnetization along $\pm\hat{\mathbf{z}}$, the torque vanishes, given the presence
of the $\sigma_h$ mirror. Once the magnetization tilts, $\sigma_h$ is no longer a symmetry of the magnetic configuration, allowing the emergence of torque ~\cite{PhysRevB.108.144422}.

The TR-odd torkances exceed the TR-even ones in peak magnitude by factors of $2.5$ (FGT) and $4.7$ (FGaT). Along the equator, the two split by symmetry: the TR-even channel resides entirely in $t_\theta^{\mathrm{even}}$ and the TR-odd channel in $t_\phi^{\mathrm{odd}}$. The TR-odd maps for FGT and FGaT are similar (the peak magnitude is ${\approx}15\%$ smaller in FGaT), which  is a bit surprising given the rather distinct band configuration at the Fermi surfaces, (Fig.~\ref{fig:bands}). However, the TR-even torkances differ markedly: the FGT peak is about twice the FGaT one, the overall sign is reversed between the compounds, and the FGT channel is strongly peaked at
$\hat{\mathbf{m}}\parallel\pm\hat{\mathbf{x}}$ and nearly
vanishing at $\pm\hat{\mathbf{y}}$, whereas the FGaT channel
remains uniformly small on the equator, about five times below the FGT peak. Both features are analyzed by the harmonic expansion of Sec.~\ref{sec:layer2_potentials}.

The total torkances [Figs.~\ref{fig:angular_maps_FGT_FGaT}(c),(f)], unlike the parity-resolved channels, are not mirror-symmetric about the $xz$ and $yz$ planes: the magnitudes of the even and odd channels are separately symmetric under these reflections of $\hat{\mathbf{m}}$, but the two channels transform with opposite signs under the antiunitary operations $\sigma_v\hat{\Theta}$ and $C_2\hat{\Theta}$, so their interplay in $|\mathbf{t}_x|$ results in the asymmetry of the total magnitude reaching ${\approx}35\%$ (FGT) and ${\approx}30\%$ (FGaT) of the peak. On the equator, the total
torkance does not vanish for any azimuthal angle---its minimum, near the diagonals, is $18\%$ of the peak in FGT and $7\%$ in FGaT. The true zeros lie off the equator: in FGT the magnitude drops to $3\%$ of the peak one grid spacing below it ($\theta\approx95^\circ$, $\phi\approx50^\circ$), resolving the out-of-equator zero-torkance points found for monolayer FGT by Xue \emph{et~al.}~\cite{PhysRevB.108.144422}; in FGaT any such zero lies within one grid spacing of the equator at our $5^\circ$ resolution.

\subsubsection{Symmetry expansion and closed forms}
\label{sec:layer2_potentials}

We decompose the torkance over the magnetization sphere in the
vector-spherical-harmonics (VSH) basis generated by two scalar potentials $\psi(\hat{\mathbf{m}})$ and $\chi(\hat{\mathbf{m}})$. Expanding these in spherical harmonics $Y_{\ell m}(\hat{\mathbf{m}})$ gives the two tangent families $\boldsymbol{\Psi}_{\ell m}=\nabla_{\!s}Y_{\ell m}$ and $\boldsymbol{\Phi}_{\ell m}=\hat{\mathbf{m}}\times\nabla_{\!s}Y_{\ell m}$, with $\nabla_{\!s}$ the surface gradient~\cite{PhysRevB.101.020407,PhysRevB.108.144422}. The torkance vectors can then be expressed as: 

\begin{equation}
\mathbf{t}_x
=\nabla_{\!s}\psi
+\hat{\mathbf{m}}\times\nabla_{\!s}\chi
=\sum_{\ell\ge1}\sum_{m=-\ell}^{\ell}
\bigl[\psi_{\ell m}\,\boldsymbol{\Psi}_{\ell m}
+\chi_{\ell m}\,\boldsymbol{\Phi}_{\ell m}\bigr],
\label{eq:hodge_vsh_main}
\end{equation}
where $\ell$ and $m$ are the degree and azimuthal order of the scalar
harmonics. Equation~\eqref{eq:hodge_vsh_main} holds for any magnet and any crystal symmetry:  for a rigid, magnetization-independent field $\mathbf{b}$, the potential $\psi=\hat{\mathbf{m}}\cdot\mathbf{b}$ generates the dampinglike form $ \hat{\mathbf{m}}\times(\hat{\mathbf{m}}\times\mathbf{b})$ and
$\chi=\hat{\mathbf{m}}\cdot\mathbf{b}$ the fieldlike $\hat{\mathbf{m}}\times\mathbf{b}$, so the $\ell=1$ channel would describe the Rashba spin accumulation effect. The TR-even torkance is described by the $\chi$ harmonics of even $\ell$, and the $\psi$ harmonics of odd $\ell$; the TR-odd torkance by the corresponding complements \cite{PhysRevB.108.144422}. 

\begin{table}[tb]
\caption{Expansion coefficients of Eq.~\eqref{eq:vsh_closed_forms} for $\mathbf{E}\parallel\hat{\mathbf{x}}$, in units of $e\,a_0$, extracted from the torkance computed on the magnetization sphere (broadening $\Gamma=10$~meV, $800\times800$ Monkhorst--Pack $k$ mesh, $37\times72$ grid in the polar and azimuthal angles). Coefficients refer to the unnormalized basis of Eq.~\eqref{eq:hodge_vsh_main}. For the unit-normalized convention of Ref.~\cite{PhysRevB.108.144422},
$C^{F}_{\ell m}=\sqrt{\ell(\ell+1)}\,\chi_{\ell m}$ and
$C^{D}_{\ell m}=\sqrt{\ell(\ell+1)}\,\psi_{\ell m}$.}
\label{tab:potential_coefficients}
\small
\setlength{\tabcolsep}{12pt}
\renewcommand{\arraystretch}{1.1}
\begin{tabular}{@{}lrr@{}}
\hline\hline
 & FGT & FGaT \\
\hline
\multicolumn{3}{@{}l}{\emph{TR-odd (Fermi surface)}}\\
$\psi^{s}_{2,2}$     & $+0.1387$ & $+0.1074$ \\
$\chi^{c}_{3,2}$     & $+0.0243$ & $+0.0129$ \\
\hline
\multicolumn{3}{@{}l}{\emph{TR-even (Fermi sea)}}\\
$\chi^{s}_{2,2}$     & $-0.0071$ & $+0.0104$ \\
$\psi^{c}_{3,2}$     & $+0.0123$ & $-0.0097$ \\
$\chi^{s}_{4,2}$     & $+0.0026$ & $-0.0024$ \\
$\psi^{c}_{5,2}$     & $-0.0045$ & $+0.0028$ \\
$\chi^{s}_{4,4}$     & $+0.0010$ & $-0.0011$ \\
$\psi^{c}_{5,4}$     & $+0.0039$ & $+0.0004$ \\
\hline\hline
\end{tabular}
\end{table}

Xue \textit{et al.}~\cite{PhysRevB.108.144422} showed that the monolayer-FGT torque exhibits vector spherical harmonics beyond the leading shell, the higher-order terms controlling the field-free switching trajectory. We recover that content as the $m=4$ pair of the TR-even channel, assign it to Fe$_{\rm I}$ (see the following section), and show that it is significantly reduced in FGaT. 

In Table~\ref{tab:potential_coefficients} we give the
lowest expansion coefficients calculated for FGT and FGaT, grouped by symmetric and  antisymmetric combinations of the spherical harmonics with $m>0$,
\begin{align}
\boldsymbol{\Psi}^{c}_{\ell,m}
&{}=\frac{1}{\sqrt{2}}\bigl(\boldsymbol{\Psi}_{\ell,m}+\boldsymbol{\Psi}_{\ell,-m}\bigr),\\
\boldsymbol{\Psi}^{s}_{\ell,m}
&{}=\frac{1}{i\sqrt{2}}\bigl(\boldsymbol{\Psi}_{\ell,m}-\boldsymbol{\Psi}_{\ell,-m}\bigr),
\end{align}
and likewise for $\boldsymbol{\Phi}^{c,s}_{\ell,m}$, which result in real $\sin$-like ($s$) and $\cos$-like ($c$) functions
of the azimuthal angle $\phi$ for the even $m$ that occur here. In terms of these lowest harmonics,
the explicit expansion reads:
\begin{equation}
\begin{aligned}
\mathbf{t}_x^{\mathrm{odd}}
\simeq{}&\psi^{s}_{2,2}\boldsymbol{\Psi}^{s}_{2,2}
+\chi^{c}_{3,2}\boldsymbol{\Phi}^{c}_{3,2},\\[4pt]
\mathbf{t}_x^{\mathrm{even}}
\simeq{}&\chi^{s}_{2,2}\boldsymbol{\Phi}^{s}_{2,2}
+\psi^{c}_{3,2}\boldsymbol{\Psi}^{c}_{3,2}\\
&+\chi^{s}_{4,2}\boldsymbol{\Phi}^{s}_{4,2}
+\psi^{c}_{5,2}\boldsymbol{\Psi}^{c}_{5,2}\\
&+\chi^{s}_{4,4}\boldsymbol{\Phi}^{s}_{4,4}
+\psi^{c}_{5,4}\boldsymbol{\Psi}^{c}_{5,4}.
\end{aligned}
\label{eq:vsh_closed_forms}
\end{equation}
These expressions can be employed to model the current-induced magnetization dynamics with the Landau--Lifshitz--Gilbert equation.
They encode both the azimuthal multipoles as well as the polar
anisotropy of the torques. We note that, due to $D_{3h}$ symmetry, the allowed torque begins at $\ell=2$ with azimuthal indices restricted to $m=6n\pm2$ \cite{PhysRevB.108.144422}.

\subsubsection{Atom- and momentum-resolved torkances}
\label{sec:layer3}

\begin{figure*}[!tp]
  \centering
  \includegraphics[width=\textwidth]{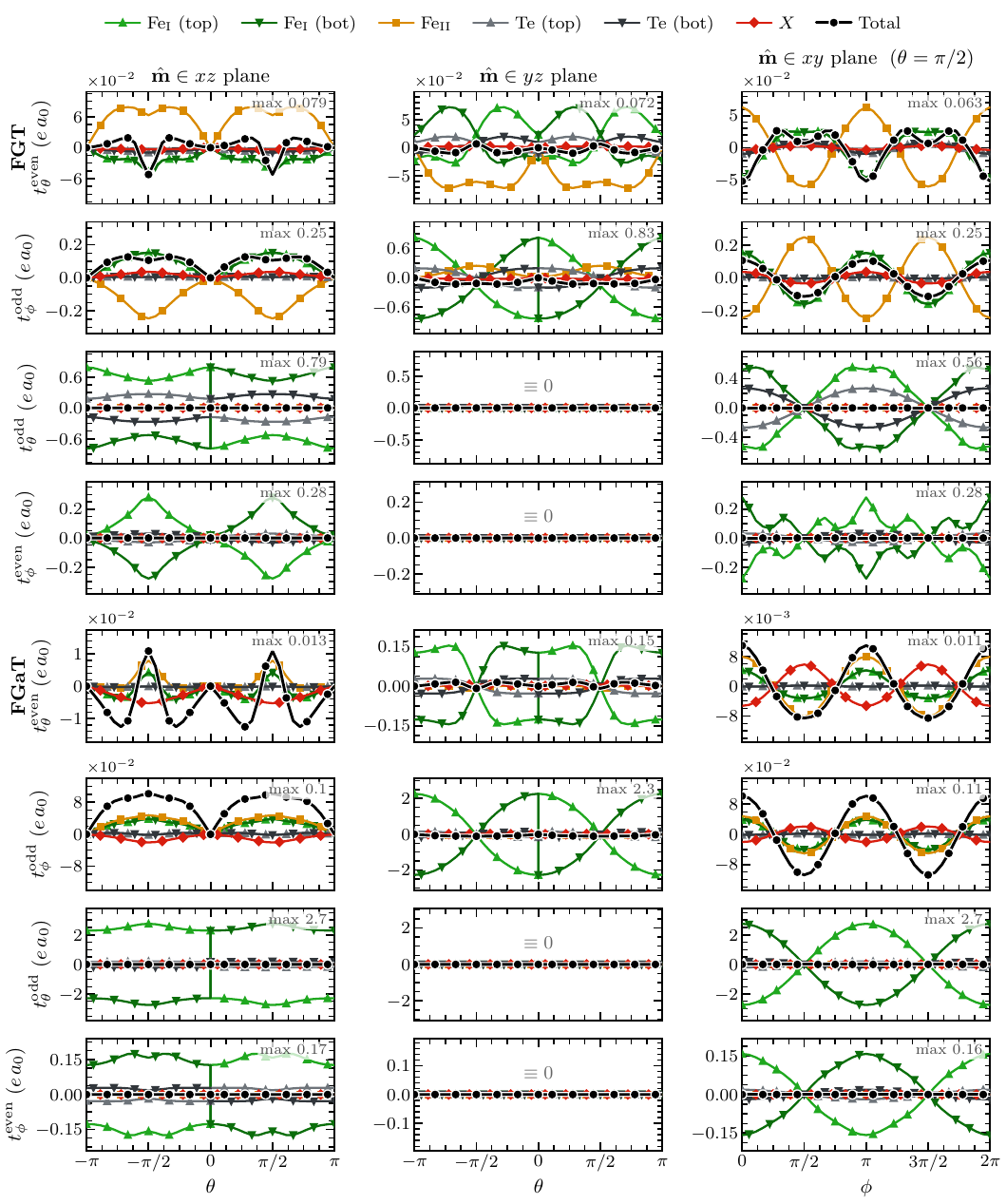}
\caption{\label{fig:atom_resolved}
Calculated atom-resolved torkance for $\mathbf{E}\parallel\hat{\mathbf{x}}$, along the $xz$ and $yz$ meridians and the equator $\theta=\pi/2$ (columns). For each compound the upper two rows are the channels the magnetic point group allows in the total, the lower two those it forbids. Curves give the sublattice groups and their sum (black). The forbidden channels vanish in the sum but not per site: $\sigma_h$ exchanges the Fe$_{\rm I}$ and Te partners and so is not a site symmetry of either, leaving their individual torques unconstrained. Fe$_{\rm II}$ and $X$ lie on the mirror plane, where $\sigma_h$ \emph{is} a site symmetry, and vanish individually. In the $yz$ plane the vertical mirror $\sigma_v$ fixes every atom, so the cancellation is site by site and those panels are identically zero. The step at $\theta=0$ is the $\hat{\mathbf{e}}_\phi\to-\hat{\mathbf{e}}_\phi$ branch change across
the pole, not a real discontinuity.}
\end{figure*}

Having analyzed the torques for the monolayers, let us now look at how the torques affect the
magnetization of the individual atoms. This is particularly useful in order to see how the
atoms work together to induce the magnetization dynamics. The atom-resolved torques are in
general richer, especially if the atomic sites are not at high-symmetry positions, which can
lead to hidden torques~\cite{PhysRevLett.113.157201,PhysRevB.105.064412}. Such an analysis could facilitate atomic design of SOT via defect and van der Waals engineering. 

We resolve the torkances onto the Fe$_{\rm I}$ pair, Fe$_{\rm II}$, Te,
and $X$ sublattices (Fig.~\ref{fig:atom_resolved}; consult
Fig.~\ref{fig:structure} for the geometry); the two Fe$_{\rm I}$ atoms are connected by $\sigma_h$. The angular dependences depend strongly on the magnetization orientation cuts through the unit spheres, but we can observe
that one of the main differences between FGT and FGaT is the opposite contributions (except for the $yz$ plane) of the Fe$_{\rm I}$ and
Fe$_{\rm II}$ atoms to the torques. It is not possible to trace this
dependence to the band structure features; we rather present this as a numerical observation. In most cases Fe$_{\rm II}$ contributes most strongly, although
the peak magnitudes for the Fe$_{\rm I}$ and Fe$_{\rm II}$ atoms are similar for the
FGaT at the $xz$ and $xy$ plane. 

As for the full monolayers, it is the odd torkances that dominate.
The same can be said of the atom-resolved quantities. Interestingly, while Te atoms also contribute, for example in FGT in $yz$ and $xz$ planes, a much stronger effect comes from Ga, see
for example the odd torkance in $xz$ and $xy$ planes.

We analyze more closely the case of the Fe$_{\rm I}$ atom pairs. The two atoms are related by the horizontal mirror. While they are equivalent for out-of-plane magnetization, as soon as $\hat{\mathbf{m}}$ tilts
to the plane, the atoms are no longer equivalent~\cite{Martin02012023}.
From a physics perspective, one could expect a hidden spin-orbit field, equal in magnitude but antiparallel, at the two atoms, induced by the electric current. 
Indeed, removing $\sigma_h$ allows the following spin accumulation at the
two atoms, top and bottom,
\begin{equation}
  \boldsymbol{\sigma}^{(0)}_{\mathrm{top}}
  = \lambda\,\hat{\mathbf{z}}\times\mathbf{E},
  \qquad
  \boldsymbol{\sigma}^{(0)}_{\mathrm{bot}}
  = -\lambda\,\hat{\mathbf{z}}\times\mathbf{E},
\end{equation}
with a real parameter $\lambda$.
Considering both atoms, the spin accumulation vanishes. However, this hidden spin accumulation induces hidden
torques, opposite at the two atoms. 

These Rashba-like spin polarizations are independent of the magnetization direction.
Their most striking consequence is the appearance of odd $\theta$-components
in the $xz$ plane, although such torques are forbidden by the global
$D_{3h}$ symmetry. They originate from hidden spin-orbit fields whose
current-induced spin polarizations are locally perpendicular to the
magnetization. In FGT the hidden $\theta$-component is about three times
larger than the symmetry-allowed $\phi$-component, while in FGaT it is
enhanced by nearly a factor of 30. As required by symmetry, the hidden
contributions from the two Fe$_{\rm I}$ sublattices cancel exactly in the total
torque. Remarkably, the two Te atoms, which are also related by 
$\sigma_h$, and are weakly magnetic due to the hybridization with Fe, have also staggered spin polarizations, though weaker than the Fe$_{\rm I}$ pair.

At higher order, the two Fe$_{\rm I}$ atoms respond differently to the
magnetization-induced spin accumulation. Based on the geometry in
Fig.~\ref{fig:structure}, the symmetry-allowed spin accumulations are
\begin{equation}
\label{eq:staggered}
\boldsymbol{\sigma}^{(1)}_{\mathrm{top}}
=
E_x
\begin{pmatrix}
a m_y + b m_z\\
a m_x\\
c m_x
\end{pmatrix},
\quad
\boldsymbol{\sigma}^{(1)}_{\mathrm{bot}}
=
E_x
\begin{pmatrix}
a m_y - b m_z\\
a m_x\\
-c m_x
\end{pmatrix},
\end{equation}
where $a$, $b$, and $c$ are real, material-dependent coefficients. Parameter $a$ describes the $D_{3h}$ allowed spin accumulation, common to both atoms, while $b$ and $c$ induce the
hidden spin-orbit fields responsible for the hidden torques. Consequently,
the two Fe$_{\rm I}$ atoms generally exhibit different, rather than merely opposite,
torques, as observed, for example, in the $yz$ plane for both FGT and FGaT. 

\begin{figure}[tbp]
  \centering
  \includegraphics[width=\columnwidth]{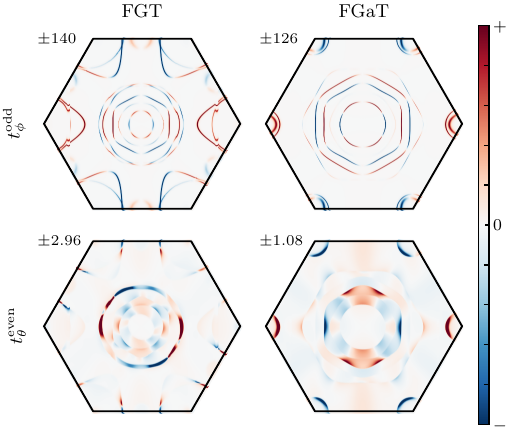}
\caption{\label{fig:kresolved_equator}%
Momentum-resolved torkance density for FGT (left) and FGaT (right) calculated for
broadening $\Gamma=10$~meV on the $800\times800$ mesh. The Brillouin-zone
average of each panel reproduces the corresponding equatorial ($xy$)
torkance of Fig.~\ref{fig:atom_resolved}. \emph{Top}: the TR-odd
(fieldlike) $t_\phi^{\mathrm{odd}}=+t_{yx}^{\mathrm{odd}}$ at
$\hat{\mathbf{m}}\parallel\hat{\mathbf{x}}$. \emph{Bottom}: the
TR-even (Fermi-sea) $t_\theta^{\mathrm{even}}=-t_{zx}^{\mathrm{even}}$ at $\phi=\pi/4$, where the $m=2$ harmonic vanishes and the panel displays the $m=4$ content. The diverging colorbar gives the sign (blue negative, red positive) and each panel is scaled independently to the value quoted at its upper left. The black hexagon marks the first Brillouin zone.}
\end{figure}

Figure~\ref{fig:kresolved_equator} resolves the torkance densities in momentum space across the Brillouin zone. Specifically, we consider the TR-odd $t_\phi^{\mathrm{odd}}$ at $\hat{\mathbf{m}}\parallel\hat{\mathbf{x}}$, and the TR-even
$t_\theta^{\mathrm{even}}$ at $\phi=\pi/4$, where $\cos2\phi$
vanishes so that the panel displays the fourth harmonics ($m=4$). The TR-odd density is confined to sharp contours tracing the Fermi-surface sheets. The TR-even integrand is instead spread over the zone, being
strongly enhanced at band anticrossings. 

The odd torkance is dominated by the Fermi-surface pockets near the K points along the $k_x$ direction, which contribute positively to $t_\phi$. Other regions of the Fermi surface exhibit alternating positive and negative torkance densities. In FGaT, the K-point pockets are smaller due to hole doping, which likely explains the somewhat reduced self-torque compared to FGT. Interestingly, these pockets also contribute to the fourth harmonic of the even torkance in FGaT, whereas their contribution is much less pronounced in FGT. However, the fourth harmonic itself, which originates predominantly from the Fe$_{\rm I}$-derived states (see the (xy) panel
in Fig. \ref{fig:atom_resolved}, where the high-harmonic behavior is most evident), is more pronounced in FGT. This suggests that, in FGaT, the momentum-space compensation is more complete, leading to a partial cancellation of the high-harmonic contributions, as can also be inferred from Fig.~\ref{fig:kresolved_equator}.

\section{Discussion}
\label{sec:discussion}

Both FGT and FGaT monolayers possess $D_{3h}$ point-group symmetry. This symmetry alone forbids a current-induced spin accumulation and, consequently, a spin--orbit torque. Formally, the magnetization-independent spin accumulation,
\begin{equation}
\sigma_i^{(0)} = C_{ij} E_j,
\end{equation}
vanishes identically, implying $C_{ij}=0$. The in-plane spin accumulation is prohibited by the horizontal mirror symmetry $\sigma_h$, whereas an out-of-plane spin accumulation is forbidden by the threefold rotational symmetry $C_3$. As a result, neither a uniform in-plane nor an out-of-plane spin polarization can be generated by an in-plane electric field in the absence of magnetization. This does not mean that there are no hidden torques. 

What is the origin of the torque we present here? The mechanism can be described as magnetization-induced spin accumulation in the presence of electric current. Indeed, a group symmetry analysis gives the spin accumulation in the first order in magnetization and electric field as
\begin{equation}
\sigma_i^{(1)} = C_{ijk} E_j m_k.
\end{equation}
The third-order tensor $C_{ijk}$ does not vanish. The mirror and rotation symmetries constrain this spin polarization to the
form (obtained also from Eq.~\eqref{eq:staggered} by setting
$b=c=0$)
\begin{equation}
\boldsymbol{\sigma}^{(1)}
=
E_x
\begin{pmatrix}
a m_y\\
a m_x\\
0
\end{pmatrix}.
\end{equation}
While this spin polarization is still confined to the plane, 
as would be expected for the Rashba effect, it can be both 
transverse and parallel to the electric current. If the
magnetization is along $y$, which lies in the vertical mirror plane, $\boldsymbol{\sigma}^{(1)}$ is parallel to the current. 
This is rather unique, not compatible with the conventional picture illustrated in Fig. \ref{fig:structure}. Certainly, being effective requires strong electric fields, but this magnetization tunability has not yet been explored in experiments. 

What about out-of-plane spin accumulations? As shown above [Eq.~\eqref{eq:staggered}], removing the horizontal mirror symmetry unlocks out-of-plane spin components. These can generate current-induced perpendicular spin polarization and dampinglike torques, providing an additional mechanism for manipulating the magnetization. This suggests that defect engineering, as well as symmetry-reducing van der Waals heterostructures, could provide an effective route toward deterministic switching of perpendicular ferromagnets by enabling out-of-plane spin polarization and the associated spin--orbit torques.

\section{Conclusion}
\label{sec:conclusion}
We have computed the complete angular dependence of the intrinsic spin--orbit torque in monolayer Fe$_3$GeTe$_2$ and Fe$_3$GaTe$_2$ from first principles and decomposed it by time-reversal parity, atomic sublattice, and momentum, into its microscopic constituents. The two ferromagnets form a rather controlled pair, having identical crystal symmetry, the same dominant Fe~$3d$ character, while differing by a single valence electron.
From the calculated band structures in the presence of out-of-plane magnetization, we extracted the valley Zeeman spin-orbit fields of about 40 meV in both FGT and FGaT.

While the Fermi-surface structures of FGT and FGaT differ primarily in the size of the $K$-point pockets, their odd torkances, which dominate the overall response, are remarkably similar. The odd torkance is only slightly weaker in FGaT, likely reflecting the reduced density of states associated with the smaller $K$-point pockets caused by hole doping. In contrast, the even torkances exhibit pronounced fourth harmonics for in-plane magnetization in FGT, whereas these higher-order harmonics are significantly suppressed in FGaT, indicating a more complete compensation of the momentum-space contributions.

The atomic decomposition also reveals 
hidden torques. In particular, we 
find that the two $\sigma_h$ related
atoms of the Fe$_{\rm I}$ pair exhibit different
torques, which result from magnetization driven in-plane and out-of-plane spin accumulations. This is fully consistent with the symmetry arguments.

We also provide a detailed description of the computational pipeline, from density functional theory including spin-orbit coupling, through carefully symmetrized Wannier functions, to the linear-response formalism, highlighting several subtleties of the methodology, such as the use of rigid spin rotation to construct the spherical torque maps. In the Appendices, we present a transparent account of the computational workflow together with convergence benchmarks, with the aim of facilitating the implementation and reproducibility of these calculations.

\begin{acknowledgments}
This project was supported by the European Union Graphene Flagship project 2DSPIN TECH (grant agreement No. 101135853) and SFB 1277 (Project-ID 314695032).
\end{acknowledgments}

\appendix
\appendix
\section{Wannier-interpolation formalism}
\label{app:wannier}
\subsection{Conventions and real-space matrix elements}
\label{app:rspace}
Dense-$\mathbf{k}$ Hamiltonians, Berry connections, and velocity operators are interpolated from the coarse \emph{ab initio} $\mathbf{q}$ mesh using the symmetry-adapted
variant~\cite{PhysRevB.87.235109,tsirkin2021a} of the maximally localized Wannier function (WF) formalism~\cite{PhysRevB.65.035109,RevModPhys.84.1419,PhysRevB.75.195121}, to which we refer for the disentanglement and gauge-rotation matrices $W(\mathbf{q})$. Here these are additionally constrained by the magnetic space group, as described in Sec.~\ref{sec:wannier}. Two conventions specific to the present calculation are needed below. 

First, the basis is fully relativistic, so the Wannier index $j$ is a composite orbital--spin label. We order the $N_w$ spinor WFs spin-interlaced, whereby time reversal acts as
$\hat{\Theta} = (\mathbb{I}_{\mathrm{orb}}\otimes i\sigma_y)\mathcal{K}$ with $\mathcal{K}$ complex conjugation. 

Second, we use Convention~II~\cite{Vanderbilt_2018,Cole_Python_Tight_Binding_2025}: the Wannier centers $\bar{\mathbf{r}}_j$ are not absorbed into the Fourier phases but enter through the position matrix elements. The matrix elements of operator $\hat{O}$ in the momentum space are expanded as $O^{W}_{ij}(\mathbf{k}) = \sum_{\mathbf{R}} O_{ij}(\mathbf{R})\, e^{i\mathbf{k}\cdot\mathbf{R}}$. In particular, the Hamiltonian and the Wannier position matrix elements read
\begin{equation}
H_{ij}(\mathbf{R}) = \langle \mathbf{0},i|H|\mathbf{R},j\rangle,
\qquad
\mathcal{A}_{\alpha,ij}(\mathbf{R}) = \langle \mathbf{0},i|\hat{r}_\alpha|\mathbf{R},j\rangle.
\label{eq:HR_AR_def}
\end{equation}
Similarly for the spin operators, see App.~\ref{app:operators}. Here $i,j$ are composite orbital--spin Wannier indices, $\mathbf{R}$ a lattice vector, and $\alpha\in\{x,y,z\}$ a Cartesian direction; $\alpha$ labels the component of the position operator
throughout, and hence of the Berry connection, the velocity, and the $\mathbf{k}$ derivative.

A practical point concerns the position matrix elements stored by Wannier90. The diagonal $\mathbf{R}=\mathbf{0}$ contribution corresponding to the Wannier centers is removed from the stored matrices. The Berry connection must therefore be reconstructed as
\begin{equation}
\mathcal{A}^{W}_{\alpha,ij}(\mathbf{k})
=
\sum_{\mathbf{R}}
\mathcal{A}^{(\mathrm{wcc})}_{\alpha,ij}(\mathbf{R})
e^{i\mathbf{k}\cdot\mathbf{R}}
+
\bar{r}_{i,\alpha}\delta_{ij},
\label{eq:Aalpha_proper}
\end{equation}
where $\bar{r}_{i,\alpha}=\langle \mathbf{0},i|\hat r_\alpha|\mathbf{0},i\rangle$ is the $\alpha$ component of the center of the $i$th Wannier function. Although the restored term is $\mathbf{k}$ independent, it generally does not commute with $H^{W}(\mathbf{k})$. Neglecting it therefore yields an incorrect velocity operator and consequently erroneous response functions.

\subsection{Velocity operator and band basis}
\label{app:velocity}
The velocity operator $\hat{v}_\alpha = (i\hbar)^{-1}[\hat{r}_\alpha,H]$ in the Wannier gauge can be written as~\cite{PhysRevB.74.195118,PhysRevB.75.195121}
\begin{equation}
v^{W}_{\alpha}(\mathbf{k})
= \frac{1}{\hbar}\,\partial_{k_\alpha} H^{W}(\mathbf{k})
+ \frac{i}{\hbar}\bigl[H^{W}(\mathbf{k}),\mathcal{A}^{W}_{\alpha}(\mathbf{k})\bigr].
\label{eq:velocity_W}
\end{equation}
The first term is the group-velocity contribution. The second,
geometric term carries the entire Wannier-center dependence through the
diagonal of Eq.~\eqref{eq:Aalpha_proper} and distinguishes the proper
Wannier velocity from the bare Peierls term
$\partial_{k_\alpha}H^{W}/\hbar$ which needs to be taken into account if the centers are inequivalent. Its role has recently been underscored by Go \emph{et~al.}~\cite{PhysRevB.109.174435}, who recast it as an
orbital-dependent anomalous position and showed it to be decisive for the intrinsic orbital Hall effect. The same gauge covariance underlies the symmetry-protected torkance zeros benchmarked in Appendix~\ref{app:rigid_validation}.

Diagonalizing the Wannier-gauge Hamiltonian with unitary matrix $U(\mathbf{k})$,
\begin{equation}
U^{\dagger}(\mathbf{k})
H^{W}(\mathbf{k})
U(\mathbf{k})
=
\varepsilon(\mathbf{k}),
\end{equation}
defines the Hamiltonian gauge. The Berry connection in this gauge is
\begin{equation}
\mathcal{A}^{H}_{\alpha}
=
U^{\dagger}\mathcal{A}^{W}_{\alpha}U
+
i\,U^{\dagger}\partial_{\alpha}U.
\end{equation}
Transforming Eq.~\eqref{eq:velocity_W} accordingly yields the standard
band-basis velocity~\cite{PhysRevB.74.195118,PhysRevB.75.195121}
\begin{equation}
  v^{H}_{\alpha,nm}(\mathbf{k}) =
  \begin{cases}
    \displaystyle \frac{1}{\hbar}\,\partial_{k_\alpha}\varepsilon_{n\mathbf{k}},
    & n = m, \\[10pt]
    \displaystyle \frac{i}{\hbar}\,
      (\varepsilon_{n\mathbf{k}}-\varepsilon_{m\mathbf{k}})\,
      \mathcal{A}^{H}_{\alpha,nm}(\mathbf{k}),
    & n \neq m,
  \end{cases}
\label{eq:vH_band}
\end{equation}
where the off-diagonal velocity matrix elements are determined by the full
Hamiltonian-gauge Berry connection. Together with the spin operator of
Appendix~\ref{app:operators}, Eqs.~\eqref{eq:velocity_W}
and~\eqref{eq:vH_band} provide the matrix elements required to evaluate
the torkance (Appendix~\ref{app:kubo}).


\section{Operator conventions and auxiliary formulas}
\label{app:operators}
\subsection{Spin operator}
\label{app:operators_spinor}
For perfectly atom-centered Wannier functions the spin operator on the
composite orbital\,$\otimes$\,spin space of Appendix~\ref{app:rspace} is
block diagonal,
\begin{equation}
  \hat{S}_a = \mathbb{I}_{\mathrm{orb}} \otimes S_a
            = \frac{\hbar}{2}\,\mathbb{I}_{\mathrm{orb}} \otimes \sigma_a,
  \qquad a = x,y,z,
\label{eq:spin_full}
\end{equation}
with $\sigma_a$ the Pauli matrices, $\mathbb{I}_{\mathrm{orb}}$ the
identity on the $N_{\mathrm{orb}}$-dimensional orbital subspace, and
$N_w = 2N_{\mathrm{orb}}$. As WFs are typically not strictly atom-centered, we use
the interpolated spin operator built from
$S_{a,ij}(\mathbf{R}) = \langle\mathbf{0},i|\hat{S}_a|\mathbf{R},j\rangle$
(in Convention~II, see Appendix~\ref{app:rspace}),
\begin{equation}
  S^{W}_{a,ij}(\mathbf{k})
  = \sum_{\mathbf{R}} S_{a,ij}(\mathbf{R})\,e^{i\mathbf{k}\cdot\mathbf{R}},
\label{eq:spin_interp}
\end{equation}
which reduces to Eq.~\eqref{eq:spin_full} in the atom-centered limit
$S_{a,ij}(\mathbf{R}) \to \delta_{\mathbf{R},\mathbf{0}}[\hat{S}_a]_{ij}$.

Resolved on the Pauli basis, $S^{W}_a = \tfrac{\hbar}{2}\sum_{b}\Sigma^{(ab)}(\mathbf{k})\otimes\sigma_b$
with $\Sigma^{(ab)} = \tfrac{1}{2}\mathrm{Tr}_{\rm spin}[\sigma_b S^{W}_a]$, where $b$ runs over $0,x,y,z$ with $\sigma_0 = \mathbb{I}_2$ and each $\Sigma^{(ab)}$ is an orbital-space matrix in units of $\hbar/2$. A pure
spin operator has $\Sigma^{(ab)} = \delta_{ab}\Sigma^{W}$, with $\Sigma^{W}$ the orbital-space spin-length matrix; Eq.~\eqref{eq:spin_full}
is the case $\Sigma^{W} = \mathbb{I}_{\rm orb}$.

Because the Wannier functions span only a finite subspace of the full
Hilbert space, Eq.~\eqref{eq:spin_interp} yields the spin operator projected
onto the Wannier manifold. In the presence of SOC, the projected
operator is no longer of the form $\mathbb{I}_{\mathrm{orb}}\otimes\sigma_a$.
The deviation $\Sigma^{W}\neq\mathbb{I}_{\mathrm{orb}}$ therefore quantifies
the SOC-induced spin mixing within the Wannier manifold.

\subsection{Time-reversal decomposition of the Hamiltonian}
\label{app:operators_evenodd}
With $\hat{\Theta} = U_T\mathcal{K}$ the time-reversal operator of Appendix~\ref{app:rspace} and $U_T=\mathbb{I}_{\mathrm{orb}}\otimes i\sigma_y$ its unitary part, the time-reversed
Bloch Hamiltonian is $H_{\mathrm{TR}}(\mathbf{k}) = U_T H^{\ast}(-\mathbf{k})U_T^{\dagger}$, and the time-reversal even and odd parts
of the Hamiltonian can be introduced as
\begin{equation}
  H_{\mathrm{even}/\mathrm{odd}}(\mathbf{k})
  = \tfrac{1}{2}\bigl[H(\mathbf{k}) \pm H_{\mathrm{TR}}(\mathbf{k})\bigr].
\label{eq:Hevenodd}
\end{equation}
Here, $H_{\mathrm{even}}$ collects the kinetic, crystal-field, and pure spin--orbit terms while $H_{\mathrm{odd}}$ carries
the entire dependence of the band structure on the magnetization
direction $\hat{\mathbf{m}}$, comprising a collinear exchange
contribution and SOC-induced noncollinear corrections, which are
disentangled by spin-sector projection below.

In the rigid-rotation treatment adopted in the paper, the decomposition of the
Hamiltonian into the even and odd terms is
performed once at the reference $\hat{\mathbf{m}}_0 = \hat{z}$ and only
the odd part is transformed 
[Eq.~\eqref{eq:m_rotation}], with $H_{\mathrm{even}}$ held fixed. The
neglected self-consistent $\hat{\mathbf{m}}$ dependence of
$H_{\mathrm{even}}$ (Sec.~\ref{sec:symmetry_bands}) is quantified against
noncollinear DFT in Appendix~\ref{app:rigid_validation}.

\subsection{Spin-sector projection of $H_{\mathrm{odd}}$}
\label{app:operators_ssproj}
Without SOC, $H_{\mathrm{odd}}$ is block diagonal in the eigenbasis of
$\hat{\mathbf{m}}\cdot\boldsymbol{\sigma}$. SOC adds TR-odd terms coupling
the two spin sectors. These are physical, but they are locked to the
lattice rather than following a rigid magnetization rotation, and they mix
with the torque operator (Sec.~\ref{app:operators_torque}). We therefore
project them out.

Angular sweeps parametrize $\hat{\mathbf{m}}$ by an axis--angle pair $(\hat{\mathbf{n}},\theta)$ relative to $\hat{z}$, so that $\hat{\mathbf{m}} = R(\hat{\mathbf{n}},\theta)\,\hat{z}$, and rotate at the
operator level,
\begin{equation}
\begin{split}
  H_{\mathrm{odd}}(\mathbf{k};\hat{\mathbf{m}})
    &= U_R\,H_{\mathrm{odd}}(\mathbf{k};\hat{z})\,U_R^{\dagger}, \\
  U_R &= \mathbb{I}_{\mathrm{orb}} \otimes
         e^{-i\theta\,\hat{\mathbf{n}}\cdot\boldsymbol{\sigma}/2}.
\end{split}
\label{eq:m_rotation}
\end{equation}
The rotation acts on the \emph{unprojected} $H_{\mathrm{odd}}$; the projection defined below is applied afterwards, at the rotated $\hat{\mathbf{m}}$.

Let $S^{W}_{\hat{\mathbf{m}}}(\mathbf{k}) =
\hat{\mathbf{m}}\cdot\mathbf{S}^{W}(\mathbf{k})$, an $N_w\times N_w$
Hermitian matrix built from the interpolated spin operator
[Eq.~\eqref{eq:spin_interp}] at the unrotated $\mathbf{k}$. Its spectral
decomposition
\begin{equation}
  S^{W}_{\hat{\mathbf{m}}}(\mathbf{k})
  = \sum_{p=1}^{N_w} s_p\,v_p^{\phantom{\dagger}} v_p^{\dagger}
\label{eq:Sm_spectral}
\end{equation}
has real eigenvalues $s_p$ and orthonormal eigenvectors $v_p$, with
$p$ labeling the eigenpairs at fixed $\mathbf{k}$ (not the Wannier index
of Appendix~\ref{app:rspace}). Although SOC shortens the spin
($\Sigma^{W}\neq\mathbb{I}_{\mathrm{orb}}$, see above), the exchange
splitting keeps the spectrum bimodal: $N_{\mathrm{orb}}$ eigenvalues are
positive and $N_{\mathrm{orb}}$ negative. These define the $\pm$ sectors
and the orthogonal projectors
\begin{equation}
  P_{\pm}(\mathbf{k};\hat{\mathbf{m}})
  = \!\!\sum_{p\,:\,s_p \gtrless 0}\!\! v_p^{\phantom{\dagger}} v_p^{\dagger},
  \qquad P_+ + P_- = \mathbb{I}_{N_w}.
\label{eq:sector_proj}
\end{equation}
The projected Hamiltonian is
\begin{equation}
H_{\mathrm{odd}}^{\mathrm{proj}}(\mathbf{k};\hat{\mathbf{m}})
= P_{+}H_{\mathrm{odd}}(\mathbf{k};\hat{\mathbf{m}})P_{+}
+ P_{-}H_{\mathrm{odd}}(\mathbf{k};\hat{\mathbf{m}})P_{-},
\label{eq:Hodd_proj}
\end{equation}
Hermitian and TR-odd, preserving all intra-sector matrix elements and
removing only the inter-sector entries.

The order matters: because $\mathbf{S}^{W}(\mathbf{k})$ is not of the form $\mathbb{I}_{\mathrm{orb}}\otimes\boldsymbol{\sigma}$, one has $U_R P_{\pm}(\hat{z}) U_R^{\dagger} \neq P_{\pm}(\hat{\mathbf{m}})$, so projecting at $\hat{z}$ and then rotating is not equivalent to Eqs.~\eqref{eq:m_rotation}--\eqref{eq:Hodd_proj}. We adopt the latter, in which the sectors are defined by the exact spin operator quantized along the physical magnetization direction.

\subsection{Exchange torque operator}
\label{app:operators_torque}
The total spin-torque operator is
$\hat{\boldsymbol{\tau}}^{\mathrm{tot}} = (i/\hbar)[H,\hat{\mathbf{S}}]$. In the
magnetization-frame SOT formulation~\cite{PhysRevB.90.174423} the
relevant torque generates magnetization rotations: under a uniform SU(2)
rotation about axis $a$ only the TR-odd part of $H$ varies, so that
\begin{equation}
  \frac{\partial H}{\partial\theta_a}\bigg|_{\theta_a=0}
  = \frac{i}{\hbar}\bigl[H_{\mathrm{odd}},\hat{S}_a\bigr].
\label{eq:dH_dtheta}
\end{equation}
This presupposes that $H_{\mathrm{odd}}$ transforms rigidly under SU(2)
magnetization rotations, which in the presence of SOC holds only for the
projected operator of Sec.~\ref{app:operators_ssproj}. Substituting
$H_{\mathrm{odd}}\to H_{\mathrm{odd}}^{\mathrm{proj}}$ and
$\hat{S}_a \to S^{W}_a(\mathbf{k})$, so that both factors sit in the same
Wannier gauge, defines the \emph{exchange} torque operator
$\hat{\tau}_a$ of Sec.~\ref{sec:LR}. Like $S^{W}_a$ and $v^{W}_\alpha$,
its Wannier-gauge matrix is written without a hat and with the
superscript $W$,
\begin{equation}
  \tau^{W}_a(\mathbf{k})
  = \frac{i}{\hbar}\bigl[H_{\mathrm{odd}}^{\mathrm{proj}}(\mathbf{k}),
    S^{W}_{a}(\mathbf{k})\bigr],
  \qquad a = x,y,z.
\label{eq:tau_ex}
\end{equation}
The discarded piece $(i/\hbar)[H_{\mathrm{even}},\hat{S}_a]$ is not a
generator of magnetization rotations but a TR-even, SOC-mediated exchange of angular momentum between spin and orbital channels and does not contribute to the magnetization-frame SOT response.

Band energies, eigenvectors, and the velocity operator are computed from
the \emph{unprojected} Hamiltonian
\begin{equation}
  H(\mathbf{k};\hat{\mathbf{m}})
  = H_{\mathrm{even}}(\mathbf{k}) + H_{\mathrm{odd}}(\mathbf{k};\hat{\mathbf{m}}).
\label{eq:H_rotated_full}
\end{equation}
We note that the spin-sector projection enters \emph{only} through
Eq.~\eqref{eq:tau_ex}, with projectors evaluated at the target
magnetization so that the projection axis tracks $\hat{\mathbf{m}}$ at
every angle.

\subsection{Velocity operator under magnetization rotation}
\label{app:operators_velocity_rotation}
The velocity operator also carries a TR-odd component that must be
rotated consistently. Because $\partial_{k_\alpha}$ reverses TR parity,
the Peierls derivative $\partial_\alpha H^{W}$ inherits a TR-odd piece
from $H_{\mathrm{even}}$ and a TR-even piece from $H_{\mathrm{odd}}$;
decomposing at $\pm\mathbf{k}$ with the unitary $U_T$ of
Sec.~\ref{app:operators_evenodd} gives
\begin{equation}
v^{\mathrm{odd}/\mathrm{even}}_\alpha(\mathbf{k})
= \tfrac{1}{2}\Bigl[\partial_\alpha H^{W}(\mathbf{k})
  \mp U_T\bigl(\partial_\alpha H^{W}(-\mathbf{k})\bigr)^{\ast}U_T^{\dagger}\Bigr],
\label{eq:v_TR_decomp}
\end{equation}
the superscript labeling the TR parity of the velocity itself. Here
$v^{\mathrm{odd}}_\alpha$ originates from $H_{\mathrm{even}}$ and is
unchanged by the magnetization rotation, while $v^{\mathrm{even}}_\alpha$
originates from $H_{\mathrm{odd}}$ and must be rotated by $U_R$.
Assembling the Berry-connection term of Eq.~\eqref{eq:velocity_W} with
the rotated Hamiltonian,
\begin{equation}
  \hbar\,v^{\mathrm{rot}}_\alpha(\mathbf{k})
  = v^{\mathrm{odd}}_\alpha + U_R\,v^{\mathrm{even}}_\alpha\,U_R^{\dagger}
    + i\bigl[H(\mathbf{k};\hat{\mathbf{m}}),\mathcal{A}^{W}_\alpha(\mathbf{k})\bigr],
\label{eq:v_rotated_full}
\end{equation}
where the rotated Peierls derivative is explicitly Hermitized to remove
residual finite-precision asymmetry. Since $\mathcal{A}^{W}_\alpha$
[Eq.~\eqref{eq:Aalpha_proper}] is a property of the Wannier basis,
independent of $\hat{\mathbf{m}}$, and does not generally commute with
$U_R$, the last term cannot be written as a rigid SU(2) rotation of an
unrotated Berry term.

\subsection{Atom- and sublattice-resolved operators}
\label{app:operators_resolved}
Sublattice and orbital contributions are extracted with orthogonal
Hermitian projectors $\{P_\nu\}$, diagonal in the Wannier index $\nu$ and
satisfying $\sum_\nu P_\nu = \mathbb{I}_{N_w}$ and
$P_\nu P_\mu = \delta_{\nu\mu}P_\nu$. Here the partition is by atomic
site---the two Fe$_{\rm I}$ members, resolved individually throughout
Sec.~\ref{sec:layer3}, Fe$_{\rm II}$, the two Te sites, and the spacer
(Ge/Ga)---with coarser groupings formed by summing site projectors.
Because $S^{W}_{\hat{\mathbf{m}}}(\mathbf{k})$ carries inter-site matrix
elements, it does not commute with the $P_\nu$, so the sublattice and
spin-sector decompositions do not commute. We fix the order by resolving
the \emph{already} spin-sector-projected torque of Eq.~\eqref{eq:tau_ex},
\begin{equation}
  \tau^{W}_{a,\nu}(\mathbf{k})
  = \tfrac{1}{2}\bigl\{P_\nu,\tau^{W}_a(\mathbf{k})\bigr\},
\label{eq:tau_resolved}
\end{equation}
the symmetrization ensuring Hermiticity of each $\tau^{W}_{a,\nu}$.
Completeness of $\{P_\nu\}$ makes the sum rule
$\sum_\nu \tau^{W}_{a,\nu} = \tau^{W}_a$ exact.

\section{Kubo formulas for the torkance}
\label{app:kubo}
We take $e > 0$ (electron charge $-e$), a uniform $N_k$-point mesh, and
write
\begin{equation}
  x_n \equiv \varepsilon_{n\mathbf{k}} - \varepsilon_F,
  \qquad
  g_n \equiv x_n^2 + \Gamma^2,
\label{eq:kubo_shorthands}
\end{equation}
with $\Gamma$ the constant phenomenological broadening and the
$\mathbf{k}$ argument suppressed. The index $a$ labels the torque
component and $b$ the electric-field direction, identified with the
velocity index $\alpha$ of Appendixes~\ref{app:wannier}
and~\ref{app:operators}. Band-basis matrix elements $\hat{\tau}^{a}_{nm}$ and
$v^{b}_{nm}$ follow by rotating the exchange torque
[Eq.~\eqref{eq:tau_ex}] and the full Wannier velocity
[Eq.~\eqref{eq:velocity_W}; Eq.~\eqref{eq:v_rotated_full} for
$\hat{\mathbf{m}} \neq \hat{z}$] with the eigenvectors of
$H(\mathbf{k};\hat{\mathbf{m}})$ [Eq.~\eqref{eq:H_rotated_full}], so both
enter the Kubo--Bastin trace in a common gauge. At zero temperature the
energy integral can be performed analytically in the eigenstate basis
with retarded Green's functions
$G^{R}_n = (\varepsilon_F - \varepsilon_{n\mathbf{k}} + i\Gamma)^{-1}$,
yielding the Fermi-sea and Fermi-surface contributions of Freimuth
\emph{et~al.}~\cite{PhysRevB.90.174423},
\begin{equation}
\begin{aligned}
t^{\mathrm{even}}_{ab}
&= \frac{e\hbar}{2\pi N_k}
   \sum_{\mathbf{k}}\sum_{n\neq m}
   \mathrm{Im}\!\bigl[\hat{\tau}^{a}_{nm}\,v^{b}_{mn}\bigr]
\\[2pt]
&\quad\times \biggl[
   \underbrace{\frac{\Gamma\,(x_m - x_n)}{g_n\,g_m}}_{\text{surface--surface}}
   \;+\;
   \underbrace{\frac{2\Gamma}{(x_n - x_m)\,g_m}}_{\text{surface--sea}}
\\[2pt]
&\qquad\quad
   \;+\;
   \underbrace{\frac{2}{(x_n - x_m)^2}\,
   \mathrm{Im}\ln\!\frac{x_m - i\Gamma}{x_n - i\Gamma}}_{\text{sea--sea}}
   \biggr],
\end{aligned}
\label{eq:freimuth_even}
\end{equation}
\begin{equation}
t^{\mathrm{odd}}_{ab}
= \frac{e\hbar}{\pi N_k}
\sum_{\mathbf{k}}\sum_{n,m}
\frac{\Gamma^2\,\mathrm{Re}\!\bigl[\hat{\tau}^{a}_{nm}\,v^{b}_{mn}\bigr]}
     {g_n\,g_m}\,,
\label{eq:freimuth_odd}
\end{equation}
which in atomic units give the torkance in units of $e a_0$.
This split follows from the structure of the integrand rather than from a
symmetry argument, and in general need not coincide with the
decomposition of the response tensor under magnetization reversal. For
the magnetization-frame torque of Sec.~\ref{app:operators_torque} the two
do coincide---$t^{\mathrm{even}}_{ab}$ is even and $t^{\mathrm{odd}}_{ab}$
odd under $\hat{\mathbf{m}} \to -\hat{\mathbf{m}}$---and we therefore use
``TR-even/Fermi-sea'' and ``TR-odd/Fermi-surface'' interchangeably.
The unrestricted sum in Eq.~\eqref{eq:freimuth_odd} retains the diagonal
$n=m$ terms, which dominate $t^{\mathrm{odd}}$ in the clean limit and
generate its $\Gamma^{-1}$ scaling. 

\subsection{Broadening dependence}
\label{app:eta_scan}
\begin{figure}[htb]
  \centering
  \includegraphics[width=\columnwidth]{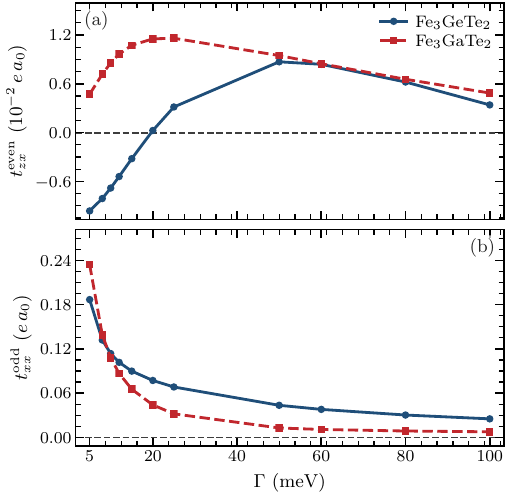}
  \caption{Broadening dependence of the spin--orbit torkance in monolayer
    $\mathrm{Fe}_3\mathrm{GeTe}_2$ (blue, solid) and
    $\mathrm{Fe}_3\mathrm{GaTe}_2$ (red, dashed) for
    $\hat{\mathbf{m}}=\hat{\mathbf{y}}$ and
    $\mathbf{E}\parallel\hat{\mathbf{x}}$: (a) Fermi-sea contribution
    $t^{\mathrm{even}}_{zx}$ and (b) Fermi-surface contribution
    $t^{\mathrm{odd}}_{xx}$ versus the phenomenological broadening
    $\Gamma$, on different ordinate scales. In
    $\mathrm{Fe}_3\mathrm{GeTe}_2$ the dampinglike torkance reverses sign
    near $\Gamma \simeq 20$~meV, whereas in
    $\mathrm{Fe}_3\mathrm{GaTe}_2$ it remains positive throughout.}
  \label{fig:gamma_scan}
\end{figure}

The constant broadening $\Gamma$ of Eqs.~\eqref{eq:freimuth_even}
and~\eqref{eq:freimuth_odd} corresponds to an effective elastic
relaxation time $\Gamma = \hbar/(2\tau_{\mathrm{rel}})$. Following
Refs.~\cite{PhysRevB.108.144422,PhysRevB.90.174423} we treat it as a
physical input encoding the impurity content. Figure~\ref{fig:gamma_scan} shows $t^{\mathrm{even}}_{zx}$ and
$t^{\mathrm{odd}}_{xx}$ versus $\Gamma$ at
$\hat{\mathbf{m}} = \hat{\mathbf{y}}$,
$\mathbf{E} \parallel \hat{\mathbf{x}}$, evaluated as outlined in  Sec.~\ref{app:operators_ssproj}, on the same $800\times800$
mesh and at the same chemical potentials as the main-text sweeps.

The Fermi-surface component decreases monotonically in both materials, approximately as $1/\Gamma$, drifting upward at larger $\Gamma$ as inter-band corrections to the dominant intra-band term of Eq.~\eqref{eq:freimuth_odd} grow. The wider window in Fe$_3$GaTe$_2$ follows from its larger inter-band splittings near $\varepsilon_F$ (Appendix~\ref{app:dos_valleys}). The Fermi-sea component is
non-monotonic in both, with a maximum in the range
$\Gamma \sim 25\text{--}50$~meV, but differs in sign structure: $t^{\mathrm{even}}_{zx}$ is negative at small $\Gamma$ in Fe$_3$GeTe$_2$, changes sign near $\Gamma \approx 20$~meV and peaks at $8.7\times10^{-3}\,ea_0$, whereas in Fe$_3$GaTe$_2$ it remains positive throughout. This reflects the competing signs of the surface--surface,
surface--sea and sea--sea terms of Eq.~\eqref{eq:freimuth_even}, whose cancellation shifts continuously with $\Gamma$. For all angular sweeps we adopt $\Gamma = 10$~meV ($\tau_{\mathrm{rel}} \approx 33$~fs).

\section{Noncollinear content of $H_{\mathrm{odd}}$ and its
response-level consequences}
\label{app:hodd_soc_contamination}
\begin{figure*}[!tp]
  \centering
  \includegraphics[width=\textwidth]{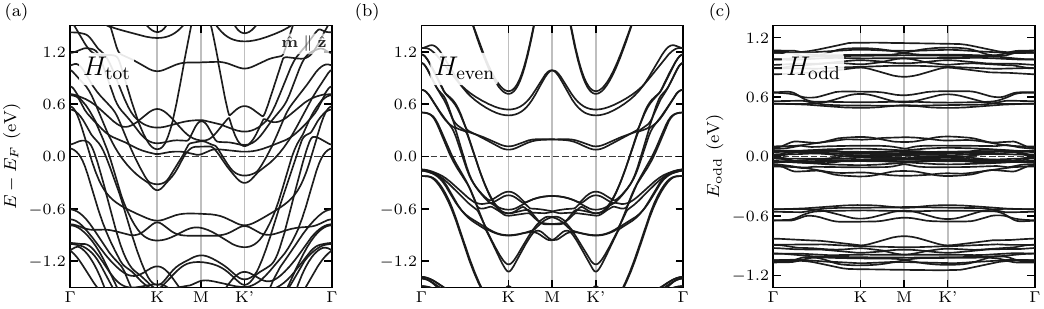}\\[0.9em]
  \includegraphics[width=\textwidth]{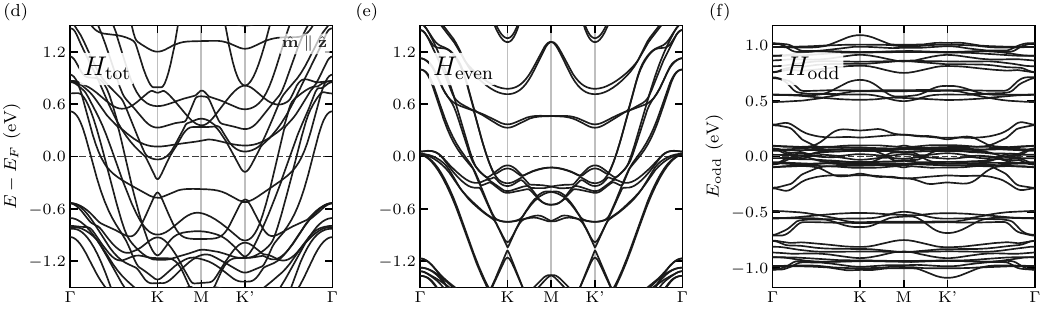}
  \caption{Band decomposition of monolayer Fe$_3$GeTe$_2$ [(a)--(c)] and
Fe$_3$GaTe$_2$ [(d)--(f)] at $\hat{\mathbf{m}} = \hat{z}$: spectra of the
full Hamiltonian, its TR-even component, and its TR-odd component
[Eq.~\eqref{eq:Hevenodd}]. The dashed line marks the Fermi level of the full Hamiltonian, repeated on the $H_{\mathrm{even}}$ panel as an energy reference; the $H_{\mathrm{odd}}$ spectrum is centered on zero and carries no filling of its own. Its narrow bandwidth [(c),(f)] reflects
the exchange scale, small compared with the full $p$--$d$ dispersion.}
  \label{fig:bands_decomposition_FGT_FGaT}
\end{figure*}

Without SOC, $H_{\mathrm{odd}}$ at the reference
$\hat{\mathbf{m}} = \hat{z}$ would be block diagonal in the two spin
sectors (Sec.~\ref{app:operators_ssproj}). Because the Wannier
Hamiltonian is built from relativistic spinors, the extracted
$H_{\mathrm{odd}}$ [Eq.~\eqref{eq:Hevenodd}] carries additional
spin-flip entries, the noncollinear content which is removed by the projection of Eq.~\eqref{eq:Hodd_proj}. We quantify this content by the 
effective deviation of the projected and unprojected odd Hamiltonians,
measured using the Frobenius norm,
$\|A\|_F=\sqrt{\mathrm{Tr}(A^\dagger A)}$,
\begin{equation}
  R_{\mathrm{nc}}
  = \frac{
    \bigl[\,\sum_{\mathbf{k}}
      \bigl\|H_{\mathrm{odd}}(\mathbf{k})
        - H_{\mathrm{odd}}^{\mathrm{proj}}(\mathbf{k})\bigr\|_F^{2}\,\bigr]^{1/2}
  }{
    \bigl[\,\sum_{\mathbf{k}}
      \bigl\|H_{\mathrm{odd}}(\mathbf{k})\bigr\|_F^{2}\,\bigr]^{1/2}
  }.
\label{eq:Rnc_hodd}
\end{equation}
The scaling of $R_{\mathrm{nc}}$ with the Wannier mesh is given in
Table~\ref{tab:hodd_soc_contamination_convergence}. Refining from $6\times6\times1$ to $12\times12\times1$ reduces $R_{\mathrm{nc}}$ from $1.31\%$ to $0.95\%$, staying below $1\%$ for finer mesh sizes. We can
thus characterize $H_{\mathrm{odd}}$ along $\hat{z}$ as a predominantly collinear-exchange Hamiltonian carrying $\lesssim 1\%$ SOC-induced noncollinear terms.

\begin{table}[t]
  \centering
  \caption{Convergence of the noncollinear content $R_{\mathrm{nc}}$
  [Eq.~\eqref{eq:Rnc_hodd}] for monolayer Fe$_3$GeTe$_2$ with respect to
  the \emph{ab initio}/Wannierization $\mathbf{k}$ mesh, converging to
  $\approx 0.9\%$. $R_{\mathrm{nc}}$ depends only on the mesh used to
  build the Wannier model.}
  \label{tab:hodd_soc_contamination_convergence}
  \begin{tabular}{c S[table-format=1.2]}
    \toprule
    DFT/Wannier mesh & {$R_{\mathrm{nc}}$ (\%)} \\
    \midrule
    $6\times6\times1$   & 1.31 \\
    $12\times12\times1$ & 0.95 \\
    $24\times24\times1$ & 0.92 \\
    $48\times48\times1$ & 0.90 \\
    \bottomrule
  \end{tabular}
\end{table}

As already stated, we project $H_{\rm odd}$ to the pure spin sectors,
removing the spin-flip content, but only in the torque operator. 
The electronic band states derive from the full rotated Hamiltonian
(Sec.~\ref{app:operators_torque}), with the spin-flip terms present. Since the spin-flip terms are not subject to the
rigid magnetization rotation, the Bloch states, and with them the response, acquire a weak rotation-path dependence. All torkance data reported here are therefore little-group symmetrized, removing the symmetry-forbidden components exactly while leaving the allowed channels unchanged.

\section{Benchmark of the rigid-exchange rotation method}
\label{app:rigid_validation}
\subsection{Symmetry validation and the Berry-connection contribution}
\label{app:rigid_validation_berry}

With $\hat{\mathbf{m}}$ in the $yz$ plane the magnetic point group contains $\sigma_{yz}\cdot\hat{\Theta}$, with $\sigma_{yz}$ the mirror with the normal $\hat{\mathbf{x}}$ and time-reversal $\hat{\Theta}$ from Sec.~\ref{app:operators_evenodd}: $\sigma_{yz}$ alone reverses any $\hat{\mathbf{m}}$ in that plane, so only the product leaves it invariant. It reverses $E_x$ and, through the TR parity of each Kubo channel, acts with opposite overall sign on the torque pseudovector in the Fermi-sea and Fermi-surface channels, so that preservation of
$\tau_a=\sum_b t_{ab}E_b$ [Eq.~\eqref{eq:torkance_def}] forces, for $\mathbf{E}\parallel\hat{\mathbf{x}}$,
\begin{equation}
  t^{\mathrm{even}}_{xx} = t^{\mathrm{odd}}_{yx} = t^{\mathrm{odd}}_{zx} = 0 .
\label{eq:forbidden_components_yz}
\end{equation}

\begin{figure}[!tbp]
  \centering
  \includegraphics[width=\columnwidth]{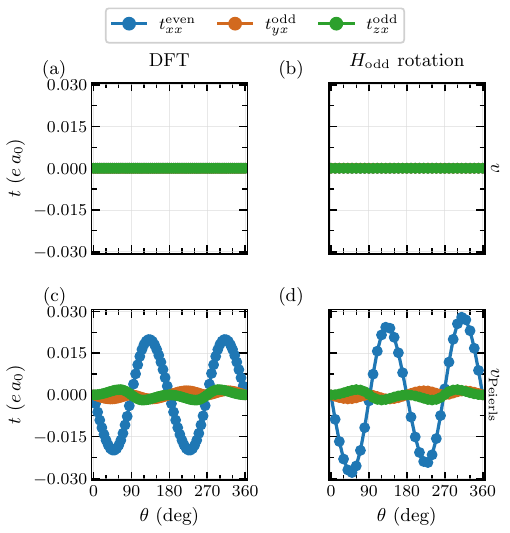}
  \caption{Symmetry-protected torkance components of monolayer Fe$_3$GeTe$_2$ versus the polar angle $\theta$ of $\hat{\mathbf{m}}$ in the $yz$ plane, at $\Gamma = 10$~meV, $\varepsilon_F = -4.2019$~eV, an $800\times800\times1$ $\mathbf{k}$ mesh, and  $\mathbf{E}\parallel\hat{\mathbf{x}}$. The components
  $t^{\mathrm{even}}_{xx}$ (blue, solid), $t^{\mathrm{odd}}_{yx}$ (orange, dashed) and $t^{\mathrm{odd}}_{zx}$ (green, dotted) are
  forbidden by $\sigma_{yz}\hat{\Theta}$
  [Eq.~\eqref{eq:forbidden_components_yz}]. They vanish with the full velocity (a,b), whereas dropping the
  Berry term (c,d) generates a spurious $t^{\mathrm{even}}_{xx}$ of
  order $2\times10^{-2}\,e\,a_0$, confirming that the
  symmetry is carried by the velocity operator and not by the
  Hamiltonian alone.}
  \label{fig:vel_diagnose}
\end{figure}

Figure~\ref{fig:vel_diagnose} monitors them against the polar angle $\theta$ for $H_{\mathrm{odd}}(\hat{\mathbf{m}})$ from self-consistent DFT and from rigid rotation, each evaluated with the full Wannier-gauge velocity of Eq.~\eqref{eq:velocity_W} and with the Peierls-only velocity $\partial_\alpha H^{W}/\hbar$. With the full velocity, all three torkances remain below $5\times10^{-5}\,ea_0$ in both schemes---two orders of magnitude below the peak allowed Fermi-sea torkance ($\approx 9\times10^{-3}\,ea_0$) and three below the fieldlike component ($\approx 0.1\,ea_0$). However, using only the Peierls velocity, without the gauge terms, would violate the symmetry, leading to $|t^{\mathrm{even}}_{xx}|$
reaching $\sim 2\times10^{-2}\,ea_0$---larger than the allowed Fermi-sea response itself---and $|t^{\mathrm{odd}}_{yx}|$, $|t^{\mathrm{odd}}_{zx}|$
reaching $\sim 2\times10^{-3}\,ea_0$.

This demonstration makes explicit the use of gauge covariant
expressions~\cite{PhysRevB.74.195118,PhysRevB.75.195121,PhysRevB.109.174435}: the Berry commutator $i[H^{W},\mathcal{A}^{W}_\alpha]$ [Eq.~\eqref{eq:Aalpha_proper}] generates the interband velocity matrix elements [Eq.~\eqref{eq:vH_band}] that carry the point-group constraints into the response, whereas the Peierls term does not include the Wannier centers and transforms covariantly only under $\mathbf{k}$-independent basis changes. All torkances in this work therefore use the full velocity of Eq.~\eqref{eq:velocity_W}.

\subsection{Equivalence with self-consistent DFT}
\label{app:rigid_validation_dft}
The rigid-rotation method holds $H_{\mathrm{even}}$ fixed and rotates only the
exchange field of $H_{\mathrm{odd}}$ [Sec.~\ref{app:operators_ssproj}];
it is valid if the $\hat{\mathbf{m}}$ dependence of
$H_{\mathrm{even}}$, entering solely through spin-orbit coupling, is much weaker than the exchange splitting. We test this method against full noncollinear
DFT, recomputed at each $\hat{\mathbf{m}}$, sweeping
\begin{equation}
  \hat{\mathbf{m}}(\theta) = (0,\,\sin\theta,\,\cos\theta)
\label{eq:m_yz_rotation}
\end{equation}
with $\Gamma = 10$~meV and an $800\times800$ mesh.

\begin{figure}[t]
  \centering
  \includegraphics[width=\columnwidth]{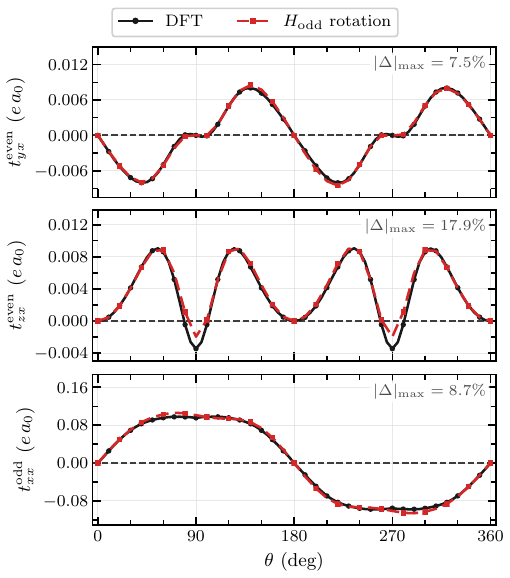}
  \caption{%
Rigid-exchange rotation versus self-consistent noncollinear DFT for the
three allowed torkance components at $\mathbf{E}\parallel\hat{\mathbf{x}}$
and $\hat{\mathbf{m}}$ in the $yz$ plane [Eq.~\eqref{eq:m_yz_rotation}].
Panels (a)--(c), top to bottom: $t^{\mathrm{even}}_{yx}$,
$t^{\mathrm{even}}_{zx}$ and $t^{\mathrm{odd}}_{xx}$ from DFT at each
angle (black circles, solid) and from rigid SU(2) rotation of a single
reference calculation (red squares, dashed). Each panel quotes the peak
relative deviation $|\Delta|_{\max}$ over the sweep, normalized to the
largest $|t^{\mathrm{DFT}}|$ of that component
[Eq.~\eqref{eq:residual_metric}]. Full velocity, $\Gamma = 10$~meV,
$\varepsilon_F = -4.2019$~eV, $800\times800$ mesh.}
  \label{fig:vel_diagnose1}
\end{figure}
The result for FGT is presented in Fig. \ref{fig:vel_diagnose1},
demonstrating that both methods agree very well for this
$yz$-plane sweep. We also plot the relative deviation
\begin{equation}
  \frac{\Delta(\theta)}{|t^{\mathrm{DFT}}|_{\max}}
  \equiv
  \frac{t^{H_{\mathrm{odd}}\text{-rot}}(\theta) - t^{\mathrm{DFT}}(\theta)}
       {\max_{\theta'}|t^{\mathrm{DFT}}(\theta')|},
\label{eq:residual_metric}
\end{equation}
normalized by the peak values. The torques in panels (a)
and (c) agree to $\lesssim 5\%$ over most of the range, with local peaks of $\sim 7\%$ near $\theta\approx150^\circ,210^\circ$ and $\sim 9\%$ near
$\theta\approx70^\circ,290^\circ$.

The even $t^{\mathrm{even}}_{zx}$ [panel (b)] is the exception: at
$\theta\approx90^\circ,270^\circ$ the deviation reaches $\sim 17\%$, an absolute $\approx 2.6\times10^{-3}\,ea_0$ against a local DFT value of $-3.45\times10^{-3}\,ea_0$.
This is not unexpected, given that the rigid rotation starts
by freezing the spin-flip components of the exchange 
Hamiltonian at $\theta = 0$.

Most importantly, the agreement of the rigid rotation and full DFT extends to the harmonic content:
$t^{\mathrm{even}}_{zx}(\theta)$ shows four extrema with symmetry-forced
zeros at the poles, and the rigid rotation reproduces their positions and relative magnitudes. We thus adopt the rigid-rotation method for all dense angular sweeps to obtain full spherical maps of the torques, while recognizing its inherent numerical inaccuracies.

\section{Spin-resolved density of states}
\label{app:dos_valleys}

\begin{figure}[tp]
  \centering
  \includegraphics[width=0.49\textwidth]{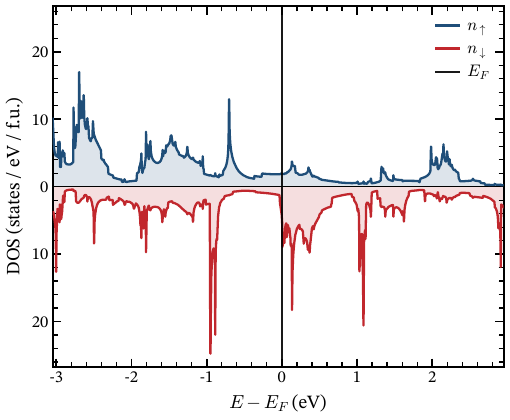}\hfill
  \includegraphics[width=0.49\textwidth]{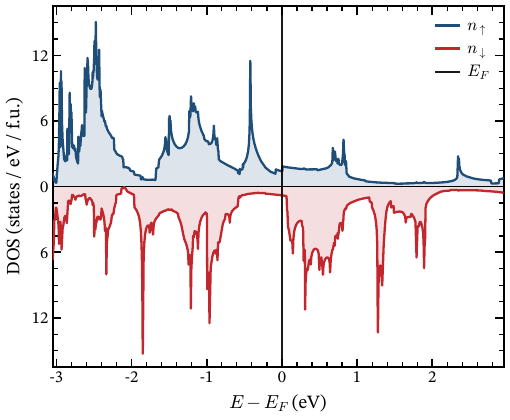}
  \caption{\label{fig:dos} Calculated spin-resolved DOS of monolayer Fe$_3$GeTe$_2$
(top) and Fe$_3$GaTe$_2$ (bottom) at
$\hat{\mathbf{m}}\parallel\hat{\mathbf{z}}$, with majority
($\langle S_z\rangle > 0$) and minority (plotted downward) channels and $E=0$ at each compound's Fermi level. In Fe$_3$GeTe$_2$ the Fermi level sits at the onset of a sharp, purely minority feature; in
Fe$_3$GaTe$_2$ the spectrum around $\varepsilon_F$ is smooth and
majority dominated.}
\end{figure}

For completeness, we also report the spin-resolved
density of states (DOS). The two ferromagnets differ at $\varepsilon_F$ in both magnitude and spin character (Fig.~\ref{fig:dos}). In FGT the total DOS is $N(\varepsilon_F) = 6.46$~eV$^{-1}$ ($n_\uparrow = 1.90$,
$n_\downarrow = 4.55$~eV$^{-1}$), a minority-like polarization $P = -0.41$, with $\varepsilon_F$ at the edge of a steep, purely minority feature: $n_\downarrow$ grows by a factor $3.8$ over the $50$~meV below $\varepsilon_F$ and peaks in slope $5$~meV above it, while $n_\uparrow$ stays flat near $1.9$~eV$^{-1}$. The feature survives the transport broadening ($P = -0.44$ at $\Gamma = 10$~meV).

In contrast, FGaT has a smaller DOS: $N(\varepsilon_F) = 2.21$~eV$^{-1}$ ($n_\uparrow = 1.41$,
$n_\downarrow = 0.80$) and majority-like spin polarization, $P = +0.28$. The substitution of Ge by Ga thus
reduces the Fermi-level DOS nearly threefold and interchanges its spin character. This should be manifested in spin-polarized tunneling experiments.

\FloatBarrier

\begin{thebibliography}{62}%
\makeatletter
\providecommand \@ifxundefined [1]{%
 \@ifx{#1\undefined}
}%
\providecommand \@ifnum [1]{%
 \ifnum #1\expandafter \@firstoftwo
 \else \expandafter \@secondoftwo
 \fi
}%
\providecommand \@ifx [1]{%
 \ifx #1\expandafter \@firstoftwo
 \else \expandafter \@secondoftwo
 \fi
}%
\providecommand \natexlab [1]{#1}%
\providecommand \enquote  [1]{``#1''}%
\providecommand \bibnamefont  [1]{#1}%
\providecommand \bibfnamefont [1]{#1}%
\providecommand \citenamefont [1]{#1}%
\providecommand \href@noop [0]{\@secondoftwo}%
\providecommand \href [0]{\begingroup \@sanitize@url \@href}%
\providecommand \@href[1]{\@@startlink{#1}\@@href}%
\providecommand \@@href[1]{\endgroup#1\@@endlink}%
\providecommand \@sanitize@url [0]{\catcode `\\12\catcode `\$12\catcode `\&12\catcode `\#12\catcode `\^12\catcode `\_12\catcode `\%12\relax}%
\providecommand \@@startlink[1]{}%
\providecommand \@@endlink[0]{}%
\providecommand \url  [0]{\begingroup\@sanitize@url \@url }%
\providecommand \@url [1]{\endgroup\@href {#1}{\urlprefix }}%
\providecommand \urlprefix  [0]{URL }%
\providecommand \Eprint [0]{\href }%
\providecommand \doibase [0]{https://doi.org/}%
\providecommand \selectlanguage [0]{\@gobble}%
\providecommand \bibinfo  [0]{\@secondoftwo}%
\providecommand \bibfield  [0]{\@secondoftwo}%
\providecommand \translation [1]{[#1]}%
\providecommand \BibitemOpen [0]{}%
\providecommand \bibitemStop [0]{}%
\providecommand \bibitemNoStop [0]{.\EOS\space}%
\providecommand \EOS [0]{\spacefactor3000\relax}%
\providecommand \BibitemShut  [1]{\csname bibitem#1\endcsname}%
\let\auto@bib@innerbib\@empty
\bibitem [{\citenamefont {\ifmmode \check{Z}\else \v{Z}\fi{}uti\ifmmode~\acute{c}\else \'{c}\fi{}}\ \emph {et~al.}(2004)\citenamefont {\ifmmode \check{Z}\else \v{Z}\fi{}uti\ifmmode~\acute{c}\else \'{c}\fi{}}, \citenamefont {Fabian},\ and\ \citenamefont {Das~Sarma}}]{RevModPhys.76.323}%
  \BibitemOpen
  \bibfield  {author} {\bibinfo {author} {\bibfnamefont {I.}~\bibnamefont {\ifmmode \check{Z}\else \v{Z}\fi{}uti\ifmmode~\acute{c}\else \'{c}\fi{}}}, \bibinfo {author} {\bibfnamefont {J.}~\bibnamefont {Fabian}},\ and\ \bibinfo {author} {\bibfnamefont {S.}~\bibnamefont {Das~Sarma}},\ }\bibfield  {title} {\bibinfo {title} {Spintronics: Fundamentals and applications},\ }\href {https://doi.org/10.1103/RevModPhys.76.323} {\bibfield  {journal} {\bibinfo  {journal} {Rev. Mod. Phys.}\ }\textbf {\bibinfo {volume} {76}},\ \bibinfo {pages} {323} (\bibinfo {year} {2004})}\BibitemShut {NoStop}%
\bibitem [{\citenamefont {Slonczewski}(1996)}]{SlONCZEWSKI1996L1}%
  \BibitemOpen
  \bibfield  {author} {\bibinfo {author} {\bibfnamefont {J.}~\bibnamefont {Slonczewski}},\ }\bibfield  {title} {\bibinfo {title} {Current-driven excitation of magnetic multilayers},\ }\href {https://doi.org/https://doi.org/10.1016/0304-8853(96)00062-5} {\bibfield  {journal} {\bibinfo  {journal} {Journal of Magnetism and Magnetic Materials}\ }\textbf {\bibinfo {volume} {159}},\ \bibinfo {pages} {L1} (\bibinfo {year} {1996})}\BibitemShut {NoStop}%
\bibitem [{\citenamefont {Berger}(1996)}]{PhysRevB.54.9353}%
  \BibitemOpen
  \bibfield  {author} {\bibinfo {author} {\bibfnamefont {L.}~\bibnamefont {Berger}},\ }\bibfield  {title} {\bibinfo {title} {Emission of spin waves by a magnetic multilayer traversed by a current},\ }\href {https://doi.org/10.1103/PhysRevB.54.9353} {\bibfield  {journal} {\bibinfo  {journal} {Phys. Rev. B}\ }\textbf {\bibinfo {volume} {54}},\ \bibinfo {pages} {9353} (\bibinfo {year} {1996})}\BibitemShut {NoStop}%
\bibitem [{\citenamefont {Katine}\ \emph {et~al.}(2000)\citenamefont {Katine}, \citenamefont {Albert}, \citenamefont {Buhrman}, \citenamefont {Myers},\ and\ \citenamefont {Ralph}}]{PhysRevLett.84.3149}%
  \BibitemOpen
  \bibfield  {author} {\bibinfo {author} {\bibfnamefont {J.~A.}\ \bibnamefont {Katine}}, \bibinfo {author} {\bibfnamefont {F.~J.}\ \bibnamefont {Albert}}, \bibinfo {author} {\bibfnamefont {R.~A.}\ \bibnamefont {Buhrman}}, \bibinfo {author} {\bibfnamefont {E.~B.}\ \bibnamefont {Myers}},\ and\ \bibinfo {author} {\bibfnamefont {D.~C.}\ \bibnamefont {Ralph}},\ }\bibfield  {title} {\bibinfo {title} {Current-driven magnetization reversal and spin-wave excitations in {Co}/{Cu}/{Co} pillars},\ }\href {https://doi.org/10.1103/PhysRevLett.84.3149} {\bibfield  {journal} {\bibinfo  {journal} {Phys. Rev. Lett.}\ }\textbf {\bibinfo {volume} {84}},\ \bibinfo {pages} {3149} (\bibinfo {year} {2000})}\BibitemShut {NoStop}%
\bibitem [{\citenamefont {Edelstein}(1990)}]{EDELSTEIN1990233}%
  \BibitemOpen
  \bibfield  {author} {\bibinfo {author} {\bibfnamefont {V.}~\bibnamefont {Edelstein}},\ }\bibfield  {title} {\bibinfo {title} {Spin polarization of conduction electrons induced by electric current in two-dimensional asymmetric electron systems},\ }\href {https://doi.org/https://doi.org/10.1016/0038-1098(90)90963-C} {\bibfield  {journal} {\bibinfo  {journal} {Solid State Communications}\ }\textbf {\bibinfo {volume} {73}},\ \bibinfo {pages} {233} (\bibinfo {year} {1990})}\BibitemShut {NoStop}%
\bibitem [{\citenamefont {Sinova}\ \emph {et~al.}(2015)\citenamefont {Sinova}, \citenamefont {Valenzuela}, \citenamefont {Wunderlich}, \citenamefont {Back},\ and\ \citenamefont {Jungwirth}}]{RevModPhys.87.1213}%
  \BibitemOpen
  \bibfield  {author} {\bibinfo {author} {\bibfnamefont {J.}~\bibnamefont {Sinova}}, \bibinfo {author} {\bibfnamefont {S.~O.}\ \bibnamefont {Valenzuela}}, \bibinfo {author} {\bibfnamefont {J.}~\bibnamefont {Wunderlich}}, \bibinfo {author} {\bibfnamefont {C.~H.}\ \bibnamefont {Back}},\ and\ \bibinfo {author} {\bibfnamefont {T.}~\bibnamefont {Jungwirth}},\ }\bibfield  {title} {\bibinfo {title} {Spin hall effects},\ }\href {https://doi.org/10.1103/RevModPhys.87.1213} {\bibfield  {journal} {\bibinfo  {journal} {Rev. Mod. Phys.}\ }\textbf {\bibinfo {volume} {87}},\ \bibinfo {pages} {1213} (\bibinfo {year} {2015})}\BibitemShut {NoStop}%
\bibitem [{\citenamefont {Mihai~Miron}\ \emph {et~al.}(2010)\citenamefont {Mihai~Miron}, \citenamefont {Gaudin}, \citenamefont {Auffret} \emph {et~al.}}]{mihai2010a}%
  \BibitemOpen
  \bibfield  {author} {\bibinfo {author} {\bibfnamefont {I.}~\bibnamefont {Mihai~Miron}}, \bibinfo {author} {\bibfnamefont {G.}~\bibnamefont {Gaudin}}, \bibinfo {author} {\bibfnamefont {S.}~\bibnamefont {Auffret}}, \emph {et~al.},\ }\bibfield  {title} {{\selectlanguage {en}\bibinfo {title} {Current-driven spin torque induced by the rashba effect in a ferromagnetic metal layer}},\ }\href {https://doi.org/10.1038/nmat2613} {\bibfield  {journal} {\bibinfo  {journal} {Nature Mater}\ }\textbf {\bibinfo {volume} {9}},\ \bibinfo {pages} {230–234} (\bibinfo {year} {2010})}\BibitemShut {NoStop}%
\bibitem [{\citenamefont {Miron}\ \emph {et~al.}(2011)\citenamefont {Miron}, \citenamefont {Garello}, \citenamefont {Gaudin} \emph {et~al.}}]{miron2011a}%
  \BibitemOpen
  \bibfield  {author} {\bibinfo {author} {\bibfnamefont {I.}~\bibnamefont {Miron}}, \bibinfo {author} {\bibfnamefont {K.}~\bibnamefont {Garello}}, \bibinfo {author} {\bibfnamefont {G.}~\bibnamefont {Gaudin}}, \emph {et~al.},\ }\bibfield  {title} {{\selectlanguage {en}\bibinfo {title} {Perpendicular switching of a single ferromagnetic layer induced by in-plane current injection}},\ }\href {https://doi.org/10.1038/nature10309} {\bibfield  {journal} {\bibinfo  {journal} {Nature}\ }\textbf {\bibinfo {volume} {476}},\ \bibinfo {pages} {189–193} (\bibinfo {year} {2011})}\BibitemShut {NoStop}%
\bibitem [{\citenamefont {Manchon}\ \emph {et~al.}(2019)\citenamefont {Manchon}, \citenamefont {\ifmmode~\check{Z}\else \v{Z}\fi{}elezn\'y}, \citenamefont {Miron}, \citenamefont {Jungwirth}, \citenamefont {Sinova}, \citenamefont {Thiaville}, \citenamefont {Garello},\ and\ \citenamefont {Gambardella}}]{RevModPhys.91.035004}%
  \BibitemOpen
  \bibfield  {author} {\bibinfo {author} {\bibfnamefont {A.}~\bibnamefont {Manchon}}, \bibinfo {author} {\bibfnamefont {J.}~\bibnamefont {\ifmmode~\check{Z}\else \v{Z}\fi{}elezn\'y}}, \bibinfo {author} {\bibfnamefont {I.~M.}\ \bibnamefont {Miron}}, \bibinfo {author} {\bibfnamefont {T.}~\bibnamefont {Jungwirth}}, \bibinfo {author} {\bibfnamefont {J.}~\bibnamefont {Sinova}}, \bibinfo {author} {\bibfnamefont {A.}~\bibnamefont {Thiaville}}, \bibinfo {author} {\bibfnamefont {K.}~\bibnamefont {Garello}},\ and\ \bibinfo {author} {\bibfnamefont {P.}~\bibnamefont {Gambardella}},\ }\bibfield  {title} {\bibinfo {title} {Current-induced spin-orbit torques in ferromagnetic and antiferromagnetic systems},\ }\href {https://doi.org/10.1103/RevModPhys.91.035004} {\bibfield  {journal} {\bibinfo  {journal} {Rev. Mod. Phys.}\ }\textbf {\bibinfo {volume} {91}},\ \bibinfo {pages} {035004} (\bibinfo {year} {2019})}\BibitemShut {NoStop}%
\bibitem [{\citenamefont {Huang}\ \emph {et~al.}(2017)\citenamefont {Huang}, \citenamefont {Clark}, \citenamefont {Navarro-Moratalla} \emph {et~al.}}]{huang2017a}%
  \BibitemOpen
  \bibfield  {author} {\bibinfo {author} {\bibfnamefont {B.}~\bibnamefont {Huang}}, \bibinfo {author} {\bibfnamefont {G.}~\bibnamefont {Clark}}, \bibinfo {author} {\bibfnamefont {E.}~\bibnamefont {Navarro-Moratalla}}, \emph {et~al.},\ }\bibfield  {title} {{\selectlanguage {en}\bibinfo {title} {Layer-dependent ferromagnetism in a van der waals crystal down to the monolayer limit}},\ }\href {https://doi.org/10.1038/nature22391} {\bibfield  {journal} {\bibinfo  {journal} {Nature}\ }\textbf {\bibinfo {volume} {546}},\ \bibinfo {pages} {270–273} (\bibinfo {year} {2017})}\BibitemShut {NoStop}%
\bibitem [{\citenamefont {Gong}\ \emph {et~al.}(2017)\citenamefont {Gong}, \citenamefont {Li}, \citenamefont {Li} \emph {et~al.}}]{gong2017a}%
  \BibitemOpen
  \bibfield  {author} {\bibinfo {author} {\bibfnamefont {C.}~\bibnamefont {Gong}}, \bibinfo {author} {\bibfnamefont {L.}~\bibnamefont {Li}}, \bibinfo {author} {\bibfnamefont {Z.}~\bibnamefont {Li}}, \emph {et~al.},\ }\bibfield  {title} {{\selectlanguage {en}\bibinfo {title} {Discovery of intrinsic ferromagnetism in two-dimensional van der waals crystals}},\ }\href {https://doi.org/10.1038/nature22060} {\bibfield  {journal} {\bibinfo  {journal} {Nature}\ }\textbf {\bibinfo {volume} {546}},\ \bibinfo {pages} {265–269} (\bibinfo {year} {2017})}\BibitemShut {NoStop}%
\bibitem [{\citenamefont {Fei}\ \emph {et~al.}(2018)\citenamefont {Fei}, \citenamefont {Huang}, \citenamefont {Malinowski} \emph {et~al.}}]{Fei2018}%
  \BibitemOpen
  \bibfield  {author} {\bibinfo {author} {\bibfnamefont {Z.}~\bibnamefont {Fei}}, \bibinfo {author} {\bibfnamefont {B.}~\bibnamefont {Huang}}, \bibinfo {author} {\bibfnamefont {P.}~\bibnamefont {Malinowski}}, \emph {et~al.},\ }\bibfield  {title} {\bibinfo {title} {Two-dimensional itinerant ferromagnetism in atomically thin {Fe$_3$GeTe$_2$}},\ }\href {https://doi.org/10.1038/s41563-018-0149-7} {\bibfield  {journal} {\bibinfo  {journal} {Nature Materials}\ }\textbf {\bibinfo {volume} {17}},\ \bibinfo {pages} {778} (\bibinfo {year} {2018})}\BibitemShut {NoStop}%
\bibitem [{\citenamefont {Deng}\ \emph {et~al.}(2018)\citenamefont {Deng}, \citenamefont {Yu}, \citenamefont {Song}, \citenamefont {Zhang}, \citenamefont {Wang}, \citenamefont {Sun}, \citenamefont {Yi}, \citenamefont {Wu}, \citenamefont {Wu}, \citenamefont {Zhu}, \citenamefont {Wang}, \citenamefont {Chen},\ and\ \citenamefont {Zhang}}]{deng2018a}%
  \BibitemOpen
  \bibfield  {author} {\bibinfo {author} {\bibfnamefont {Y.}~\bibnamefont {Deng}}, \bibinfo {author} {\bibfnamefont {Y.}~\bibnamefont {Yu}}, \bibinfo {author} {\bibfnamefont {Y.}~\bibnamefont {Song}}, \bibinfo {author} {\bibfnamefont {J.}~\bibnamefont {Zhang}}, \bibinfo {author} {\bibfnamefont {N.~Z.}\ \bibnamefont {Wang}}, \bibinfo {author} {\bibfnamefont {Z.}~\bibnamefont {Sun}}, \bibinfo {author} {\bibfnamefont {Y.}~\bibnamefont {Yi}}, \bibinfo {author} {\bibfnamefont {Y.~Z.}\ \bibnamefont {Wu}}, \bibinfo {author} {\bibfnamefont {S.}~\bibnamefont {Wu}}, \bibinfo {author} {\bibfnamefont {J.}~\bibnamefont {Zhu}}, \bibinfo {author} {\bibfnamefont {J.}~\bibnamefont {Wang}}, \bibinfo {author} {\bibfnamefont {X.~H.}\ \bibnamefont {Chen}},\ and\ \bibinfo {author} {\bibfnamefont {Y.}~\bibnamefont {Zhang}},\ }\bibfield  {title} {{\selectlanguage {en}\bibinfo {title} {Gate-tunable room-temperature ferromagnetism in two-dimensional {Fe$_3$GeTe$_2$}}},\ }\href {https://doi.org/10.1038/s41586-018-0626-9} {\bibfield
  {journal} {\bibinfo  {journal} {Nature}\ }\textbf {\bibinfo {volume} {563}},\ \bibinfo {pages} {94–99} (\bibinfo {year} {2018})}\BibitemShut {NoStop}%
\bibitem [{\citenamefont {MacNeill}\ \emph {et~al.}(2017)\citenamefont {MacNeill}, \citenamefont {Stiehl}, \citenamefont {Guimar{\~a}es}, \citenamefont {Buhrman}, \citenamefont {Park},\ and\ \citenamefont {Ralph}}]{macneill2017a}%
  \BibitemOpen
  \bibfield  {author} {\bibinfo {author} {\bibfnamefont {D.}~\bibnamefont {MacNeill}}, \bibinfo {author} {\bibfnamefont {G.~M.}\ \bibnamefont {Stiehl}}, \bibinfo {author} {\bibfnamefont {M.~H.~D.}\ \bibnamefont {Guimar{\~a}es}}, \bibinfo {author} {\bibfnamefont {R.~A.}\ \bibnamefont {Buhrman}}, \bibinfo {author} {\bibfnamefont {J.}~\bibnamefont {Park}},\ and\ \bibinfo {author} {\bibfnamefont {D.~C.}\ \bibnamefont {Ralph}},\ }\bibfield  {title} {{\selectlanguage {en}\bibinfo {title} {Control of spin–orbit torques through crystal symmetry in {WTe$_2$}/ferromagnet bilayers}},\ }\href {https://doi.org/10.1038/nphys3933} {\bibfield  {journal} {\bibinfo  {journal} {Nature Phys}\ }\textbf {\bibinfo {volume} {13}},\ \bibinfo {pages} {300–305} (\bibinfo {year} {2017})}\BibitemShut {NoStop}%
\bibitem [{\citenamefont {Kao}\ \emph {et~al.}(2022)\citenamefont {Kao}, \citenamefont {Muzzio}, \citenamefont {Zhang}, \citenamefont {Zhu}, \citenamefont {Gobbo}, \citenamefont {Yuan}, \citenamefont {Weber}, \citenamefont {Rao}, \citenamefont {Li}, \citenamefont {Edgar}, \citenamefont {Goldberger}, \citenamefont {Yan}, \citenamefont {Mandrus}, \citenamefont {Hwang}, \citenamefont {Cheng}, \citenamefont {Katoch},\ and\ \citenamefont {Singh}}]{kao2022a}%
  \BibitemOpen
  \bibfield  {author} {\bibinfo {author} {\bibfnamefont {I.-H.}\ \bibnamefont {Kao}}, \bibinfo {author} {\bibfnamefont {R.}~\bibnamefont {Muzzio}}, \bibinfo {author} {\bibfnamefont {H.}~\bibnamefont {Zhang}}, \bibinfo {author} {\bibfnamefont {M.}~\bibnamefont {Zhu}}, \bibinfo {author} {\bibfnamefont {J.}~\bibnamefont {Gobbo}}, \bibinfo {author} {\bibfnamefont {S.}~\bibnamefont {Yuan}}, \bibinfo {author} {\bibfnamefont {D.}~\bibnamefont {Weber}}, \bibinfo {author} {\bibfnamefont {R.}~\bibnamefont {Rao}}, \bibinfo {author} {\bibfnamefont {J.}~\bibnamefont {Li}}, \bibinfo {author} {\bibfnamefont {J.~H.}\ \bibnamefont {Edgar}}, \bibinfo {author} {\bibfnamefont {J.~E.}\ \bibnamefont {Goldberger}}, \bibinfo {author} {\bibfnamefont {J.}~\bibnamefont {Yan}}, \bibinfo {author} {\bibfnamefont {D.~G.}\ \bibnamefont {Mandrus}}, \bibinfo {author} {\bibfnamefont {J.}~\bibnamefont {Hwang}}, \bibinfo {author} {\bibfnamefont {R.}~\bibnamefont {Cheng}}, \bibinfo {author} {\bibfnamefont {J.}~\bibnamefont {Katoch}},\ and\
  \bibinfo {author} {\bibfnamefont {S.}~\bibnamefont {Singh}},\ }\bibfield  {title} {{\selectlanguage {en}\bibinfo {title} {Deterministic switching of a perpendicularly polarized magnet using unconventional spin–orbit torques in {WTe$_2$}}},\ }\href {https://doi.org/10.1038/s41563-022-01275-5} {\bibfield  {journal} {\bibinfo  {journal} {Nat. Mater}\ }\textbf {\bibinfo {volume} {21}},\ \bibinfo {pages} {1029–1034} (\bibinfo {year} {2022})}\BibitemShut {NoStop}%
\bibitem [{\citenamefont {Roemer}\ \emph {et~al.}(2020)\citenamefont {Roemer}, \citenamefont {Liu},\ and\ \citenamefont {Zou}}]{roemer2020a}%
  \BibitemOpen
  \bibfield  {author} {\bibinfo {author} {\bibfnamefont {R.}~\bibnamefont {Roemer}}, \bibinfo {author} {\bibfnamefont {C.}~\bibnamefont {Liu}},\ and\ \bibinfo {author} {\bibfnamefont {K.}~\bibnamefont {Zou}},\ }\bibfield  {title} {{\selectlanguage {da}\bibinfo {title} {Robust ferromagnetism in wafer-scale monolayer and multilayer {Fe$_3$GeTe$_2$}}},\ }\href {https://doi.org/10.1038/s41699-020-00167-z} {\bibfield  {journal} {\bibinfo  {journal} {npj 2D Mater Appl}\ }\textbf {\bibinfo {volume} {4}},\ \bibinfo {pages} {33} (\bibinfo {year} {2020})}\BibitemShut {NoStop}%
\bibitem [{\citenamefont {Zhang}\ \emph {et~al.}(2022)\citenamefont {Zhang}, \citenamefont {Guo}, \citenamefont {Wu}, \citenamefont {Wen}, \citenamefont {Yang}, \citenamefont {Jin}, \citenamefont {Zhang},\ and\ \citenamefont {Chang}}]{zhang2022a}%
  \BibitemOpen
  \bibfield  {author} {\bibinfo {author} {\bibfnamefont {G.}~\bibnamefont {Zhang}}, \bibinfo {author} {\bibfnamefont {F.}~\bibnamefont {Guo}}, \bibinfo {author} {\bibfnamefont {H.}~\bibnamefont {Wu}}, \bibinfo {author} {\bibfnamefont {X.}~\bibnamefont {Wen}}, \bibinfo {author} {\bibfnamefont {L.}~\bibnamefont {Yang}}, \bibinfo {author} {\bibfnamefont {W.}~\bibnamefont {Jin}}, \bibinfo {author} {\bibfnamefont {W.}~\bibnamefont {Zhang}},\ and\ \bibinfo {author} {\bibfnamefont {H.}~\bibnamefont {Chang}},\ }\bibfield  {title} {{\selectlanguage {en}\bibinfo {title} {Above-room-temperature strong intrinsic ferromagnetism in 2d van der waals {Fe$_3$GaTe$_2$} with large perpendicular magnetic anisotropy}},\ }\href {https://doi.org/10.1038/s41467-022-32605-5} {\bibfield  {journal} {\bibinfo  {journal} {Nat Commun}\ }\textbf {\bibinfo {volume} {13}},\ \bibinfo {pages} {5067} (\bibinfo {year} {2022})}\BibitemShut {NoStop}%
\bibitem [{\citenamefont {Ruiz}\ \emph {et~al.}(2024)\citenamefont {Ruiz}, \citenamefont {Esteras}, \citenamefont {López-Alcalá},\ and\ \citenamefont {Baldoví}}]{doi:10.1021/acs.nanolett.4c01019}%
  \BibitemOpen
  \bibfield  {author} {\bibinfo {author} {\bibfnamefont {A.~M.}\ \bibnamefont {Ruiz}}, \bibinfo {author} {\bibfnamefont {D.~L.}\ \bibnamefont {Esteras}}, \bibinfo {author} {\bibfnamefont {D.}~\bibnamefont {López-Alcalá}},\ and\ \bibinfo {author} {\bibfnamefont {J.~J.}\ \bibnamefont {Baldoví}},\ }\bibfield  {title} {\bibinfo {title} {On the origin of the above-room-temperature magnetism in the 2d van der waals ferromagnet {Fe$_3$GaTe$_2$}},\ }\href {https://doi.org/10.1021/acs.nanolett.4c01019} {\bibfield  {journal} {\bibinfo  {journal} {Nano Letters}\ }\textbf {\bibinfo {volume} {24}},\ \bibinfo {pages} {7886} (\bibinfo {year} {2024})}\BibitemShut {NoStop}%
\bibitem [{\citenamefont {Alghamdi}\ \emph {et~al.}(2019)\citenamefont {Alghamdi}, \citenamefont {Lohmann}, \citenamefont {Li}, \citenamefont {Jothi}, \citenamefont {Shao}, \citenamefont {Aldosary}, \citenamefont {Su}, \citenamefont {Fokwa},\ and\ \citenamefont {Shi}}]{doi:10.1021/acs.nanolett.9b01043}%
  \BibitemOpen
  \bibfield  {author} {\bibinfo {author} {\bibfnamefont {M.}~\bibnamefont {Alghamdi}}, \bibinfo {author} {\bibfnamefont {M.}~\bibnamefont {Lohmann}}, \bibinfo {author} {\bibfnamefont {J.}~\bibnamefont {Li}}, \bibinfo {author} {\bibfnamefont {P.~R.}\ \bibnamefont {Jothi}}, \bibinfo {author} {\bibfnamefont {Q.}~\bibnamefont {Shao}}, \bibinfo {author} {\bibfnamefont {M.}~\bibnamefont {Aldosary}}, \bibinfo {author} {\bibfnamefont {T.}~\bibnamefont {Su}}, \bibinfo {author} {\bibfnamefont {B.~P.~T.}\ \bibnamefont {Fokwa}},\ and\ \bibinfo {author} {\bibfnamefont {J.}~\bibnamefont {Shi}},\ }\bibfield  {title} {\bibinfo {title} {Highly efficient spin–orbit torque and switching of layered ferromagnet {Fe$_3$GeTe$_2$}},\ }\href {https://doi.org/10.1021/acs.nanolett.9b01043} {\bibfield  {journal} {\bibinfo  {journal} {Nano Letters}\ }\textbf {\bibinfo {volume} {19}},\ \bibinfo {pages} {4400} (\bibinfo {year} {2019})}\BibitemShut {NoStop}%
\bibitem [{\citenamefont {Wang}\ \emph {et~al.}(2019)\citenamefont {Wang}, \citenamefont {Tang}, \citenamefont {Xia}, \citenamefont {He}, \citenamefont {Zhang}, \citenamefont {Liu}, \citenamefont {Wan}, \citenamefont {Fang}, \citenamefont {Guo}, \citenamefont {Yang}, \citenamefont {Guang}, \citenamefont {Zhang}, \citenamefont {Xu}, \citenamefont {Wei}, \citenamefont {Liao}, \citenamefont {Lu}, \citenamefont {Feng}, \citenamefont {Li}, \citenamefont {Peng}, \citenamefont {Wei}, \citenamefont {Yang}, \citenamefont {Shi}, \citenamefont {Zhang}, \citenamefont {Han}, \citenamefont {Zhang}, \citenamefont {Zhang}, \citenamefont {Yu},\ and\ \citenamefont {Han}}]{doi:10.1126/sciadv.aaw8904}%
  \BibitemOpen
  \bibfield  {author} {\bibinfo {author} {\bibfnamefont {X.}~\bibnamefont {Wang}}, \bibinfo {author} {\bibfnamefont {J.}~\bibnamefont {Tang}}, \bibinfo {author} {\bibfnamefont {X.}~\bibnamefont {Xia}}, \bibinfo {author} {\bibfnamefont {C.}~\bibnamefont {He}}, \bibinfo {author} {\bibfnamefont {J.}~\bibnamefont {Zhang}}, \bibinfo {author} {\bibfnamefont {Y.}~\bibnamefont {Liu}}, \bibinfo {author} {\bibfnamefont {C.}~\bibnamefont {Wan}}, \bibinfo {author} {\bibfnamefont {C.}~\bibnamefont {Fang}}, \bibinfo {author} {\bibfnamefont {C.}~\bibnamefont {Guo}}, \bibinfo {author} {\bibfnamefont {W.}~\bibnamefont {Yang}}, \bibinfo {author} {\bibfnamefont {Y.}~\bibnamefont {Guang}}, \bibinfo {author} {\bibfnamefont {X.}~\bibnamefont {Zhang}}, \bibinfo {author} {\bibfnamefont {H.}~\bibnamefont {Xu}}, \bibinfo {author} {\bibfnamefont {J.}~\bibnamefont {Wei}}, \bibinfo {author} {\bibfnamefont {M.}~\bibnamefont {Liao}}, \bibinfo {author} {\bibfnamefont {X.}~\bibnamefont {Lu}}, \bibinfo {author} {\bibfnamefont
  {J.}~\bibnamefont {Feng}}, \bibinfo {author} {\bibfnamefont {X.}~\bibnamefont {Li}}, \bibinfo {author} {\bibfnamefont {Y.}~\bibnamefont {Peng}}, \bibinfo {author} {\bibfnamefont {H.}~\bibnamefont {Wei}}, \bibinfo {author} {\bibfnamefont {R.}~\bibnamefont {Yang}}, \bibinfo {author} {\bibfnamefont {D.}~\bibnamefont {Shi}}, \bibinfo {author} {\bibfnamefont {X.}~\bibnamefont {Zhang}}, \bibinfo {author} {\bibfnamefont {Z.}~\bibnamefont {Han}}, \bibinfo {author} {\bibfnamefont {Z.}~\bibnamefont {Zhang}}, \bibinfo {author} {\bibfnamefont {G.}~\bibnamefont {Zhang}}, \bibinfo {author} {\bibfnamefont {G.}~\bibnamefont {Yu}},\ and\ \bibinfo {author} {\bibfnamefont {X.}~\bibnamefont {Han}},\ }\bibfield  {title} {\bibinfo {title} {Current-driven magnetization switching in a van der waals ferromagnet {Fe$_3$GeTe$_2$}},\ }\href {https://doi.org/10.1126/sciadv.aaw8904} {\bibfield  {journal} {\bibinfo  {journal} {Science Advances}\ }\textbf {\bibinfo {volume} {5}},\ \bibinfo {pages} {eaaw8904} (\bibinfo {year}
  {2019})}\BibitemShut {NoStop}%
\bibitem [{\citenamefont {Wang}\ \emph {et~al.}(2023)\citenamefont {Wang}, \citenamefont {Wu},\ and\ \citenamefont {Zhang}}]{wang2023a}%
  \BibitemOpen
  \bibfield  {author} {\bibinfo {author} {\bibfnamefont {H.}~\bibnamefont {Wang}}, \bibinfo {author} {\bibfnamefont {H.}~\bibnamefont {Wu}},\ and\ \bibinfo {author} {\bibfnamefont {J.}~\bibnamefont {Zhang}},\ }\bibfield  {title} {{\selectlanguage {en}\bibinfo {title} {Room temperature energy-efficient spin-orbit torque switching in two-dimensional van der waals {Fe$_3$GeTe$_2$} induced by topological insulators}},\ }\href {https://doi.org/10.1038/s41467-023-40714-y} {\bibfield  {journal} {\bibinfo  {journal} {Nat Commun}\ }\textbf {\bibinfo {volume} {14}},\ \bibinfo {pages} {5173} (\bibinfo {year} {2023})}\BibitemShut {NoStop}%
\bibitem [{\citenamefont {Kajale}\ \emph {et~al.}(2024)\citenamefont {Kajale}, \citenamefont {Nguyen}, \citenamefont {Chao}, \citenamefont {Bono}, \citenamefont {Boonkird}, \citenamefont {Li},\ and\ \citenamefont {Sarkar}}]{Kajale2024}%
  \BibitemOpen
  \bibfield  {author} {\bibinfo {author} {\bibfnamefont {S.~N.}\ \bibnamefont {Kajale}}, \bibinfo {author} {\bibfnamefont {T.}~\bibnamefont {Nguyen}}, \bibinfo {author} {\bibfnamefont {C.~A.}\ \bibnamefont {Chao}}, \bibinfo {author} {\bibfnamefont {D.~C.}\ \bibnamefont {Bono}}, \bibinfo {author} {\bibfnamefont {A.}~\bibnamefont {Boonkird}}, \bibinfo {author} {\bibfnamefont {M.}~\bibnamefont {Li}},\ and\ \bibinfo {author} {\bibfnamefont {D.}~\bibnamefont {Sarkar}},\ }\bibfield  {title} {\bibinfo {title} {Current-induced switching of a van der waals ferromagnet at room temperature},\ }\href {https://doi.org/10.1038/s41467-024-45586-4} {\bibfield  {journal} {\bibinfo  {journal} {Nature Communications}\ }\textbf {\bibinfo {volume} {15}},\ \bibinfo {pages} {1485} (\bibinfo {year} {2024})}\BibitemShut {NoStop}%
\bibitem [{\citenamefont {Sharma}\ \emph {et~al.}(2025)\citenamefont {Sharma}, \citenamefont {Jang}, \citenamefont {Zhang}, \citenamefont {Jin}, \citenamefont {Chang},\ and\ \citenamefont {Hong}}]{SHARMA2025100847}%
  \BibitemOpen
  \bibfield  {author} {\bibinfo {author} {\bibfnamefont {P.~R.}\ \bibnamefont {Sharma}}, \bibinfo {author} {\bibfnamefont {B.}~\bibnamefont {Jang}}, \bibinfo {author} {\bibfnamefont {G.}~\bibnamefont {Zhang}}, \bibinfo {author} {\bibfnamefont {W.}~\bibnamefont {Jin}}, \bibinfo {author} {\bibfnamefont {H.}~\bibnamefont {Chang}},\ and\ \bibinfo {author} {\bibfnamefont {J.}~\bibnamefont {Hong}},\ }\bibfield  {title} {\bibinfo {title} {Field-free highly efficient spin-orbit torque switching in {Fe$_3$GaTe$_2$} at room temperature enabled by a unique distorted crystal symmetry of {WTe$_2$}},\ }\href {https://doi.org/https://doi.org/10.1016/j.apsadv.2025.100847} {\bibfield  {journal} {\bibinfo  {journal} {Applied Surface Science Advances}\ }\textbf {\bibinfo {volume} {29}},\ \bibinfo {pages} {100847} (\bibinfo {year} {2025})}\BibitemShut {NoStop}%
\bibitem [{\citenamefont {Zhang}\ \emph {et~al.}(2025)\citenamefont {Zhang}, \citenamefont {Wei},\ and\ \citenamefont {Duan}}]{zhang2025a}%
  \BibitemOpen
  \bibfield  {author} {\bibinfo {author} {\bibfnamefont {D.}~\bibnamefont {Zhang}}, \bibinfo {author} {\bibfnamefont {H.}~\bibnamefont {Wei}},\ and\ \bibinfo {author} {\bibfnamefont {J.}~\bibnamefont {Duan}},\ }\bibfield  {title} {{\selectlanguage {nl}\bibinfo {title} {Orbital torque switching of room temperature two-dimensional van der waals ferromagnet {Fe$_3$GaTe$_2$}}},\ }\href {https://doi.org/10.1038/s41467-025-62333-5} {\bibfield  {journal} {\bibinfo  {journal} {Nat Commun}\ }\textbf {\bibinfo {volume} {16}},\ \bibinfo {pages} {7047} (\bibinfo {year} {2025})}\BibitemShut {NoStop}%
\bibitem [{\citenamefont {Pandey}\ \emph {et~al.}(2025)\citenamefont {Pandey}, \citenamefont {Zhao},\ and\ \citenamefont {Tenzin}}]{pandey2025a}%
  \BibitemOpen
  \bibfield  {author} {\bibinfo {author} {\bibfnamefont {L.}~\bibnamefont {Pandey}}, \bibinfo {author} {\bibfnamefont {B.}~\bibnamefont {Zhao}},\ and\ \bibinfo {author} {\bibfnamefont {K.}~\bibnamefont {Tenzin}},\ }\bibfield  {title} {{\selectlanguage {nl}\bibinfo {title} {Tunable unconventional spin orbit torque magnetization dynamics in van der waals heterostructures}},\ }\href {https://doi.org/10.1038/s41467-025-64109-3} {\bibfield  {journal} {\bibinfo  {journal} {Nat Commun}\ }\textbf {\bibinfo {volume} {16}},\ \bibinfo {pages} {8722} (\bibinfo {year} {2025})}\BibitemShut {NoStop}%
\bibitem [{\citenamefont {Johansen}\ \emph {et~al.}(2019)\citenamefont {Johansen}, \citenamefont {Risingg\aa{}rd}, \citenamefont {Sudb\o{}}, \citenamefont {Linder},\ and\ \citenamefont {Brataas}}]{PhysRevLett.122.217203}%
  \BibitemOpen
  \bibfield  {author} {\bibinfo {author} {\bibfnamefont {O.}~\bibnamefont {Johansen}}, \bibinfo {author} {\bibfnamefont {V.}~\bibnamefont {Risingg\aa{}rd}}, \bibinfo {author} {\bibfnamefont {A.}~\bibnamefont {Sudb\o{}}}, \bibinfo {author} {\bibfnamefont {J.}~\bibnamefont {Linder}},\ and\ \bibinfo {author} {\bibfnamefont {A.}~\bibnamefont {Brataas}},\ }\bibfield  {title} {\bibinfo {title} {Current control of magnetism in two-dimensional {Fe$_3$GeTe$_2$}},\ }\href {https://doi.org/10.1103/PhysRevLett.122.217203} {\bibfield  {journal} {\bibinfo  {journal} {Phys. Rev. Lett.}\ }\textbf {\bibinfo {volume} {122}},\ \bibinfo {pages} {217203} (\bibinfo {year} {2019})}\BibitemShut {NoStop}%
\bibitem [{\citenamefont {Martin}\ \emph {et~al.}(2023)\citenamefont {Martin}, \citenamefont {Lee}, \citenamefont {Schmitt}, \citenamefont {Liedtke}, \citenamefont {Shahee}, \citenamefont {Simensen}, \citenamefont {Scholz}, \citenamefont {Saunderson}, \citenamefont {Go}, \citenamefont {Gradhand}, \citenamefont {Mokrousov}, \citenamefont {Denneulin}, \citenamefont {Kovács}, \citenamefont {Lotsch}, \citenamefont {Brataas},\ and\ \citenamefont {Kläui}}]{Martin02012023}%
  \BibitemOpen
  \bibfield  {author} {\bibinfo {author} {\bibfnamefont {F.}~\bibnamefont {Martin}}, \bibinfo {author} {\bibfnamefont {K.}~\bibnamefont {Lee}}, \bibinfo {author} {\bibfnamefont {M.}~\bibnamefont {Schmitt}}, \bibinfo {author} {\bibfnamefont {A.}~\bibnamefont {Liedtke}}, \bibinfo {author} {\bibfnamefont {A.}~\bibnamefont {Shahee}}, \bibinfo {author} {\bibfnamefont {H.~T.}\ \bibnamefont {Simensen}}, \bibinfo {author} {\bibfnamefont {T.}~\bibnamefont {Scholz}}, \bibinfo {author} {\bibfnamefont {T.~G.}\ \bibnamefont {Saunderson}}, \bibinfo {author} {\bibfnamefont {D.}~\bibnamefont {Go}}, \bibinfo {author} {\bibfnamefont {M.}~\bibnamefont {Gradhand}}, \bibinfo {author} {\bibfnamefont {Y.}~\bibnamefont {Mokrousov}}, \bibinfo {author} {\bibfnamefont {T.}~\bibnamefont {Denneulin}}, \bibinfo {author} {\bibfnamefont {A.}~\bibnamefont {Kovács}}, \bibinfo {author} {\bibfnamefont {B.}~\bibnamefont {Lotsch}}, \bibinfo {author} {\bibfnamefont {A.}~\bibnamefont {Brataas}},\ and\ \bibinfo {author} {\bibfnamefont
  {M.}~\bibnamefont {Kläui}},\ }\bibfield  {title} {\bibinfo {title} {Strong bulk spin–orbit torques quantified in the van der waals ferromagnet {Fe$_3$GeTe$_2$}},\ }\href {https://doi.org/10.1080/21663831.2022.2119108} {\bibfield  {journal} {\bibinfo  {journal} {Materials Research Letters}\ }\textbf {\bibinfo {volume} {11}},\ \bibinfo {pages} {84} (\bibinfo {year} {2023})}\BibitemShut {NoStop}%
\bibitem [{\citenamefont {Saunderson}\ \emph {et~al.}(2022)\citenamefont {Saunderson}, \citenamefont {Go}, \citenamefont {Bl\"ugel}, \citenamefont {Kl\"aui},\ and\ \citenamefont {Mokrousov}}]{PhysRevResearch.4.L042022}%
  \BibitemOpen
  \bibfield  {author} {\bibinfo {author} {\bibfnamefont {T.~G.}\ \bibnamefont {Saunderson}}, \bibinfo {author} {\bibfnamefont {D.}~\bibnamefont {Go}}, \bibinfo {author} {\bibfnamefont {S.}~\bibnamefont {Bl\"ugel}}, \bibinfo {author} {\bibfnamefont {M.}~\bibnamefont {Kl\"aui}},\ and\ \bibinfo {author} {\bibfnamefont {Y.}~\bibnamefont {Mokrousov}},\ }\bibfield  {title} {\bibinfo {title} {Hidden interplay of current-induced spin and orbital torques in bulk {Fe$_3$GeTe$_2$}},\ }\href {https://doi.org/10.1103/PhysRevResearch.4.L042022} {\bibfield  {journal} {\bibinfo  {journal} {Phys. Rev. Res.}\ }\textbf {\bibinfo {volume} {4}},\ \bibinfo {pages} {L042022} (\bibinfo {year} {2022})}\BibitemShut {NoStop}%
\bibitem [{\citenamefont {\ifmmode~\check{Z}\else \v{Z}\fi{}elezn\'y}\ \emph {et~al.}(2014)\citenamefont {\ifmmode~\check{Z}\else \v{Z}\fi{}elezn\'y}, \citenamefont {Gao}, \citenamefont {V\'yborn\'y}, \citenamefont {Zemen}, \citenamefont {Ma\ifmmode~\check{s}\else \v{s}\fi{}ek}, \citenamefont {Manchon}, \citenamefont {Wunderlich}, \citenamefont {Sinova},\ and\ \citenamefont {Jungwirth}}]{PhysRevLett.113.157201}%
  \BibitemOpen
  \bibfield  {author} {\bibinfo {author} {\bibfnamefont {J.}~\bibnamefont {\ifmmode~\check{Z}\else \v{Z}\fi{}elezn\'y}}, \bibinfo {author} {\bibfnamefont {H.}~\bibnamefont {Gao}}, \bibinfo {author} {\bibfnamefont {K.}~\bibnamefont {V\'yborn\'y}}, \bibinfo {author} {\bibfnamefont {J.}~\bibnamefont {Zemen}}, \bibinfo {author} {\bibfnamefont {J.}~\bibnamefont {Ma\ifmmode~\check{s}\else \v{s}\fi{}ek}}, \bibinfo {author} {\bibfnamefont {A.}~\bibnamefont {Manchon}}, \bibinfo {author} {\bibfnamefont {J.}~\bibnamefont {Wunderlich}}, \bibinfo {author} {\bibfnamefont {J.}~\bibnamefont {Sinova}},\ and\ \bibinfo {author} {\bibfnamefont {T.}~\bibnamefont {Jungwirth}},\ }\bibfield  {title} {\bibinfo {title} {Relativistic n\'eel-order fields induced by electrical current in antiferromagnets},\ }\href {https://doi.org/10.1103/PhysRevLett.113.157201} {\bibfield  {journal} {\bibinfo  {journal} {Phys. Rev. Lett.}\ }\textbf {\bibinfo {volume} {113}},\ \bibinfo {pages} {157201} (\bibinfo {year} {2014})}\BibitemShut {NoStop}%
\bibitem [{\citenamefont {Fang}\ and\ \citenamefont {Belashchenko}(2022)}]{PhysRevB.105.064412}%
  \BibitemOpen
  \bibfield  {author} {\bibinfo {author} {\bibfnamefont {W.}~\bibnamefont {Fang}}\ and\ \bibinfo {author} {\bibfnamefont {K.~D.}\ \bibnamefont {Belashchenko}},\ }\bibfield  {title} {\bibinfo {title} {First-principles calculations of spin-orbit torques in {Mn$_2$Au}/heavy-metal bilayers},\ }\href {https://doi.org/10.1103/PhysRevB.105.064412} {\bibfield  {journal} {\bibinfo  {journal} {Phys. Rev. B}\ }\textbf {\bibinfo {volume} {105}},\ \bibinfo {pages} {064412} (\bibinfo {year} {2022})}\BibitemShut {NoStop}%
\bibitem [{\citenamefont {Xue}\ and\ \citenamefont {Haney}(2021)}]{PhysRevB.104.224414}%
  \BibitemOpen
  \bibfield  {author} {\bibinfo {author} {\bibfnamefont {F.}~\bibnamefont {Xue}}\ and\ \bibinfo {author} {\bibfnamefont {P.~M.}\ \bibnamefont {Haney}},\ }\bibfield  {title} {\bibinfo {title} {Intrinsic staggered spin-orbit torque for the electrical control of antiferromagnets: Application to {CrI$_3$}},\ }\href {https://doi.org/10.1103/PhysRevB.104.224414} {\bibfield  {journal} {\bibinfo  {journal} {Phys. Rev. B}\ }\textbf {\bibinfo {volume} {104}},\ \bibinfo {pages} {224414} (\bibinfo {year} {2021})}\BibitemShut {NoStop}%
\bibitem [{\citenamefont {Xue}\ \emph {et~al.}(2023)\citenamefont {Xue}, \citenamefont {Stiles},\ and\ \citenamefont {Haney}}]{PhysRevB.108.144422}%
  \BibitemOpen
  \bibfield  {author} {\bibinfo {author} {\bibfnamefont {F.}~\bibnamefont {Xue}}, \bibinfo {author} {\bibfnamefont {M.~D.}\ \bibnamefont {Stiles}},\ and\ \bibinfo {author} {\bibfnamefont {P.~M.}\ \bibnamefont {Haney}},\ }\bibfield  {title} {\bibinfo {title} {Angular dependence of spin-orbit torque in monolayer {Fe$_3$GeTe$_2$}},\ }\href {https://doi.org/10.1103/PhysRevB.108.144422} {\bibfield  {journal} {\bibinfo  {journal} {Phys. Rev. B}\ }\textbf {\bibinfo {volume} {108}},\ \bibinfo {pages} {144422} (\bibinfo {year} {2023})}\BibitemShut {NoStop}%
\bibitem [{\citenamefont {Freimuth}\ \emph {et~al.}(2014)\citenamefont {Freimuth}, \citenamefont {Bl\"ugel},\ and\ \citenamefont {Mokrousov}}]{PhysRevB.90.174423}%
  \BibitemOpen
  \bibfield  {author} {\bibinfo {author} {\bibfnamefont {F.}~\bibnamefont {Freimuth}}, \bibinfo {author} {\bibfnamefont {S.}~\bibnamefont {Bl\"ugel}},\ and\ \bibinfo {author} {\bibfnamefont {Y.}~\bibnamefont {Mokrousov}},\ }\bibfield  {title} {\bibinfo {title} {Spin-orbit torques in {Co}/{Pt}(111) and {Mn}/{W}(001) magnetic bilayers from first principles},\ }\href {https://doi.org/10.1103/PhysRevB.90.174423} {\bibfield  {journal} {\bibinfo  {journal} {Phys. Rev. B}\ }\textbf {\bibinfo {volume} {90}},\ \bibinfo {pages} {174423} (\bibinfo {year} {2014})}\BibitemShut {NoStop}%
\bibitem [{\citenamefont {Giannozzi}\ \emph {et~al.}(2009)\citenamefont {Giannozzi}, \citenamefont {Baroni}, \citenamefont {Bonini}, \citenamefont {Calandra}, \citenamefont {Car}, \citenamefont {Cavazzoni}, \citenamefont {Ceresoli}, \citenamefont {Chiarotti}, \citenamefont {Cococcioni}, \citenamefont {Dabo}, \citenamefont {Dal~Corso}, \citenamefont {de~Gironcoli}, \citenamefont {Fabris}, \citenamefont {Fratesi}, \citenamefont {Gebauer}, \citenamefont {Gerstmann}, \citenamefont {Gougoussis}, \citenamefont {Kokalj}, \citenamefont {Lazzeri}, \citenamefont {Martin-Samos}, \citenamefont {Marzari}, \citenamefont {Mauri}, \citenamefont {Mazzarello}, \citenamefont {Paolini}, \citenamefont {Pasquarello}, \citenamefont {Paulatto}, \citenamefont {Sbraccia}, \citenamefont {Scandolo}, \citenamefont {Sclauzero}, \citenamefont {Seitsonen}, \citenamefont {Smogunov}, \citenamefont {Umari},\ and\ \citenamefont {Wentzcovitch}}]{Giannozzi_2009}%
  \BibitemOpen
  \bibfield  {author} {\bibinfo {author} {\bibfnamefont {P.}~\bibnamefont {Giannozzi}}, \bibinfo {author} {\bibfnamefont {S.}~\bibnamefont {Baroni}}, \bibinfo {author} {\bibfnamefont {N.}~\bibnamefont {Bonini}}, \bibinfo {author} {\bibfnamefont {M.}~\bibnamefont {Calandra}}, \bibinfo {author} {\bibfnamefont {R.}~\bibnamefont {Car}}, \bibinfo {author} {\bibfnamefont {C.}~\bibnamefont {Cavazzoni}}, \bibinfo {author} {\bibfnamefont {D.}~\bibnamefont {Ceresoli}}, \bibinfo {author} {\bibfnamefont {G.~L.}\ \bibnamefont {Chiarotti}}, \bibinfo {author} {\bibfnamefont {M.}~\bibnamefont {Cococcioni}}, \bibinfo {author} {\bibfnamefont {I.}~\bibnamefont {Dabo}}, \bibinfo {author} {\bibfnamefont {A.}~\bibnamefont {Dal~Corso}}, \bibinfo {author} {\bibfnamefont {S.}~\bibnamefont {de~Gironcoli}}, \bibinfo {author} {\bibfnamefont {S.}~\bibnamefont {Fabris}}, \bibinfo {author} {\bibfnamefont {G.}~\bibnamefont {Fratesi}}, \bibinfo {author} {\bibfnamefont {R.}~\bibnamefont {Gebauer}}, \bibinfo {author} {\bibfnamefont
  {U.}~\bibnamefont {Gerstmann}}, \bibinfo {author} {\bibfnamefont {C.}~\bibnamefont {Gougoussis}}, \bibinfo {author} {\bibfnamefont {A.}~\bibnamefont {Kokalj}}, \bibinfo {author} {\bibfnamefont {M.}~\bibnamefont {Lazzeri}}, \bibinfo {author} {\bibfnamefont {L.}~\bibnamefont {Martin-Samos}}, \bibinfo {author} {\bibfnamefont {N.}~\bibnamefont {Marzari}}, \bibinfo {author} {\bibfnamefont {F.}~\bibnamefont {Mauri}}, \bibinfo {author} {\bibfnamefont {R.}~\bibnamefont {Mazzarello}}, \bibinfo {author} {\bibfnamefont {S.}~\bibnamefont {Paolini}}, \bibinfo {author} {\bibfnamefont {A.}~\bibnamefont {Pasquarello}}, \bibinfo {author} {\bibfnamefont {L.}~\bibnamefont {Paulatto}}, \bibinfo {author} {\bibfnamefont {C.}~\bibnamefont {Sbraccia}}, \bibinfo {author} {\bibfnamefont {S.}~\bibnamefont {Scandolo}}, \bibinfo {author} {\bibfnamefont {G.}~\bibnamefont {Sclauzero}}, \bibinfo {author} {\bibfnamefont {A.~P.}\ \bibnamefont {Seitsonen}}, \bibinfo {author} {\bibfnamefont {A.}~\bibnamefont {Smogunov}}, \bibinfo {author}
  {\bibfnamefont {P.}~\bibnamefont {Umari}},\ and\ \bibinfo {author} {\bibfnamefont {R.~M.}\ \bibnamefont {Wentzcovitch}},\ }\bibfield  {title} {\bibinfo {title} {Quantum espresso: a modular and open-source software project for quantum simulations of materials},\ }\href {https://doi.org/10.1088/0953-8984/21/39/395502} {\bibfield  {journal} {\bibinfo  {journal} {Journal of Physics: Condensed Matter}\ }\textbf {\bibinfo {volume} {21}},\ \bibinfo {pages} {395502} (\bibinfo {year} {2009})}\BibitemShut {NoStop}%
\bibitem [{\citenamefont {Giannozzi}\ \emph {et~al.}(2017)\citenamefont {Giannozzi}, \citenamefont {Andreussi}, \citenamefont {Brumme}, \citenamefont {Bunau}, \citenamefont {Buongiorno~Nardelli}, \citenamefont {Calandra}, \citenamefont {Car}, \citenamefont {Cavazzoni}, \citenamefont {Ceresoli}, \citenamefont {Cococcioni}, \citenamefont {Colonna}, \citenamefont {Carnimeo}, \citenamefont {Dal~Corso}, \citenamefont {de~Gironcoli}, \citenamefont {Delugas}, \citenamefont {DiStasio}, \citenamefont {Ferretti}, \citenamefont {Floris}, \citenamefont {Fratesi}, \citenamefont {Fugallo}, \citenamefont {Gebauer}, \citenamefont {Gerstmann}, \citenamefont {Giustino}, \citenamefont {Gorni}, \citenamefont {Jia}, \citenamefont {Kawamura}, \citenamefont {Ko}, \citenamefont {Kokalj}, \citenamefont {Küçükbenli}, \citenamefont {Lazzeri}, \citenamefont {Marsili}, \citenamefont {Marzari}, \citenamefont {Mauri}, \citenamefont {Nguyen}, \citenamefont {Nguyen}, \citenamefont {Otero-de-la Roza}, \citenamefont {Paulatto},
  \citenamefont {Poncé}, \citenamefont {Rocca}, \citenamefont {Sabatini}, \citenamefont {Santra}, \citenamefont {Schlipf}, \citenamefont {Seitsonen}, \citenamefont {Smogunov}, \citenamefont {Timrov}, \citenamefont {Thonhauser}, \citenamefont {Umari}, \citenamefont {Vast}, \citenamefont {Wu},\ and\ \citenamefont {Baroni}}]{Giannozzi_2017}%
  \BibitemOpen
  \bibfield  {author} {\bibinfo {author} {\bibfnamefont {P.}~\bibnamefont {Giannozzi}}, \bibinfo {author} {\bibfnamefont {O.}~\bibnamefont {Andreussi}}, \bibinfo {author} {\bibfnamefont {T.}~\bibnamefont {Brumme}}, \bibinfo {author} {\bibfnamefont {O.}~\bibnamefont {Bunau}}, \bibinfo {author} {\bibfnamefont {M.}~\bibnamefont {Buongiorno~Nardelli}}, \bibinfo {author} {\bibfnamefont {M.}~\bibnamefont {Calandra}}, \bibinfo {author} {\bibfnamefont {R.}~\bibnamefont {Car}}, \bibinfo {author} {\bibfnamefont {C.}~\bibnamefont {Cavazzoni}}, \bibinfo {author} {\bibfnamefont {D.}~\bibnamefont {Ceresoli}}, \bibinfo {author} {\bibfnamefont {M.}~\bibnamefont {Cococcioni}}, \bibinfo {author} {\bibfnamefont {N.}~\bibnamefont {Colonna}}, \bibinfo {author} {\bibfnamefont {I.}~\bibnamefont {Carnimeo}}, \bibinfo {author} {\bibfnamefont {A.}~\bibnamefont {Dal~Corso}}, \bibinfo {author} {\bibfnamefont {S.}~\bibnamefont {de~Gironcoli}}, \bibinfo {author} {\bibfnamefont {P.}~\bibnamefont {Delugas}}, \bibinfo {author} {\bibfnamefont
  {R.~A.}\ \bibnamefont {DiStasio}}, \bibinfo {author} {\bibfnamefont {A.}~\bibnamefont {Ferretti}}, \bibinfo {author} {\bibfnamefont {A.}~\bibnamefont {Floris}}, \bibinfo {author} {\bibfnamefont {G.}~\bibnamefont {Fratesi}}, \bibinfo {author} {\bibfnamefont {G.}~\bibnamefont {Fugallo}}, \bibinfo {author} {\bibfnamefont {R.}~\bibnamefont {Gebauer}}, \bibinfo {author} {\bibfnamefont {U.}~\bibnamefont {Gerstmann}}, \bibinfo {author} {\bibfnamefont {F.}~\bibnamefont {Giustino}}, \bibinfo {author} {\bibfnamefont {T.}~\bibnamefont {Gorni}}, \bibinfo {author} {\bibfnamefont {J.}~\bibnamefont {Jia}}, \bibinfo {author} {\bibfnamefont {M.}~\bibnamefont {Kawamura}}, \bibinfo {author} {\bibfnamefont {H.-Y.}\ \bibnamefont {Ko}}, \bibinfo {author} {\bibfnamefont {A.}~\bibnamefont {Kokalj}}, \bibinfo {author} {\bibfnamefont {E.}~\bibnamefont {Küçükbenli}}, \bibinfo {author} {\bibfnamefont {M.}~\bibnamefont {Lazzeri}}, \bibinfo {author} {\bibfnamefont {M.}~\bibnamefont {Marsili}}, \bibinfo {author} {\bibfnamefont
  {N.}~\bibnamefont {Marzari}}, \bibinfo {author} {\bibfnamefont {F.}~\bibnamefont {Mauri}}, \bibinfo {author} {\bibfnamefont {N.~L.}\ \bibnamefont {Nguyen}}, \bibinfo {author} {\bibfnamefont {H.-V.}\ \bibnamefont {Nguyen}}, \bibinfo {author} {\bibfnamefont {A.}~\bibnamefont {Otero-de-la Roza}}, \bibinfo {author} {\bibfnamefont {L.}~\bibnamefont {Paulatto}}, \bibinfo {author} {\bibfnamefont {S.}~\bibnamefont {Poncé}}, \bibinfo {author} {\bibfnamefont {D.}~\bibnamefont {Rocca}}, \bibinfo {author} {\bibfnamefont {R.}~\bibnamefont {Sabatini}}, \bibinfo {author} {\bibfnamefont {B.}~\bibnamefont {Santra}}, \bibinfo {author} {\bibfnamefont {M.}~\bibnamefont {Schlipf}}, \bibinfo {author} {\bibfnamefont {A.~P.}\ \bibnamefont {Seitsonen}}, \bibinfo {author} {\bibfnamefont {A.}~\bibnamefont {Smogunov}}, \bibinfo {author} {\bibfnamefont {I.}~\bibnamefont {Timrov}}, \bibinfo {author} {\bibfnamefont {T.}~\bibnamefont {Thonhauser}}, \bibinfo {author} {\bibfnamefont {P.}~\bibnamefont {Umari}}, \bibinfo {author}
  {\bibfnamefont {N.}~\bibnamefont {Vast}}, \bibinfo {author} {\bibfnamefont {X.}~\bibnamefont {Wu}},\ and\ \bibinfo {author} {\bibfnamefont {S.}~\bibnamefont {Baroni}},\ }\bibfield  {title} {\bibinfo {title} {Advanced capabilities for materials modelling with quantum espresso},\ }\href {https://doi.org/10.1088/1361-648X/aa8f79} {\bibfield  {journal} {\bibinfo  {journal} {Journal of Physics: Condensed Matter}\ }\textbf {\bibinfo {volume} {29}},\ \bibinfo {pages} {465901} (\bibinfo {year} {2017})}\BibitemShut {NoStop}%
\bibitem [{\citenamefont {Giannozzi}\ \emph {et~al.}(2020)\citenamefont {Giannozzi}, \citenamefont {Baseggio}, \citenamefont {Bonfà}, \citenamefont {Brunato}, \citenamefont {Car}, \citenamefont {Carnimeo}, \citenamefont {Cavazzoni}, \citenamefont {de~Gironcoli}, \citenamefont {Delugas}, \citenamefont {Ferrari~Ruffino}, \citenamefont {Ferretti}, \citenamefont {Marzari}, \citenamefont {Timrov}, \citenamefont {Urru},\ and\ \citenamefont {Baroni}}]{10.1063/5.0005082}%
  \BibitemOpen
  \bibfield  {author} {\bibinfo {author} {\bibfnamefont {P.}~\bibnamefont {Giannozzi}}, \bibinfo {author} {\bibfnamefont {O.}~\bibnamefont {Baseggio}}, \bibinfo {author} {\bibfnamefont {P.}~\bibnamefont {Bonfà}}, \bibinfo {author} {\bibfnamefont {D.}~\bibnamefont {Brunato}}, \bibinfo {author} {\bibfnamefont {R.}~\bibnamefont {Car}}, \bibinfo {author} {\bibfnamefont {I.}~\bibnamefont {Carnimeo}}, \bibinfo {author} {\bibfnamefont {C.}~\bibnamefont {Cavazzoni}}, \bibinfo {author} {\bibfnamefont {S.}~\bibnamefont {de~Gironcoli}}, \bibinfo {author} {\bibfnamefont {P.}~\bibnamefont {Delugas}}, \bibinfo {author} {\bibfnamefont {F.}~\bibnamefont {Ferrari~Ruffino}}, \bibinfo {author} {\bibfnamefont {A.}~\bibnamefont {Ferretti}}, \bibinfo {author} {\bibfnamefont {N.}~\bibnamefont {Marzari}}, \bibinfo {author} {\bibfnamefont {I.}~\bibnamefont {Timrov}}, \bibinfo {author} {\bibfnamefont {A.}~\bibnamefont {Urru}},\ and\ \bibinfo {author} {\bibfnamefont {S.}~\bibnamefont {Baroni}},\ }\bibfield  {title} {\bibinfo {title}
  {Quantum espresso toward the exascale},\ }\href {https://doi.org/10.1063/5.0005082} {\bibfield  {journal} {\bibinfo  {journal} {The Journal of Chemical Physics}\ }\textbf {\bibinfo {volume} {152}},\ \bibinfo {pages} {154105} (\bibinfo {year} {2020})}\BibitemShut {NoStop}%
\bibitem [{\citenamefont {Hamann}(2013)}]{PhysRevB.88.085117}%
  \BibitemOpen
  \bibfield  {author} {\bibinfo {author} {\bibfnamefont {D.~R.}\ \bibnamefont {Hamann}},\ }\bibfield  {title} {\bibinfo {title} {Optimized norm-conserving vanderbilt pseudopotentials},\ }\href {https://doi.org/10.1103/PhysRevB.88.085117} {\bibfield  {journal} {\bibinfo  {journal} {Phys. Rev. B}\ }\textbf {\bibinfo {volume} {88}},\ \bibinfo {pages} {085117} (\bibinfo {year} {2013})}\BibitemShut {NoStop}%
\bibitem [{\citenamefont {{van Setten}}\ \emph {et~al.}(2018)\citenamefont {{van Setten}}, \citenamefont {Giantomassi}, \citenamefont {Bousquet}, \citenamefont {Verstraete}, \citenamefont {Hamann}, \citenamefont {Gonze},\ and\ \citenamefont {Rignanese}}]{VANSETTEN201839}%
  \BibitemOpen
  \bibfield  {author} {\bibinfo {author} {\bibfnamefont {M.}~\bibnamefont {{van Setten}}}, \bibinfo {author} {\bibfnamefont {M.}~\bibnamefont {Giantomassi}}, \bibinfo {author} {\bibfnamefont {E.}~\bibnamefont {Bousquet}}, \bibinfo {author} {\bibfnamefont {M.}~\bibnamefont {Verstraete}}, \bibinfo {author} {\bibfnamefont {D.}~\bibnamefont {Hamann}}, \bibinfo {author} {\bibfnamefont {X.}~\bibnamefont {Gonze}},\ and\ \bibinfo {author} {\bibfnamefont {G.-M.}\ \bibnamefont {Rignanese}},\ }\bibfield  {title} {\bibinfo {title} {The pseudodojo: Training and grading a 85 element optimized norm-conserving pseudopotential table},\ }\href {https://doi.org/https://doi.org/10.1016/j.cpc.2018.01.012} {\bibfield  {journal} {\bibinfo  {journal} {Computer Physics Communications}\ }\textbf {\bibinfo {volume} {226}},\ \bibinfo {pages} {39} (\bibinfo {year} {2018})}\BibitemShut {NoStop}%
\bibitem [{\citenamefont {Ismail-Beigi}(2006)}]{PhysRevB.73.233103}%
  \BibitemOpen
  \bibfield  {author} {\bibinfo {author} {\bibfnamefont {S.}~\bibnamefont {Ismail-Beigi}},\ }\bibfield  {title} {\bibinfo {title} {Truncation of periodic image interactions for confined systems},\ }\href {https://doi.org/10.1103/PhysRevB.73.233103} {\bibfield  {journal} {\bibinfo  {journal} {Phys. Rev. B}\ }\textbf {\bibinfo {volume} {73}},\ \bibinfo {pages} {233103} (\bibinfo {year} {2006})}\BibitemShut {NoStop}%
\bibitem [{\citenamefont {Sohier}\ \emph {et~al.}(2017)\citenamefont {Sohier}, \citenamefont {Calandra},\ and\ \citenamefont {Mauri}}]{PhysRevB.96.075448}%
  \BibitemOpen
  \bibfield  {author} {\bibinfo {author} {\bibfnamefont {T.}~\bibnamefont {Sohier}}, \bibinfo {author} {\bibfnamefont {M.}~\bibnamefont {Calandra}},\ and\ \bibinfo {author} {\bibfnamefont {F.}~\bibnamefont {Mauri}},\ }\bibfield  {title} {\bibinfo {title} {Density functional perturbation theory for gated two-dimensional heterostructures: Theoretical developments and application to flexural phonons in graphene},\ }\href {https://doi.org/10.1103/PhysRevB.96.075448} {\bibfield  {journal} {\bibinfo  {journal} {Phys. Rev. B}\ }\textbf {\bibinfo {volume} {96}},\ \bibinfo {pages} {075448} (\bibinfo {year} {2017})}\BibitemShut {NoStop}%
\bibitem [{\citenamefont {Perdew}\ \emph {et~al.}(1996)\citenamefont {Perdew}, \citenamefont {Burke},\ and\ \citenamefont {Ernzerhof}}]{PhysRevLett.77.3865}%
  \BibitemOpen
  \bibfield  {author} {\bibinfo {author} {\bibfnamefont {J.~P.}\ \bibnamefont {Perdew}}, \bibinfo {author} {\bibfnamefont {K.}~\bibnamefont {Burke}},\ and\ \bibinfo {author} {\bibfnamefont {M.}~\bibnamefont {Ernzerhof}},\ }\bibfield  {title} {\bibinfo {title} {Generalized gradient approximation made simple},\ }\href {https://doi.org/10.1103/PhysRevLett.77.3865} {\bibfield  {journal} {\bibinfo  {journal} {Phys. Rev. Lett.}\ }\textbf {\bibinfo {volume} {77}},\ \bibinfo {pages} {3865} (\bibinfo {year} {1996})}\BibitemShut {NoStop}%
\bibitem [{\citenamefont {Deiseroth}\ \emph {et~al.}(2006)\citenamefont {Deiseroth}, \citenamefont {Aleksandrov}, \citenamefont {Reiner}, \citenamefont {Kienle},\ and\ \citenamefont {Kremer}}]{https://doi.org/10.1002/ejic.200501020}%
  \BibitemOpen
  \bibfield  {author} {\bibinfo {author} {\bibfnamefont {H.-J.}\ \bibnamefont {Deiseroth}}, \bibinfo {author} {\bibfnamefont {K.}~\bibnamefont {Aleksandrov}}, \bibinfo {author} {\bibfnamefont {C.}~\bibnamefont {Reiner}}, \bibinfo {author} {\bibfnamefont {L.}~\bibnamefont {Kienle}},\ and\ \bibinfo {author} {\bibfnamefont {R.~K.}\ \bibnamefont {Kremer}},\ }\bibfield  {title} {\bibinfo {title} {{Fe$_3$GeTe$_2$} and {Ni$_3$GeTe$_2$} – two new layered transition-metal compounds: Crystal structures, hrtem investigations, and magnetic and electrical properties},\ }\href {https://doi.org/https://doi.org/10.1002/ejic.200501020} {\bibfield  {journal} {\bibinfo  {journal} {European Journal of Inorganic Chemistry}\ }\textbf {\bibinfo {volume} {2006}},\ \bibinfo {pages} {1561} (\bibinfo {year} {2006})}\BibitemShut {NoStop}%
\bibitem [{\citenamefont {May}\ \emph {et~al.}(2016)\citenamefont {May}, \citenamefont {Calder}, \citenamefont {Cantoni}, \citenamefont {Cao},\ and\ \citenamefont {McGuire}}]{PhysRevB.93.014411}%
  \BibitemOpen
  \bibfield  {author} {\bibinfo {author} {\bibfnamefont {A.~F.}\ \bibnamefont {May}}, \bibinfo {author} {\bibfnamefont {S.}~\bibnamefont {Calder}}, \bibinfo {author} {\bibfnamefont {C.}~\bibnamefont {Cantoni}}, \bibinfo {author} {\bibfnamefont {H.}~\bibnamefont {Cao}},\ and\ \bibinfo {author} {\bibfnamefont {M.~A.}\ \bibnamefont {McGuire}},\ }\bibfield  {title} {\bibinfo {title} {Magnetic structure and phase stability of the van der waals bonded ferromagnet {Fe$_{3-x}$GeTe$_2$}},\ }\href {https://doi.org/10.1103/PhysRevB.93.014411} {\bibfield  {journal} {\bibinfo  {journal} {Phys. Rev. B}\ }\textbf {\bibinfo {volume} {93}},\ \bibinfo {pages} {014411} (\bibinfo {year} {2016})}\BibitemShut {NoStop}%
\bibitem [{\citenamefont {Wu}\ \emph {et~al.}(2024)\citenamefont {Wu}, \citenamefont {He}, \citenamefont {Gu} \emph {et~al.}}]{wu2024a}%
  \BibitemOpen
  \bibfield  {author} {\bibinfo {author} {\bibfnamefont {S.}~\bibnamefont {Wu}}, \bibinfo {author} {\bibfnamefont {Z.}~\bibnamefont {He}}, \bibinfo {author} {\bibfnamefont {M.}~\bibnamefont {Gu}}, \emph {et~al.},\ }\bibfield  {title} {{\selectlanguage {en}\bibinfo {title} {Robust ferromagnetism in wafer-scale {Fe$_3$GaTe$_2$} above room-temperature}},\ }\href {https://doi.org/10.1038/s41467-024-54936-1} {\bibfield  {journal} {\bibinfo  {journal} {Nat Commun}\ }\textbf {\bibinfo {volume} {15}},\ \bibinfo {pages} {10765} (\bibinfo {year} {2024})}\BibitemShut {NoStop}%
\bibitem [{\citenamefont {Perdew}\ and\ \citenamefont {Wang}(1992)}]{PhysRevB.45.13244}%
  \BibitemOpen
  \bibfield  {author} {\bibinfo {author} {\bibfnamefont {J.~P.}\ \bibnamefont {Perdew}}\ and\ \bibinfo {author} {\bibfnamefont {Y.}~\bibnamefont {Wang}},\ }\bibfield  {title} {\bibinfo {title} {Accurate and simple analytic representation of the electron-gas correlation energy},\ }\href {https://doi.org/10.1103/PhysRevB.45.13244} {\bibfield  {journal} {\bibinfo  {journal} {Phys. Rev. B}\ }\textbf {\bibinfo {volume} {45}},\ \bibinfo {pages} {13244} (\bibinfo {year} {1992})}\BibitemShut {NoStop}%
\bibitem [{\citenamefont {Zhuang}\ \emph {et~al.}(2016)\citenamefont {Zhuang}, \citenamefont {Kent},\ and\ \citenamefont {Hennig}}]{PhysRevB.93.134407}%
  \BibitemOpen
  \bibfield  {author} {\bibinfo {author} {\bibfnamefont {H.~L.}\ \bibnamefont {Zhuang}}, \bibinfo {author} {\bibfnamefont {P.~R.~C.}\ \bibnamefont {Kent}},\ and\ \bibinfo {author} {\bibfnamefont {R.~G.}\ \bibnamefont {Hennig}},\ }\bibfield  {title} {\bibinfo {title} {Strong anisotropy and magnetostriction in the two-dimensional stoner ferromagnet {Fe$_3$GeTe$_2$}},\ }\href {https://doi.org/10.1103/PhysRevB.93.134407} {\bibfield  {journal} {\bibinfo  {journal} {Phys. Rev. B}\ }\textbf {\bibinfo {volume} {93}},\ \bibinfo {pages} {134407} (\bibinfo {year} {2016})}\BibitemShut {NoStop}%
\bibitem [{\citenamefont {Zhu}\ \emph {et~al.}(2016)\citenamefont {Zhu}, \citenamefont {Janoschek}, \citenamefont {Chaves}, \citenamefont {Cezar}, \citenamefont {Durakiewicz}, \citenamefont {Ronning}, \citenamefont {Sassa}, \citenamefont {Mansson}, \citenamefont {Scott}, \citenamefont {Wakeham}, \citenamefont {Bauer},\ and\ \citenamefont {Thompson}}]{PhysRevB.93.144404}%
  \BibitemOpen
  \bibfield  {author} {\bibinfo {author} {\bibfnamefont {J.-X.}\ \bibnamefont {Zhu}}, \bibinfo {author} {\bibfnamefont {M.}~\bibnamefont {Janoschek}}, \bibinfo {author} {\bibfnamefont {D.~S.}\ \bibnamefont {Chaves}}, \bibinfo {author} {\bibfnamefont {J.~C.}\ \bibnamefont {Cezar}}, \bibinfo {author} {\bibfnamefont {T.}~\bibnamefont {Durakiewicz}}, \bibinfo {author} {\bibfnamefont {F.}~\bibnamefont {Ronning}}, \bibinfo {author} {\bibfnamefont {Y.}~\bibnamefont {Sassa}}, \bibinfo {author} {\bibfnamefont {M.}~\bibnamefont {Mansson}}, \bibinfo {author} {\bibfnamefont {B.~L.}\ \bibnamefont {Scott}}, \bibinfo {author} {\bibfnamefont {N.}~\bibnamefont {Wakeham}}, \bibinfo {author} {\bibfnamefont {E.~D.}\ \bibnamefont {Bauer}},\ and\ \bibinfo {author} {\bibfnamefont {J.~D.}\ \bibnamefont {Thompson}},\ }\bibfield  {title} {\bibinfo {title} {Electronic correlation and magnetism in the ferromagnetic metal {Fe$_3$GeTe$_2$}},\ }\href {https://doi.org/10.1103/PhysRevB.93.144404} {\bibfield  {journal} {\bibinfo  {journal}
  {Phys. Rev. B}\ }\textbf {\bibinfo {volume} {93}},\ \bibinfo {pages} {144404} (\bibinfo {year} {2016})}\BibitemShut {NoStop}%
\bibitem [{\citenamefont {Methfessel}\ and\ \citenamefont {Paxton}(1989)}]{PhysRevB.40.3616}%
  \BibitemOpen
  \bibfield  {author} {\bibinfo {author} {\bibfnamefont {M.}~\bibnamefont {Methfessel}}\ and\ \bibinfo {author} {\bibfnamefont {A.~T.}\ \bibnamefont {Paxton}},\ }\bibfield  {title} {\bibinfo {title} {High-precision sampling for brillouin-zone integration in metals},\ }\href {https://doi.org/10.1103/PhysRevB.40.3616} {\bibfield  {journal} {\bibinfo  {journal} {Phys. Rev. B}\ }\textbf {\bibinfo {volume} {40}},\ \bibinfo {pages} {3616} (\bibinfo {year} {1989})}\BibitemShut {NoStop}%
\bibitem [{\citenamefont {Sakuma}(2013)}]{PhysRevB.87.235109}%
  \BibitemOpen
  \bibfield  {author} {\bibinfo {author} {\bibfnamefont {R.}~\bibnamefont {Sakuma}},\ }\bibfield  {title} {\bibinfo {title} {Symmetry-adapted wannier functions in the maximal localization procedure},\ }\href {https://doi.org/10.1103/PhysRevB.87.235109} {\bibfield  {journal} {\bibinfo  {journal} {Phys. Rev. B}\ }\textbf {\bibinfo {volume} {87}},\ \bibinfo {pages} {235109} (\bibinfo {year} {2013})}\BibitemShut {NoStop}%
\bibitem [{\citenamefont {Tsirkin}(2021)}]{tsirkin2021a}%
  \BibitemOpen
  \bibfield  {author} {\bibinfo {author} {\bibfnamefont {S.}~\bibnamefont {Tsirkin}},\ }\bibfield  {title} {{\selectlanguage {en}\bibinfo {title} {High performance wannier interpolation of berry curvature and related quantities with wannierberri code}},\ }\bibfield  {journal} {\bibinfo  {journal} {npj Comput Mater}\ }\textbf {\bibinfo {volume} {7, 33}},\ \href {https://doi.org/10.1038/s41524-021-00498-5} {10.1038/s41524-021-00498-5} (\bibinfo {year} {2021})\BibitemShut {NoStop}%
\bibitem [{\citenamefont {Marzari}\ and\ \citenamefont {Vanderbilt}(1997)}]{PhysRevB.56.12847}%
  \BibitemOpen
  \bibfield  {author} {\bibinfo {author} {\bibfnamefont {N.}~\bibnamefont {Marzari}}\ and\ \bibinfo {author} {\bibfnamefont {D.}~\bibnamefont {Vanderbilt}},\ }\bibfield  {title} {\bibinfo {title} {Maximally localized generalized wannier functions for composite energy bands},\ }\href {https://doi.org/10.1103/PhysRevB.56.12847} {\bibfield  {journal} {\bibinfo  {journal} {Phys. Rev. B}\ }\textbf {\bibinfo {volume} {56}},\ \bibinfo {pages} {12847} (\bibinfo {year} {1997})}\BibitemShut {NoStop}%
\bibitem [{\citenamefont {Souza}\ \emph {et~al.}(2001)\citenamefont {Souza}, \citenamefont {Marzari},\ and\ \citenamefont {Vanderbilt}}]{PhysRevB.65.035109}%
  \BibitemOpen
  \bibfield  {author} {\bibinfo {author} {\bibfnamefont {I.}~\bibnamefont {Souza}}, \bibinfo {author} {\bibfnamefont {N.}~\bibnamefont {Marzari}},\ and\ \bibinfo {author} {\bibfnamefont {D.}~\bibnamefont {Vanderbilt}},\ }\bibfield  {title} {\bibinfo {title} {Maximally localized wannier functions for entangled energy bands},\ }\href {https://doi.org/10.1103/PhysRevB.65.035109} {\bibfield  {journal} {\bibinfo  {journal} {Phys. Rev. B}\ }\textbf {\bibinfo {volume} {65}},\ \bibinfo {pages} {035109} (\bibinfo {year} {2001})}\BibitemShut {NoStop}%
\bibitem [{\citenamefont {Iraola}\ \emph {et~al.}(2022)\citenamefont {Iraola}, \citenamefont {Mañes}, \citenamefont {Bradlyn}, \citenamefont {Horton}, \citenamefont {Neupert}, \citenamefont {Vergniory},\ and\ \citenamefont {Tsirkin}}]{IRAOLA2022108226}%
  \BibitemOpen
  \bibfield  {author} {\bibinfo {author} {\bibfnamefont {M.}~\bibnamefont {Iraola}}, \bibinfo {author} {\bibfnamefont {J.~L.}\ \bibnamefont {Mañes}}, \bibinfo {author} {\bibfnamefont {B.}~\bibnamefont {Bradlyn}}, \bibinfo {author} {\bibfnamefont {M.~K.}\ \bibnamefont {Horton}}, \bibinfo {author} {\bibfnamefont {T.}~\bibnamefont {Neupert}}, \bibinfo {author} {\bibfnamefont {M.~G.}\ \bibnamefont {Vergniory}},\ and\ \bibinfo {author} {\bibfnamefont {S.~S.}\ \bibnamefont {Tsirkin}},\ }\bibfield  {title} {\bibinfo {title} {Irrep: Symmetry eigenvalues and irreducible representations of ab initio band structures},\ }\href {https://doi.org/https://doi.org/10.1016/j.cpc.2021.108226} {\bibfield  {journal} {\bibinfo  {journal} {Computer Physics Communications}\ }\textbf {\bibinfo {volume} {272}},\ \bibinfo {pages} {108226} (\bibinfo {year} {2022})}\BibitemShut {NoStop}%
\bibitem [{\citenamefont {Ghosh}\ \emph {et~al.}(2023)\citenamefont {Ghosh}, \citenamefont {Ershadrad}, \citenamefont {Borisov} \emph {et~al.}}]{ghosh2023a}%
  \BibitemOpen
  \bibfield  {author} {\bibinfo {author} {\bibfnamefont {S.}~\bibnamefont {Ghosh}}, \bibinfo {author} {\bibfnamefont {S.}~\bibnamefont {Ershadrad}}, \bibinfo {author} {\bibfnamefont {V.}~\bibnamefont {Borisov}}, \emph {et~al.},\ }\bibfield  {title} {{\selectlanguage {en}\bibinfo {title} {Unraveling effects of electron correlation in two-dimensional {Fe$_n$GeTe$_2$} ($n = 3, 4, 5$) by dynamical mean field theory}},\ }\href {https://doi.org/10.1038/s41524-023-01024-5} {\bibfield  {journal} {\bibinfo  {journal} {npj Comput Mater}\ }\textbf {\bibinfo {volume} {9}},\ \bibinfo {pages} {86} (\bibinfo {year} {2023})}\BibitemShut {NoStop}%
\bibitem [{\citenamefont {Chen}\ \emph {et~al.}(2013)\citenamefont {Chen}, \citenamefont {Yang}, \citenamefont {Wang}, \citenamefont {Imai}, \citenamefont {Ohta}, \citenamefont {Michioka}, \citenamefont {Yoshimura},\ and\ \citenamefont {Fang}}]{doi:10.7566/JPSJ.82.124711}%
  \BibitemOpen
  \bibfield  {author} {\bibinfo {author} {\bibfnamefont {B.}~\bibnamefont {Chen}}, \bibinfo {author} {\bibfnamefont {J.}~\bibnamefont {Yang}}, \bibinfo {author} {\bibfnamefont {H.}~\bibnamefont {Wang}}, \bibinfo {author} {\bibfnamefont {M.}~\bibnamefont {Imai}}, \bibinfo {author} {\bibfnamefont {H.}~\bibnamefont {Ohta}}, \bibinfo {author} {\bibfnamefont {C.}~\bibnamefont {Michioka}}, \bibinfo {author} {\bibfnamefont {K.}~\bibnamefont {Yoshimura}},\ and\ \bibinfo {author} {\bibfnamefont {M.}~\bibnamefont {Fang}},\ }\bibfield  {title} {\bibinfo {title} {Magnetic properties of layered itinerant electron ferromagnet {Fe$_3$GeTe$_2$}},\ }\href {https://doi.org/10.7566/JPSJ.82.124711} {\bibfield  {journal} {\bibinfo  {journal} {Journal of the Physical Society of Japan}\ }\textbf {\bibinfo {volume} {82}},\ \bibinfo {pages} {124711} (\bibinfo {year} {2013})}\BibitemShut {NoStop}%
\bibitem [{\citenamefont {Belashchenko}\ \emph {et~al.}(2020)\citenamefont {Belashchenko}, \citenamefont {Kovalev},\ and\ \citenamefont {van Schilfgaarde}}]{PhysRevB.101.020407}%
  \BibitemOpen
  \bibfield  {author} {\bibinfo {author} {\bibfnamefont {K.~D.}\ \bibnamefont {Belashchenko}}, \bibinfo {author} {\bibfnamefont {A.~A.}\ \bibnamefont {Kovalev}},\ and\ \bibinfo {author} {\bibfnamefont {M.}~\bibnamefont {van Schilfgaarde}},\ }\bibfield  {title} {\bibinfo {title} {Interfacial contributions to spin-orbit torque and magnetoresistance in ferromagnet/heavy-metal bilayers},\ }\href {https://doi.org/10.1103/PhysRevB.101.020407} {\bibfield  {journal} {\bibinfo  {journal} {Phys. Rev. B}\ }\textbf {\bibinfo {volume} {101}},\ \bibinfo {pages} {020407(R)} (\bibinfo {year} {2020})}\BibitemShut {NoStop}%
\bibitem [{\citenamefont {Marzari}\ \emph {et~al.}(2012)\citenamefont {Marzari}, \citenamefont {Mostofi}, \citenamefont {Yates}, \citenamefont {Souza},\ and\ \citenamefont {Vanderbilt}}]{RevModPhys.84.1419}%
  \BibitemOpen
  \bibfield  {author} {\bibinfo {author} {\bibfnamefont {N.}~\bibnamefont {Marzari}}, \bibinfo {author} {\bibfnamefont {A.~A.}\ \bibnamefont {Mostofi}}, \bibinfo {author} {\bibfnamefont {J.~R.}\ \bibnamefont {Yates}}, \bibinfo {author} {\bibfnamefont {I.}~\bibnamefont {Souza}},\ and\ \bibinfo {author} {\bibfnamefont {D.}~\bibnamefont {Vanderbilt}},\ }\bibfield  {title} {\bibinfo {title} {Maximally localized wannier functions: Theory and applications},\ }\href {https://doi.org/10.1103/RevModPhys.84.1419} {\bibfield  {journal} {\bibinfo  {journal} {Rev. Mod. Phys.}\ }\textbf {\bibinfo {volume} {84}},\ \bibinfo {pages} {1419} (\bibinfo {year} {2012})}\BibitemShut {NoStop}%
\bibitem [{\citenamefont {Yates}\ \emph {et~al.}(2007)\citenamefont {Yates}, \citenamefont {Wang}, \citenamefont {Vanderbilt},\ and\ \citenamefont {Souza}}]{PhysRevB.75.195121}%
  \BibitemOpen
  \bibfield  {author} {\bibinfo {author} {\bibfnamefont {J.~R.}\ \bibnamefont {Yates}}, \bibinfo {author} {\bibfnamefont {X.}~\bibnamefont {Wang}}, \bibinfo {author} {\bibfnamefont {D.}~\bibnamefont {Vanderbilt}},\ and\ \bibinfo {author} {\bibfnamefont {I.}~\bibnamefont {Souza}},\ }\bibfield  {title} {\bibinfo {title} {Spectral and fermi surface properties from wannier interpolation},\ }\href {https://doi.org/10.1103/PhysRevB.75.195121} {\bibfield  {journal} {\bibinfo  {journal} {Phys. Rev. B}\ }\textbf {\bibinfo {volume} {75}},\ \bibinfo {pages} {195121} (\bibinfo {year} {2007})}\BibitemShut {NoStop}%
\bibitem [{\citenamefont {Vanderbilt}(2018)}]{Vanderbilt_2018}%
  \BibitemOpen
  \bibfield  {author} {\bibinfo {author} {\bibfnamefont {D.}~\bibnamefont {Vanderbilt}},\ }\href@noop {} {\emph {\bibinfo {title} {Berry Phases in Electronic Structure Theory: Electric Polarization, Orbital Magnetization and Topological Insulators}}}\ (\bibinfo  {publisher} {Cambridge University Press},\ \bibinfo {year} {2018})\BibitemShut {NoStop}%
\bibitem [{\citenamefont {Cole}\ \emph {et~al.}(2025)\citenamefont {Cole}, \citenamefont {Coh},\ and\ \citenamefont {Vanderbilt}}]{Cole_Python_Tight_Binding_2025}%
  \BibitemOpen
  \bibfield  {author} {\bibinfo {author} {\bibfnamefont {T.}~\bibnamefont {Cole}}, \bibinfo {author} {\bibfnamefont {S.}~\bibnamefont {Coh}},\ and\ \bibinfo {author} {\bibfnamefont {D.}~\bibnamefont {Vanderbilt}},\ }\href {https://doi.org/10.5281/zenodo.12721315} {\bibinfo {title} {{Python Tight Binding (PythTB)}}} (\bibinfo {year} {2025})\BibitemShut {NoStop}%
\bibitem [{\citenamefont {Wang}\ \emph {et~al.}(2006)\citenamefont {Wang}, \citenamefont {Yates}, \citenamefont {Souza},\ and\ \citenamefont {Vanderbilt}}]{PhysRevB.74.195118}%
  \BibitemOpen
  \bibfield  {author} {\bibinfo {author} {\bibfnamefont {X.}~\bibnamefont {Wang}}, \bibinfo {author} {\bibfnamefont {J.~R.}\ \bibnamefont {Yates}}, \bibinfo {author} {\bibfnamefont {I.}~\bibnamefont {Souza}},\ and\ \bibinfo {author} {\bibfnamefont {D.}~\bibnamefont {Vanderbilt}},\ }\bibfield  {title} {\bibinfo {title} {Ab initio calculation of the anomalous hall conductivity by wannier interpolation},\ }\href {https://doi.org/10.1103/PhysRevB.74.195118} {\bibfield  {journal} {\bibinfo  {journal} {Phys. Rev. B}\ }\textbf {\bibinfo {volume} {74}},\ \bibinfo {pages} {195118} (\bibinfo {year} {2006})}\BibitemShut {NoStop}%
\bibitem [{\citenamefont {Go}\ \emph {et~al.}(2024)\citenamefont {Go}, \citenamefont {Lee}, \citenamefont {Oppeneer}, \citenamefont {Bl\"ugel},\ and\ \citenamefont {Mokrousov}}]{PhysRevB.109.174435}%
  \BibitemOpen
  \bibfield  {author} {\bibinfo {author} {\bibfnamefont {D.}~\bibnamefont {Go}}, \bibinfo {author} {\bibfnamefont {H.-W.}\ \bibnamefont {Lee}}, \bibinfo {author} {\bibfnamefont {P.~M.}\ \bibnamefont {Oppeneer}}, \bibinfo {author} {\bibfnamefont {S.}~\bibnamefont {Bl\"ugel}},\ and\ \bibinfo {author} {\bibfnamefont {Y.}~\bibnamefont {Mokrousov}},\ }\bibfield  {title} {\bibinfo {title} {First-principles calculation of orbital hall effect by wannier interpolation: Role of orbital dependence of the anomalous position},\ }\href {https://doi.org/10.1103/PhysRevB.109.174435} {\bibfield  {journal} {\bibinfo  {journal} {Phys. Rev. B}\ }\textbf {\bibinfo {volume} {109}},\ \bibinfo {pages} {174435} (\bibinfo {year} {2024})}\BibitemShut {NoStop}%
\end{thebibliography}
%

\end{document}